\documentclass[12pt]{article}
\pdfoutput=1

\usepackage{putex}
\usepackage{graphicx}
\usepackage{caption}
\usepackage{amsmath}
\usepackage{array}
\usepackage{subcaption}
\usepackage{epstopdf}
\usepackage{enumerate}
\usepackage{cite}
\usepackage{youngtab}
\usepackage{tensor}
\usepackage{slashed}
\usepackage[aligntableaux=center]{ytableau}
\usepackage[utf8]{inputenc}
\usepackage{rotating} 
\usepackage{multirow}
\usepackage{bigfoot}
\usepackage[
      colorlinks=true,
      linkcolor=blue,
      urlcolor=blue,
      filecolor=black,
      citecolor=red,
      ]{hyperref}
\usepackage{dsfont}
\usepackage{simpler-wick}
\usepackage{braket}

\newcommand{\abs}[1]{\left\lvert #1 \right\rvert}

\newcommand {\be} {\begin {equation}}
\newcommand {\ee} {\end {equation}}

\newcommand {\bes} {\begin {equation*}}
\newcommand {\ees} {\end {equation*}}

\newcommand{\es}[2] {\begin{equation} \label{#1} \begin{split} #2 \end{split} \end{equation}}

\newcommand{\Z}{\mathbb{Z}}

\newcommand{\R}{\mathbb{R}}

\newcommand{\cD}{{\mathcal D}}

\newcommand{\cO}{{\mathcal O}}

\newcommand{\cM}{{\mathcal M}}

\newcommand{\vphi}{\varphi}
\newcommand{\tX}{\widetilde{X}}

\newcommand{\beq}{\begin{equation}}
\newcommand{\eeq}{\end{equation}}

\newcommand{\red}[1]{\textcolor{red}{#1}}

\def\ie{\begin{equation}\begin{aligned}}
\def\fe{\end{aligned}\end{equation}}

\numberwithin{equation}{section}

\def\<{\langle}
\def\>{\rangle}

\allowdisplaybreaks

\begin{document}

\preprint{PUPT-2597}

\institution{HU}{Jefferson Physical Laboratory, Harvard University, Cambridge, MA 02138, USA }
\institution{Exile}{Department of Particle Physics and Astrophysics, Weizmann Institute of Science, Rehovot, Israel}
\institution{PU}{Joseph Henry Laboratories, Princeton University, Princeton, NJ 08544, USA}

\title{
The M-theory Archipelago
}

\authors{Nathan B.~Agmon,\worksat{\HU} Shai M.~Chester,\worksat{\Exile} and Silviu S.~Pufu\worksat{\PU}}

\abstract{

We combine supersymmetric localization results and the numerical conformal bootstrap technique to study the 3d maximally supersymmetric (${\cal N} = 8$) CFT on $N$ coincident M2-branes (the $U(N)_k \times U(N)_{-k}$ ABJM theory at Chern-Simons level $k=1$).  In particular, we perform a mixed correlator bootstrap study of the superconformal primaries of the stress tensor multiplet and of the next possible lowest-dimension half-BPS multiplet that is allowed by 3d ${\cal N} = 8$ superconformal symmetry.  Of all known 3d ${\cal N} = 8$ SCFTs,  the $k=1$ ABJM theory is the only one that contains both types of multiplets in its operator spectrum.  By imposing the values of the short OPE coefficients that can be computed exactly using supersymmetric localization, we are able to derive precise islands in the space of semi-short OPE coefficients for an infinite number of such coefficients.  We find that these islands decrease in size with increasing $N$.  More generally, we also analyze 3d ${\cal N} = 8$ SCFT that contain both aforementioned multiplets in their operator spectra without inputing any additional information that is specific to ABJM theory.  For such theories, we compute upper and lower bounds on the semi-short OPE coefficients as well as upper bounds on the scaling dimension of the lowest unprotected scalar operator. These latter bounds are more constraining than the analogous bounds previously derived from a single correlator bootstrap of the stress tensor multiplet.  This leads us to conjecture that the $U(N)_2 \times U(N+1)_{-2}$ ABJ theory, and not the $k=1$ ABJM theory, saturates the single correlator bounds.
}
\date{July 2019}

\maketitle

\tableofcontents

\section{Introduction and summary}

The conformal bootstrap \cite{Polyakov:1974gs,Ferrara:1973yt,Mack:1975jr} is a non-perturbative method that places rigorous numerical bounds \cite{Rattazzi:2008pe} on scaling dimensions and operator product expansion (OPE) coefficients that appear in a given four-point function of a conformal field theory (CFT) (for reviews, see~\cite{Rychkov:2016iqz,Simmons-Duffin:2016gjk,Poland:2018epd,Chester:2019wfx,Qualls:2015qjb}). In the best scenario, these bounds form precise islands in the space of a small number of CFT data that contain the values of a single known CFT, as was originally shown for the 3d Ising model \cite{Kos:2014bka,Simmons-Duffin:2015qma,Kos:2016ysd} and later generalized to the 3d $O(N)$ models \cite{Kos:2015mba,Kos:2016ysd} and $\mathcal{N}=1$ Ising model \cite{Rong:2018okz,Atanasov:2018kqw}. In these cases, the small allowed islands were found in the space of scaling dimensions of the two lowest-lying scalar operators which, in these theories, are the only relevant operators in their corresponding global symmetry charge sectors.  For instance, in the case of the Ising model, these operators were the lowest $\Z_2$-even and $\Z_2$-odd operators. Performing a mixed correlator bootstrap study was necessary because it is not possible to access both charge sectors from the four-point function of a single relevant operator.\footnote{Scaling dimension islands for these and other theories can be found even from single correlators by imposing gaps in multiple operators that appear in the OPE \cite{Li:2017kck}, such as spin 1 operators. Islands have also been found for other theories by considering mixed correlators and imposing gaps on both relevant and irrelevant operators \cite{Li:2016wdp,Kousvos:2018rhl}. However, all these cases involve gaps that, while in many cases plausible, are not rigorously justified.}

In this paper, we obtain islands and (in principle) infinite amount of CFT data for 3d maximally supersymmetric ($\mathcal{N}=8$) superconformal field theories (SCFTs) by also studying a mixed correlator system.  (Before doing so, we will also review the kinds of islands that can be obtained from a single correlator bootstrap study.)  All ${\cal N} = 8$ SCFTs may contain $1/2$-BPS scalar operators $S_p$ with scaling dimension $\Delta = p/2$ that transform in the $[00p0]$ irreducible representation of the $\mathfrak{so}(8)$ R-symmetry.  Out of them, $S_1$ must be a free scalar and it thus must decouple from the rest of the SCFT, and $S_2$ represents the bottom component of the stress tensor multiplet.  In general, a given ${\cal N} = 8$ SCFT may not have operators belonging to all of these multiplets, or it may have multiple linearly independent $1/2$-BPS operators with the same value of $p$.  All local SCFTs, however, do have a unique stress tensor operator and thus a unique $S_2$ operator.\footnote{It can be shown by taking OPEs of $S_2$ with itself any number of times, one can also construct operators $S_p$ for all even $p$.}  In this paper we will restrict our attention to ${\cal N} = 8$ SCFTs that contain an $S_3$ operator and examine the mixed four-point functions of $S_2$ and $S_3$.  As will be explained in more detail shortly, the $S_3$ operators do not exist in all known 3d ${\cal N} = 8$ SCFTs, but they do exist in the theories on $N$ coincident M2-branes in flat space.

Since $S_2$ and $S_3$ have fixed scaling dimensions $1$ and $3/2$, respectively, we cannot find islands in the space of scaling dimensions as was done in the other cases mentioned above.  Instead, we find small islands in the space of the OPE coefficients of low-lying protected eighth and sixteenth-BPS semishort operators that appear in the $S_2 \times S_2$, $S_2 \times S_3$, and $S_3 \times S_3$ OPEs.   
Both upper and lower bounds can be computed for these quantities because the scaling dimensions of these operators are fixed to distinct values by supersymmetry, unlike the theories discussed above where scaling dimensions are not protected so only upper bounds on OPE coefficients could be computed without additional assumptions.\footnote{Lower bounds can be computed for the OPE coefficients of the operators that have been assumed to be relevant, since there is now a gap between them and the continuum of irrelevant operators. Islands in the space of these OPE coefficients are reported in \cite{Kos:2016ysd,Atanasov:2018kqw}.} These operators appear in the OPE for an infinite number of spins, so islands for an infinite number of such operators can be derived in principle.

More generally, protected operators appear generically in four point functions of BPS operators in SCFTs with at least four supercharges, and upper and lower bounds on their OPE coefficients can be computed by considering just a single correlator, as was shown originally for the 4d $\mathcal{N}=1$ bootstrap \cite{Poland:2011ey} and later for bootstrap studies in other dimensions with different amounts of supersymmetry \cite{Bobev:2015jxa,Baggio:2017mas,Chester:2015lej,Berkooz:2014yda,Li:2017ddj,Chang:2017xmr,Chang:2017cdx,Beem:2015aoa,Chester:2015qca,Beem:2014zpa,Lemos:2015awa,Cornagliotto:2017snu,Liendo:2018ukf,Gimenez-Grau:2019hez}.\footnote{Note that for 4d SCFTs with $\mathcal{N}\geq3$, all the protected operators that appear in the four-point function of half-BPS operator are already fixed by the 2d chiral algebra \cite{Beem:2013sza}, so the bootstrap studies \cite{Beem:2013qxa,Beem:2016wfs,Lemos:2016xke} could not compute any upper bounds on OPE coefficients.} In these cases, the bounds are not very constraining, however, and so do not lead to precise islands such as the 3d Ising or $O(N)$ islands in scaling dimension space mentioned above.  For 3d $\mathcal{N}=8$ theories, on the other hand, the upper and lower bounds that were computed for semishort OPE coefficients previously using only the single correlator $\langle S_2S_2S_2S_2\rangle$ \cite{Chester:2014mea} are already very constraining. For instance, the bounds on the OPE coefficients of the spin 0 and 2 eighth-BPS semishort operators for the $k=3$ BLG theory that were computed in that study with $\Lambda=19$ have a percent error\footnote{Percent error as computed, for instance, in \eqref{error}.} of $0.5\%$ and $0.3\%$, respectively, which is comparable to the 0.12\% and 0.23\% percent error reported for the island in the smallest scaling dimension operator for the $O(2)$ and $O(3)$ models, as computed using $\Lambda=35$ and scanning over OPE coefficients in the best available study in \cite{Kos:2016ysd}.\footnote{Here, $\Lambda$ is a parameter counting the number of derivatives in the functionals used for numerical bootstrap.  See \cite{Chester:2014fya} for the precise definition.}  This infinite number of islands could be computed for infinitely many different 3d $\mathcal{N}=8$ theories.

While the single correlator OPE coefficient islands are already very precise, the mixed correlator system that is the primary focus of this paper has three advantages:
\begin{enumerate}
\item In principle, the free multiplet $S_1$ is allowed to appear in the mixed OPE $S_2\times S_3$, so we can exclude the presence of a free sector by setting its OPE coefficient to zero.
\item As mentioned above, only certain 3d $\mathcal{N}=8$ SCFTs contain the operator $S_3$, unlike $S_2$ which appears in all theories with a stress tensor, so we automatically restrict the space of theories by considering correlators of $S_3$.
\item Ten protected half and quarter-BPS short scalar operators appear in the mixed system, while only three appear in $\langle S_2S_2S_2S_2\rangle$. The OPE coefficients of these operators can be computed for a given theory using the 1d topological sector \cite{Dedushenko:2016jxl} and then imposed on the 3d bootstrap, so having access to more such operators makes the bootstrap much more constraining.
\end{enumerate}

Let us now provide more background and summarize our results.  There are only a few known infinite families of ${\cal N}=8$ SCFTs, and they can all be realized (in ${\cal N } = 3$ SUSY notation) as Chern-Simons (CS) theories with a product gauge group $G_1\times G_2$ coupled to two matter hypermultiplets transforming in the bifundamental representation. These families are:  the $SU(2)_k \times SU(2)_{-k}$ and $(SU(2)_k \times SU(2)_{-k})/\mathbb{Z}_2$ reformulations \cite{VanRaamsdonk:2008ft, Bandres:2008vf} of the theories of Bagger-Lambert-Gustavsson (BLG) \cite{Bagger:2007vi,Bagger:2007jr,Bagger:2006sk, Gustavsson:2007vu}, which are indexed by an arbitrary integer Chern-Simons level $k$;  the $U(N)_k \times U(N)_{-k}$ theories of Aharony-Bergman-Jafferis-Maldacena (ABJM) \cite{Aharony:2008ug}, which we denote as ABJM$_{N,k}$ for integer $N$ and $k = 1, 2$; and the $U(N+1)_2 \times U(N)_{-2}$ theories \cite{Bashkirov:2011pt} of Aharony-Bergman-Jafferis (ABJ) \cite{Aharony:2008gk}, which are labeled by the integer $N$. The ABJ(M) theories can be interpreted as effective theories on $N$ coincident M2-branes placed at a $\mathbb{C}^4/\mathbb{Z}_k$ singularity in the transverse directions, so that when $N\to\infty$ they contain a sector described by weakly coupled supergravity.\footnote{The BLG theories, in contrast, do not have a known M-theory interpretation except when $k\leq4$.  For each $k \leq 4$, BLG theory is dual to a theory of ABJ(M) type \cite{Agmon:2017lga,Lambert:2010ji,Bashkirov:2011pt,Bashkirov:2012rf}.} When $N>1$, ABJM$_{N, 1}$ flows to two decoupled SCFTs in the infrared: a free SCFT with eight massless real scalars and eight Majorana fermions that is isomorphic to ABJM$_{1, 1}$, and a strongly coupled interacting SCFT that we denote as ABJM$_{N,1}^\text{int}$. The only known interacting theories that contain an $S_3$ operator are ABJM$_{N,1}^\text{int}$ with $N\geq3$, ABJM$_{3,2}$, and the $(SU(2)_3 \times SU(2)_{-3})/\mathbb{Z}_2$ BLG theory, where the last two theories are dual to ABJM$_{4,1}^\text{int}$ \cite{Gang:2011xp} and ABJM$_{3,1}^\text{int}$ \cite{Agmon:2017lga}, respectively.  Thus, we need only consider ABJM$_{N,1}^\text{int}$ in this work. 

In general, 3d $\mathcal{N}=8$ SCFTs are most conveniently parameterized by $c_T$, which is defined as the coefficient appearing in the two-point function of the canonically-normalized stress-tensor, 
 \es{CanStress}{
  \langle T_{\mu\nu}(\vec{x}) T_{\rho \sigma}(0) \rangle = \frac{c_T}{64} \left(P_{\mu\rho} P_{\nu \sigma} + P_{\nu \rho} P_{\mu \sigma} - P_{\mu\nu} P_{\rho\sigma} \right) \frac{1}{16 \pi^2 \vec{x}^2} \,, \qquad P_{\mu\nu} \equiv \eta_{\mu\nu} \nabla^2 - \partial_\mu \partial_\nu \,.
 }
This coefficient can be computed exactly using supersymmetric localization for any ${\cal N} \geq 2$ SCFT with a Lagrangian description (see \cite{Closset:2012ru} and \cite{Imamura:2011wg}).  In \eqref{CanStress}, $c_T$ is normalized such that it equals $1$ for a (non-supersymmetric) free massless real scalar or a free massless Majorana fermion.   Thus, $c_T = 16$ for the free ${\cal N} = 8$ theory of eight massless real scalars and eight massless Majorana fermions (equivalent to ABJM$_{1, 1}$), and 
\es{cTlargeN}{
  c_T \approx \frac{64}{3\pi}\sqrt{2k}N^{3/2}
}
for ABJ(M) theory at large $N$.  We will use $N$ and $c_T$ interchangeably to specify ABJM$_{N,1}^\text{int}$. 

\begin{figure}[]
\begin{center}
        \includegraphics[width=\textwidth]{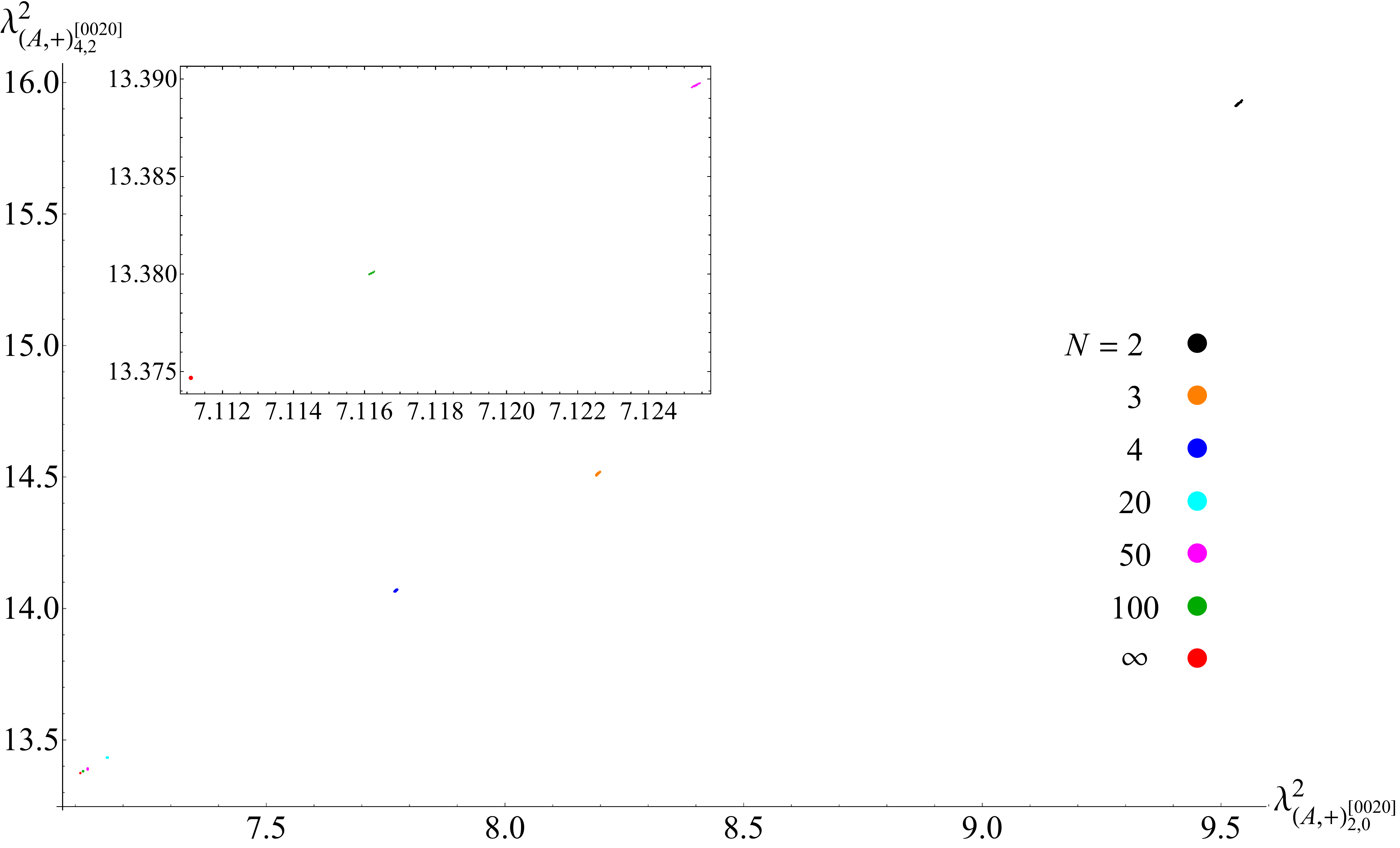}
\caption{Islands in the space of the semi-short OPE coefficients $\lambda_{(A, +)_{2,0}^{[0020]}}^2$, $\lambda_{(A, +)_{4,2}^{[0020]}}^2$ (to be defined precisely later) for ABJM$_{N,1}^\text{int}$ for $N=2,3,4,20,50,100$, where orange is allowed.   These bounds are derived from the single correlator $\langle S_2 S_2 S_2 S_2\rangle$ with  the short OPE coefficients fixed to their ABJM$_{N,1}^\text{int}$ values using the 1d theory in Section \ref{1d} and \cite{Chester:2014mea} for $N=2,3,4$, and from the all orders in $1/N$ formulae in \cite{Agmon:2017xes} for $N=20,50,100$. The red denotes the $N\to\infty$ GFFT values in Table \ref{Avalues}. Zoomed-in plots will be presented in Figure~\ref{islandsS}.}
\label{islandsCombined}
\end{center}
\end{figure}

The conformal bootstrap was first applied to 3d $\mathcal{N}=8$ SCFTs in \cite{Chester:2014fya}, which used the single correlator $\langle S_2S_2S_2S_2\rangle$ to derive bounds on CFT data as a function of $c_T$, which itself was bounded as $c_T\gtrsim15.35$ at $\Lambda = 19$, so that in the infinite $\Lambda$ limit this bound is likely saturated by the free theory. Upper bounds were derived on scaling dimensions of unprotected operators, which showed a kink at the value $c_T\approx22.8$ that, to our knowledge, does not correspond to any known theory. Upper and lower bounds were also derived for OPE coefficients of the two short scalars with $\Delta=2$, both of which also showed kinks at $c_T\approx22.8$, and one of which went to zero at this value. Next, in \cite{Chester:2014mea} it was shown that these scalar short operators are described by a 1d theory that relates their OPE coefficients, so that there is only one independent short OPE coefficient in terms of $c_T$, and in fact the numerical bounds on the two short scalar operators are exactly related by the 1d theory relation. Upper and lower bounds were also computed for several of infinitely many semishort operators, and these bounds also showed a kink at $c_T\approx22.8$. Most recently, in \cite{Agmon:2017xes} the unique independent short operator OPE coefficient was computed for the ABJ(M) and BLG theories by relating it to derivatives of the mass deformed sphere free energy, which was computed using localization \cite{Kapustin:2009kz} exactly at small $N$ and to all orders in $1/N$ using the Fermi gas method \cite{Marino:2011eh,Nosaka:2015iiw}. This analytic formula was found to come close to saturating the numerical bounds for all interacting ABJ(M) theories. For whichever of these theories saturates the bound in the infinite precision limit, all the CFT data in $\langle S_2S_2S_2S_2\rangle$, both protected and unprotected, could then be numerically determined using the extremal functional method \cite{Poland:2010wg,ElShowk:2012hu,El-Showk:2014dwa}. To leading order in $1/c_T$, where all ABJ(M) theories are indistinguishable, the bootstrap predictions were verified by supergravity calculations of lowest twist unprotected operators \cite{Zhou:2017zaw,Chester:2018lbz} and even more non-trivially subleading twist unprotected operators \cite{Chester:2018lbz}.  Lastly, inputting the values of $c_T$ and of the short OPE coefficient that can be computed from the 1d theory, one can obtain very precise islands of allowed regions for semi-short OPE coefficients, as we show in Figure~\ref{islandsCombined}.  (See also Figure~\ref{islandsS} for zoomed-in versions of this plot.)

As mentioned already, in this work we study mixed correlators of $S_2$ and $S_3$. We find that the 1d theory for these correlators again relates the OPE coefficients of short scalar operators, so that there are five such independent quantities in terms of $c_T$, instead of just the one that appeared for $\langle S_2S_2S_2S_2\rangle$. After imposing these relations and setting the free theory OPE coefficient to zero, we numerically bound CFT data as a function of $c_T$ for a general interacting 3d $\mathcal{N}=8$ SCFT. We find that $c_T$ itself is now bounded as $c_T\gtrsim22.8$, which is curiously the same value were we observed a kink in the $\langle S_2S_2S_2S_2\rangle$ bounds. We compute upper and lower bounds for OPE coefficients of short and semishort operators that did not appear in $\langle S_2S_2S_2S_2\rangle$, as well as upper bounds on the scaling dimension of the lowest dimension unprotected scalar operator that also appeared in $\langle S_2S_2S_2S_2\rangle$. For the latter, we find that the mixed bound is more restrictive than the $\langle S_2S_2S_2S_2\rangle$ bound, which suggests that the ABJ(M) theory that conjecturally saturated that bound, and could thus be studied using the extremal functional method, must be one of the $k=2$ theories that does not appear in our mixed correlator study. This leaves two possibilities: the $U(N)_2\times U(N)_{-2}$ ABJM theory or the $U(N+1)_2\times U(N)_{-2}$ ABJ theory. When $N=2$, the former becomes a product of two ABJ with $N=1$ theories, and so does cannot lie on the boundary of the allowed region \cite{Agmon:2017xes}, while when $N=4$, this theory is dual to ABJM$_{4,1}^\text{int}$ and so also cannot saturate the bound, which suggests that ABJ is the theory saturating the bound. We also found that the mixed correlator scaling dimension bound now goes to $\Delta=1$ at $c_T\sim22.8$, which suggests that if any theory lives at that point it must include a free sector. 

Next, we restrict to ABJM$_{N,1}^\text{int}$ by imposing the values of OPE coefficients of short scalar operators. We cannot relate correlators of $S_3$ to the mass deformed free energy, as we did for $\langle S_2S_2S_2S_2\rangle$, so instead we compute these OPE coefficients using the 1d theory Lagrangian of ABJM$_{N,1}$ \cite{Dedushenko:2016jxl}, as was previously done for correlators of $S_2$ and $S_4$ in \cite{Agmon:2017lga}. The calculation involves a number of integrals that grow with $N$, so we only give results for $N=3,4$. After imposing these values, we compute precise islands in the space of OPE coefficients of semishort operators, and in several cases are able to determine these values to less than a percent!

The rest of this paper is organized as follows. In Section~\ref{four3d}, we discuss the constraints of 3d $\mathcal{N}=8$ superconformal symmetry on four point functions of half-BPS operators, and derive explicit superblocks for correlators of $S_2$ and $S_3$ (which we include in an attached \texttt{Mathematica} file).  In Section~\ref{1d}, we derive the 1d theory relations for these correlators for general 3d $\mathcal{N}=8$ SCFTs, and then compute the short scalar OPE coefficients for ABJM$_{3,1}^\text{int}$ and ABJM$_{4,1}^\text{int}$. In Section~\ref{numerics}, we derive the crossing relations for our mixed systems and compute numerical bounds both for general interacting 3d $\mathcal{N}=8$ SCFTs, and for ABJM$_{3,1}^\text{int}$ and ABJM$_{4,1}^\text{int}$ by imposing the OPE coefficients of the previous section. Finally, in Section~\ref{disc}, we end with a discussion of our results and of future directions.

\section{Half-BPS four-point functions in 3d}
\label{four3d}

We begin by discussing the constraints of the 3d $\mathcal{N}=8$ superconformal algebra $\mathfrak{osp}(8|4)\supset\mathfrak{so}(5)\oplus\mathfrak{so}(8)_R$ on 4-point functions of the bottom component of half-BPS supermultiplets. The results of this section are are not restricted to ABJM theory, but apply to any $\mathcal{N}=8$ SCFT that contains these half-BPS operators.

\subsection{Setup}
\label{setup}

We consider half-BPS superconformal primaries $S_k$ in 3d $\mathcal{N}=8$ SCFTs that are scalars with $\Delta=\frac{k}{2}$ and transform in the $[00k0]$ of $\mathfrak{so}(8)_R$,\footnote{The convention we use in defining these multiplets is that the supercharges transform in the ${\bf 8}_v = [1000]$ irrep of $\mathfrak{so}(8)_R$.}  where $k=1,2,\dots$. We will focus on the lowest three operator $S_k$ with $k=1,2,3$. $S_1$ is a free scalar with $\Delta=\frac12$ in the ${\bf8}_c$ of $\mathfrak{so}(8)_R$, which does not exist in an interacting theory. $S_2$ is a scalar with $\Delta=1$ in the ${\bf35}_c$ of $\mathfrak{so}(8)_R$, which is the bottom of the conserved stress tensor multiplet and so must exist in all local 3d $\mathcal{N}=8$ SCFTs. Lastly, $S_3$ is a scalar with $\Delta=\frac32$ in the ${\bf112}_c$ of $\mathfrak{so}(8)_R$, which need not exist in a general 3d $\mathcal{N}=8$ SCFT. 

We are interested in 4-point functions of $S_{k_i}$ for $i=1,2,3,4$.   Since the $[00k_i 0]$ irrep of $\mathfrak{so}(8)$ can be viewed as a rank-$k_i$ symmetric traceless product of the ${\bf8}_c$, we can represent the corresponding operators as traceless symmetric tensors $S_{I_1\dots I_{k_i}}( \vec x_i)$ of $\mathfrak{so}(8)_R$, where $I=1,\dots8$. It is convenient to contract with an auxiliary polarization vector $Y_i^I$ that is constrained to be null, $Y_i\cdot Y_i=0$, so that
\es{S}{
S_{k_i}( \vec x_i,Y_i)\equiv S_{I_1\dots I_{k_i}}(\vec x_i)Y^{I_1}\cdots Y_i^{I_{k_i}}\,.
}
Conformal and $\mathfrak{so}(8)_R$ symmetry then imply that four point functions of $S_{k_i}( \vec x_i,Y_i)$ take the form
 \es{FourPoint2}{
  &\langle S_{k_1}(\vec{x}_1, Y_1) S_{k_2}(\vec{x}_2, Y_2) S_{k_3}(\vec{x}_3, Y_3) S_{k_4}(\vec{x}_4, Y_4)\rangle =
    \frac{( Y_1\cdot Y_2 )^{\frac{k_1+k_2}{2}}  ( Y_3\cdot Y_4 )^{\frac{k_3+k_4}{2}} }
     {\abs{\vec{x}_{12}}^{\frac{k_1 + k_2}{2}}\abs{\vec{x}_{34}}^{\frac{k_3+ k_4}{2}} }\\
     &\times
     \left(\frac{Y_1\cdot Y_4}{Y_2\cdot Y_4}\right)^{\frac{k_{12}}{2}}
     \left(\frac{(Y_1\cdot Y_2)(Y_3\cdot Y_4)}{(Y_1\cdot Y_4)(Y_2\cdot Y_4)}\right)^{\frac{k_{34}}{2}}
      \left( \frac{\abs{\vec{x}_{14}}\abs{\vec{x}_{24}}}{\abs{\vec{x}_{12}} \abs{\vec{x}_{34}}}\right)^{\frac{k_{34}}{2}} \mathcal{G}_{k_1k_2k_3k_4}(U,V;\sigma,\tau)\,,
 } 
where $k_{ij}\equiv k_i-k_j$, $\vec x_{ij}\equiv \vec x_i-\vec x_j$, and the conformally-invariant cross ratios $U,V$ and $\mathfrak{so}(8)_R$ invariants $\sigma,\tau$ are 
 \es{uvsigmatauDefs}{
  U \equiv \frac{{x}_{12}^2 {x}_{34}^2}{{x}_{13}^2 {x}_{24}^2} \,, \qquad
   V \equiv \frac{{x}_{14}^2 {x}_{23}^2}{{x}_{13}^2 {x}_{24}^2}  \,, \qquad
   \sigma\equiv\frac{(Y_1\cdot Y_3)(Y_2\cdot Y_4)}{(Y_1\cdot Y_2)(Y_3\cdot Y_4)}\,,\qquad \tau\equiv\frac{(Y_1\cdot Y_4)(Y_2\cdot Y_3)}{(Y_1\cdot Y_2)(Y_3\cdot Y_4)} \,.
 }
Since \eqref{FourPoint2} is a degree $k_i$ polynomial in each $Y_i$, the quantity $\mathcal{G}_{k_1k_2k_3k_4}(U,V;\sigma,\tau)$ is a degree $\min\{k_i\}$ polynomial in $\sigma,\tau$. It is convenient to parametrize these polynomials in terms of eigenfunctions $Y^{k_{12},k_{34}}_{nm}(\sigma, \tau)$ of the $\mathfrak{so}(8)_R$ quadratic Casimir for irreps $[0\,n-m\, 2m\,0]$ of maximal degree $\min\{k_i\}$ that appear in the $s$-channel of \eqref{FourPoint2}. These irreps must appear in both $[00k_1 0] \otimes [00k_20]$ and $[00k_3 0] \otimes [00k_40]$, where each tensor product is given by
\es{tens0}{
[00k_10]\otimes[00k_20]=\bigoplus^{\frac{k_1+k_2}{2}}_{n=\frac{|k_{12}|}{2}}\bigoplus_{m=\frac{|k_{12}|}{2}}^{n}[0\,n-m\,2m\,0]\,.
} 
The polynomials $Y^{k_{12},k_{34}}_{nm}(\sigma, \tau)$ can be computed explicitly as shown in Appendix D of \cite{Nirschl:2004pa}; we give a few examples in Appendix \ref{Ys}. We can then expand $\mathcal{G}_{k_1k_2k_3k_4}(U,V;\sigma,\tau)$ in terms of this basis as
\es{Ybasis}{
\mathcal{G}_{k_1k_2k_3k_4}(U,V;\sigma,\tau)=\sum^{\frac{k_1+k_2}{2}}_{n=\frac{|k_{12}|}{2}}\sum_{m=\frac{|k_{12}|}{2}}^{n}Y^{k_{12},k_{34}}_{nm}(\sigma, \tau) A^{k_{12},k_{34}}_{nm}(U,V)\,.
}
By taking the $s$-channel OPEs $S_{k_1}\times S_{k_2}$ and $S_{k_3}\times S_{k_4}$ in \eqref{FourPoint2}, we can expand $A^{k_{12},k_{34}}_{nm}(U,V)$ in conformal blocks $G_{\Delta,\ell}(U,V)$ as
\es{blockExp}{
A^{k_{12},k_{34}}_{nm}(U,V)=U^{\frac{k_{34}}{4}}\sum_{\Delta,\ell}\lambda_{k_1k_2\cO_{\Delta,\ell,nm}}\lambda_{k_3k_4\cO_{\Delta,\ell,nm}}G_{\Delta,\ell}^{k_{12},k_{34}}(U,V)\,,
}
where $\cO_{\Delta,\ell,nm}$ are conformal primaries with scaling dimension $\Delta$ and spin $\ell$ in irrep $[0\,n-m\, 2m\,0]$ that appear in both OPEs $S_{k_1}\times S_{k_2}$ and $S_{k_3}\times S_{k_4}$ with OPE coefficients $\lambda_{k_1k_2\cO_{\Delta,\ell,nm}}$ and $\lambda_{k_3k_4\cO_{\Delta,\ell,nm}}$, respectively. The factor $U^{\frac{k_{34}}{4}}$ is included to match the definition of the conformal blocks in \cite{Baggio:2017mas}, which can be computed recursively as in \cite{Kos:2014bka}\footnote{Our blocks are normalized as $
G^{k_{12},k_{34}}_{\Delta,\ell}(r,\eta)\sim r^{\Delta} P_{\ell}(\eta)$ as $r\to0$, where $P_{\ell}(\eta)$ is a Legendre polynomial. This differs from the normalization of \cite{Kos:2014bka} by a factor of $(-1)^\ell$.} in terms of the variables $r,\eta$  in a small $r$ expansion, where $r,\eta$ are defined as \cite{Hogervorst:2013sma}
\es{reta}{
U=\frac{16r^2}{(1+r^2+2r\eta)^2}\,,\qquad V=\frac{(1+r^2-2r\eta)^2}{(1+r^2+2r\eta)^2}\,.
}

So far, we have imposed the bosonic subgroups of the $\mathfrak{osp}(8|4)$ algebra. The constraints from the fermionic subgroups are captured by the superconformal Ward identities \cite{Dolan:2004mu}:
\es{ward}{
\left[z\partial_z -  \frac\alpha2 \partial_\alpha\right] \mathcal{G}_{k_1k_2k_3k_4}(z,\bar{z};\alpha, \bar{\alpha})|_{\alpha = \frac1z} =
\left[\bar{z}\partial_{\bar{z}} -  \frac{\bar\alpha}{2} \partial_{\bar{\alpha}}\right] \mathcal{G}_{k_1k_2k_3k_4}(z,\bar{z};\alpha, \bar{\alpha})|_{\bar{\alpha}=\frac{1}{\bar{z}}} &= 0\,,
}
where $z,\bar z$ and $\alpha,\bar\alpha$ are written in terms of $U,V$ and $\sigma,\tau$, respectively, as
\es{UVtozzbar}{
U=z\bar z\,,\quad V=(1-z)(1-\bar z)\,,\qquad\qquad \sigma=\alpha\bar\alpha\,,\quad \tau=(1-\alpha)(1-\bar\alpha)\,.
}
These constraints can be satisfied by expanding $ \mathcal{G}_{k_1k_2k_3k_4}$ in superconformal blocks as
\es{SBDecomp}{
      \mathcal{G}_{k_1k_2k_3k_4}(U,V;\sigma,\tau)=\sum_{\mathcal{M}\in(S_{k_1}\times S_{k_2})\cap (S_{k_3}\times S_{k_4})}\lambda_{k_1k_2\mathcal{M}}\lambda_{k_3k_4\mathcal{M}}\mathfrak{G}^{k_{12},k_{34}}_{\mathcal{M}}(U,V;\sigma,\tau)\,,
}
where $\mathfrak{G}^{k_{12},k_{34}}_{\mathcal{M}}$ are superblocks for each supermultiplet $\mathcal{M}$ that appears in both OPEs $S_{k_1}\times S_{k_2}$ and $S_{k_3}\times S_{k_4}$ with OPE coefficients $\lambda_{k_1k_2\mathcal{M}}$ and $\lambda_{k_3k_4\mathcal{M}}$, respectively. Comparing \eqref{SBDecomp} to \eqref{Ybasis} and \eqref{blockExp}, we see that the superblocks are finite linear combination of conformal blocks 
\es{GExpansion}{
\mathfrak{G}^{k_{12},k_{34}}_{\mathcal{M}}=U^{\frac{k_{34}}{4}}\sum^{\frac{k_1+k_2}{2}}_{n=\frac{|k_{12}|}{2}}\sum_{m=\frac{|k_{12}|}{2}}^{n}Y^{k_{12},k_{34}}_{nm}(\sigma, \tau)  \sum_{\cO\in\mathcal{M}}\frac{\lambda_{k_1k_2\cO_{\Delta,\ell,nm}}\lambda_{k_3k_4\cO_{\Delta,\ell,nm}}}{\lambda_{k_1k_2\mathcal{M}}\lambda_{k_3k_4\mathcal{M}}}G_{\Delta,\ell}^{k_{12},k_{34}}(U,V)\,,
}
where $\cO_{\Delta,\ell,nm}$ are conformal primaries that appear in $\mathcal{M}$. For the four-point functions of $\frac12$-BPS operators that we consider, all of the pairs of OPE coefficients $\lambda_{k_1k_2\cO_{\Delta,\ell,nm}}\lambda_{k_3k_4\cO_{\Delta,\ell,nm}}$ can be fixed in terms of the single pair of OPE coefficients $\lambda_{k_1k_2\mathcal{M}}\lambda_{k_3k_4\mathcal{M}}$ in \eqref{SBDecomp} using the Ward identities \eqref{ward}. In the next subsection, we will derive these superblocks explicitly for four-point functions involving $S_2$ and $S_3$. Before we do that, we will first discuss which $\mathcal{M}$ can appear in the OPEs for general $S_{k_1}\times S_{k_2}$.

In general, there are twelve different types of $\mathfrak{osp}(8|4)$ supermultiplets \cite{Dolan:2008vc} that we list in Table~\ref{Multiplets}. 
\begin{table}[htpbp]
\begin{center}
\begin{tabular}{|l|c|c|c|c|}
\hline
 Type     & BPS    & $\Delta$             & Spin & $\mathfrak{so}(8)_R$  \\
 \hline 
 $(A,0)$ (long)      & $0$    & $\ge \Delta_0 + \ell+1$ & $\ell$  & $[a_1 a_2 a_3 a_4]$  \\
 $(A, 1)$  & $1/16$ & $\Delta_0 + \ell +1$    & $\ell$  & $[a_1 a_2 a_3 a_4]$  \\
 $(A, 2)$  & $1/8$  & $\Delta_0 + \ell +1$    & $\ell$  & $[0 a_2 a_3 a_4]$   \\
 $(A, 3)$  & $3/16$  & $\Delta_0 + \ell +1$    & $\ell$  & $[0 0 a_3 a_4]$     \\
 $(A, +)$  & $1/4$  & $\Delta_0 + \ell +1$    & $\ell$  & $[0 0 a_3 0]$       \\
 $(A, -)$  & $1/4$  & $\Delta_0 + \ell +1$    & $\ell$  & $[0 0 0 a_4]$       \\
 $(B, 1)$  & $1/8$  & $\Delta_0$           & $0$  & $[a_1 a_2 a_3 a_4]$ \\
 $(B, 2)$  & $1/4$  & $\Delta_0$           & $0$  & $[0 a_2 a_3 a_4]$   \\
 $(B, 3)$  & $3/8$  & $\Delta_0$                  & $0$  & $[0 0 a_3 a_4]$     \\
 $(B, +)$  & $1/2$  & $\Delta_0$           & $0$  & $[0 0 a_3 0]$       \\
 $(B, -)$  & $1/2$  & $\Delta_0$           & $0$  & $[0 0 0 a_4]$       \\
 conserved & $5/16$  & $\ell+1$                & $\ell$  & $[0 0 0 0]$         \\
 identity & $1$  & $0$                & $0$  & $[0 0 0 0]$         \\
 \hline
\end{tabular}
\end{center}
\caption{Multiplets of $\mathfrak{osp}(8|4)$ and the quantum numbers of their corresponding superconformal primary operator. The conformal dimension $\Delta$ is written in terms of $\Delta_0 \equiv a_1 + a_2 + (a_3 + a_4)/2$.  The Lorentz spin can take the values $\ell=0, 1/2, 1, 3/2, \ldots$.  Representations of the $\mathfrak{so}(8)$ R-symmetry are given in terms of the four $\mathfrak{so}(8)$ Dynkin labels, which are non-negative integers.}
\label{Multiplets}
\end{table}
There are two types of shortening conditions denoted by the $A$ and $B$ families.  The multiplet denoted by $(A, 0)$ is a long multiplet and does not obey any shortening conditions.  The other multiplets of type $A$ have the property that certain $\mathfrak{so}(2, 1)$ irreps of spin $\ell-1/2$ are absent from the product between the supercharges and the superconformal primary.  The multiplets of type $B$ have the property that certain $\mathfrak{so}(2, 1)$ irreps of spin $\ell \pm 1/2$ are absent from this product, and consequently, the multiplets of type $B$ are smaller.\footnote{This description is correct only when $\ell>0$.  When $\ell=0$, the definition of the multiplets also requires various conditions when acting on the primary with two supercharges.}  The half-BPS multiplets that we discussed above with bottom component $S_k$ are of $(B, +)$ type and have $a_3 = k$. The conserved current multiplet appears in the decomposition of the long multiplet at unitarity: $\Delta\to \ell+1$. This multiplet contains higher-spin conserved currents, and therefore can only appear in the free theory \cite{Maldacena:2011jn}. We will sometimes denote the superconformal multiplets by $X_{\Delta,\ell}^{[a_1\, a_2 \, a_3 \, a_4]}$, with $(\Delta,\ell)$ and $[a_1\, a_2 \, a_3 \, a_4]$ representing the $\mathfrak{so}(3, 2)$ and $\mathfrak{so}(8)_R$ quantum numbers of the superconformal primary, and the subscript $X$ denoting the type of shortening condition (for instance, $X = (A, 2)$ or $X = (B, +)$). 

Only certain $\mathcal{M}$ can appear in the OPE of half-BPS multiplets. Comparing the $\mathfrak{so}(8)_R$ irreps \eqref{tens0} that can appear in $S_{k_1}\times S_{k_2}$ to Table~\ref{Multiplets}, we see that $S_{k_1}\times S_{k_2}$ can contain only the following series of multiplets
\es{AandB}{
&(A,0)^{[0\,n-m\,2m\,0]}_{\Delta\geq n+\ell+1,\ell}\,,\qquad (A,2)^{[0\,n-m\,2m\,0]}_{ n+\ell+1,\ell}\,,\qquad (A,+)^{[0\,0\,2n\,0]}_{ n+\ell+1,\ell}\,,\\
& (B,2)^{[0\,n-m\,2n\,0]}_{ n,0}\,,\qquad (B,+)^{[0\,0\,2n\,0]}_{n,0}\,.
}
Furthermore, it was shown in  \cite{Ferrara:2001uj} that for $n=\frac12(k_1+k_2),\frac12(k_1+k_2)-1$ we have the following extra constraints:
\es{n01}{
&n=\frac12(k_1+k_2):\qquad\text{$(A,0)$, $(A,2)$, and $(A,+)$ do not appear}\,,\\
&n=\frac12(k_1+k_2)-1:\qquad\text{$(A,0)$ does not appear}\,.
}
One final constraint is that the irreps that appear in the (anti)-symmetric product of the OPE of identical operators can contain only (odd) even spins. If the spins set in \eqref{AandB} and \eqref{n01} conflict with this constraint, then the multiplet is forbidden.

\subsection{Four point functions of $S_2$ and $S_3$}
\label{S2S3}

We will now restrict our attention to the following four-point functions:
\es{4points}{
\langle 2222\rangle\,,\qquad \langle 3333\rangle\,,\qquad \langle 2233\rangle\,,\qquad\langle 2323\rangle\,,\qquad\langle 3223\rangle\,,
}
which we consider in the $s$-channel as discussed above, and we use the notation $\langle ijkl\rangle\equiv \langle S_iS_jS_kS_l\rangle$. The $\mathfrak{so}(8)_R$ tensor products that we must consider for the $S_2\times S_2$, $S_3\times S_3$, and $S_2\times S_3$ OPEs are
\es{tensProds}{
  {\bf 35}_c \otimes {\bf 35}_c &= {\bf 1} \oplus {\bf 28} \oplus {\bf 35}_c \oplus {\bf 300} \oplus {\bf 567}_c \oplus {\bf 294}_c \,,\\
    {\bf 35}_c \otimes {\bf 112}_c &= {\bf8}_c\oplus{\bf112}_c\oplus\bf{160}_c\oplus{\bf672'}_c\oplus{1400}_c\oplus{1568}_c\,,\\
      {\bf 112}_c \otimes {\bf 112}_c &= {\bf 1} \oplus {\bf 28} \oplus {\bf 35}_c \oplus {\bf 300} \oplus {\bf 567}_c \oplus {\bf 294}_c\oplus {\bf1386}_c\oplus{\bf1925}\oplus{\bf3696}_c\oplus{\bf4312}_c \,.\\
}
The supermultiplets $\mathcal{M}$ that can then appear in these OPEs according to the rules described above are listed in Tables \ref{opemult4} and \ref{opemult5}. Note that $ \langle 2233\rangle$ and $ \langle 2323\rangle$ contain the same supermultiplets as $ \langle 2222\rangle$ and $ \langle 3223\rangle$, respectively.

\begin{table}
\centering
\begin{tabular}{|c|c|r|c|}
\hline
Type    & $(\Delta,\ell)$     & $\mathfrak{so}(8)_R$ irrep&spin   \\
\hline
\red{$(B,+)$} &  \red{$(3,0)$}         & $\red{{\bf 1386}_c = [0060]}$ &\red{even}\\
\red{$(B,2)$} &  \red{$(3,0)$}         & \red{${\bf 4312}_c = [0220]$} &\red{even}\\
$(B,+)$ &  $(2,0)$         & ${\bf 294}_c = [0040]$ &even\\
\red{$(A,+)$} & \red{ $(\ell+3,\ell)$}       & \red{${\bf 294}_c = [0040]$ }&\red{even}\\
\red{$(A,2)$} &  \red{$(\ell+3,\ell)$}       & \red{${\bf 567}_c = [0120]$} &\red{odd}\\
$(B,2)$ &  $(2,0)$         & ${\bf 300} = [0200]$ &even\\
\red{$(A,2)$} & \red{$(\ell+3,\ell)$  }     & \red{${\bf 300} = [0200]$ }&\red{even}\\
$(B,+)$ &  $(1,0)$         & ${\bf 35}_c = [0020]$ &even\\
\red{$(A,0)$} & \red{ $\Delta\geq \ell+2$   }    & \red{${\bf 35}_c = [0020]$} &\red{even}\\
$(A,+)$ &  $(\ell+2,\ell)$       & ${\bf 35}_c = [0020]$ &even\\
\red{$(A,0)$} & \red{ $\Delta\geq \ell+2$  }     &\red{ ${\bf 28} = [0100]$ }&\red{odd}\\
$(A,2)$ &  $(\ell+2,\ell)$       & ${\bf 28} = [0100]$ &odd\\
$(A,0)$ &  $\Delta\geq \ell+1$       & ${\bf 1} = [0000]$ &even\\
Id &  $(0,0)$       & ${\bf 1} = [0000]$ &even\\
\hline
\end{tabular}
\caption{The possible superconformal multiplets in the $S_2\times S_2$ and $S_3\times S_3$ OPEs, where red denotes the multiplets that do not appear in $S_2\times S_2$.  The $\mathfrak{so}(3, 2) \oplus \mathfrak{so}(8)_R$ quantum numbers are those of the superconformal primary in each multiplet.}
\label{opemult4}
\end{table}

\begin{table}
\centering
\begin{tabular}{|c|c|r|c|}
\hline
Type    & $(\Delta,\ell)$     & $\mathfrak{so}(8)_R$ irrep&spin   \\
\hline
$(B,+)$ &  $(5/2,0)$         & ${\bf 672}^\prime_c = [0050]$&all \\ 
${(B,2)}$ &  ${(5/2,0)}$         & {${\bf 1568}_c = [0130]$} &{all}\\ 
$(B,2)$ &  $(5/2,0)$         & ${\bf 1400}_c = [0210]$ &all\\ 
$(B,+)$ &  $(3/2,0)$         & ${\bf 112}_c = [0030]$ &all\\
$(A,+)$ &  $(\ell+5/2,\ell)$       & ${\bf 112}_c = [0030]$ &all\\
{$(B,2)$} &  ${(3/2,0)}$         & {${\bf 160}_c = [0110]$} &{all}\\
$(A,2)$ &  $(\ell+5/2,\ell)$       & ${\bf 160}_c = [0110]$ &all\\
$(B,+)$ &  $(1/2,0)$         & ${\bf 8}_c = [0010]$ &all\\
$(A,0)$ &  $\Delta\geq \ell+3/2$       & ${\bf 8}_c = [0010]$ &all\\
\hline
\end{tabular}
\caption{The possible superconformal multiplets in the $S_2\times S_3$ OPE.  The $\mathfrak{so}(3, 2) \oplus \mathfrak{so}(8)_R$ quantum numbers are those of the superconformal primary in each multiplet.}
\label{opemult5}
\end{table}

We can compute the superblocks for each of these $\mathcal{M}$ in two steps, following the computation for $\langle 2222\rangle$ in \cite{Chester:2014fya}. First, we must determine which conformal primaries appear in each supermultiplet. This can be done by decomposing the character of each $\mathfrak{osp}(8|4)$ supermultiplet under the bosonic subgroup $\mathfrak{so}(5)\oplus\mathfrak{so}(8)_R$, using the formulae in Appendix B of \cite{Chester:2014fya}, based on \cite{Dolan:2008vc}. The results are written in Appendix \ref{superMults}. Second, we plug the conformal block expansion \eqref{GExpansion} of each superblock into the Ward identities \eqref{ward}, using the explicit expressions for the relevant $Y^{k_{12},k_{34}}_{nm}(\sigma,\tau)$ that we give in Appendix \ref{Ys}, and expand these constraints to high enough order in $r$ until we have fixed all the pairs of OPE coefficients $\lambda_{k_1k_2\cO_{\Delta,\ell,nm}}\lambda_{k_3k_4\cO_{\Delta,\ell,nm}}$ of the superconformal descendents in terms of the superconformal primary OPE coefficients $\lambda_{k_1k_2\mathcal{M}}\lambda_{k_3k_4\mathcal{M}}$. For instance, the stress tensor multiplet $(B,+)_{1,0}^{[0020]}$ has a superblock that can be written in the $A_{nm}^{k_{12},k_{34}}(U,V)$ basis of \eqref{Ybasis} in terms of conformal blocks as
\es{stressBlock}{
A^{0,0}_{11}(U,V)=G_{1,0}^{0,0}(U,V)\,,\quad A_{10}^{0,0}(U,V)=-G_{2,1}^{0,0}(U,V)\,,\quad A^{0,0}_{00}(U,V)=\frac14G_{3,2}^{0,0}(U,V)\,, 
}
where note that the relative sign of the odd spin conformal primary is opposite that of the even spins, as expected in general for a four-point function of scalars. The analogous expressions for the other superconformal blocks will have linear combinations of many conformal blocks for each $A_{nm}^{k_{12},k_{34}}(U,V)$ and are quite unwieldy, so we relegate them to an attached \texttt{Mathematica} file. Note that the superblocks for $\langle 2323\rangle$ and $\langle 3223\rangle$ have coefficients that are related as
\es{Arel}{
\lambda_{23\cO_{\Delta,\ell,nm}}\lambda_{23\cO_{\Delta,\ell,nm}}=(-1)^\ell \lambda_{32\cO_{\Delta,\ell,nm}}\lambda_{23\cO_{\Delta,\ell,nm}}
}
which follows from the specific form of the $\mathfrak{so}(8)$ polynomials in Appendix~\ref{Ys} and from how OPE coefficient of two scalars and a spin $\ell$ operator transforms under exchanging the scalars.  

We find that the relative coefficients of the super-descendents obtained by acting on the superconformal primary with $\epsilon^{\alpha\beta}Q_{a\alpha}Q_{b\beta}$ an odd number of times vanish, where $Q_{a\alpha}$ is a supercharge with $\alpha,\beta$ spinor indices and $a,b$ $\mathfrak{so}(8)_R$ indices. These conformal primaries are denoted in red in the tables of Appendix \ref{superMults}. This combination of superchargers is odd under parity, so these super-descendents have the opposite parity as the superconformal primary,  which motivates a `bonus' parity for 4-point functions in 3d $\mathcal{N}=8$ theories, as  was originally conjectured in \cite{Chester:2014fya} based on the superblocks for $\langle2222\rangle$.

We can use conformal Ward identities to relate OPE coefficients of two stress tensor operators, and by supersymmetry any two operators in the stress tensor multiplet, to $1/\sqrt{c_T}$ \cite{Osborn:1993cr}. For the four-point function we consider, this implies that
\es{stressOps}{
\lambda_{22(B,+)^{[0020]}_{1,0}}\propto \lambda_{33(B,+)^{[0020]}_{1,0}}\propto \lambda_{23(B,+)^{[0030]}_{\frac32,0}}\propto\frac{1}{\sqrt{c_T}}\,.
}
To fix the proportionality constants in our conventions, we can consider a free theory of eight real scalars $X_I$ and eight Majorana fermions mentioned in the Introduction, which has $c_T=16$. The half-BPS operators $S_2$ and $S_3$ in this case are given by 
 \es{OIJFree}{
 S^\text{free}_ {I_1I_2}&=\frac{1}{\sqrt{2}}\left[ X_{I_1} X_{I_2} - \frac{\delta_{I_1I_2}}{8} X_{I_3} X^{I_3}\right]\,,\\
 S^\text{free}_ {I_1I_2I_3}&=\frac{1}{\sqrt{6}}\left[ X_{I_1} X_{I_2}X_{I_3} - \frac{X_{(I_1}\delta_{I_2I_3)}}{20} X_{I_4} X^{I_4}\right]\,,\\
 }
Performing Wick contractions with the propagator $\langle X_I(\vec{x}) X_J(0) \rangle = \frac{\delta_{IJ}}{ \abs{\vec{x}}}$, we then find that the 4-point functions in \eqref{4points} in the notation of \eqref{FourPoint2} equal:
\es{free4}{
\mathcal{G}^\text{free}_{2222}&=
1+U\sigma^2+\frac{U}{V}\tau^2+4\sqrt{U}\sigma+
    4\sqrt{\frac{U}{V}}\tau+4\frac{U}{\sqrt{V}}\sigma\tau\,,\\
    \mathcal{G}^\text{free}_{3333}&=
1+ U^{\frac32}\sigma ^3+\frac{ U^{\frac32}}{V^{\frac32}}\tau ^3+9   \sqrt{U}\sigma+9\sqrt{\frac UV}\tau+36\frac{  U}{\sqrt{V}} \sigma  \tau  +9\frac{ U^{\frac32}}{\sqrt{V}} \sigma ^2 \tau +9 \frac{U^{\frac32}}{V} \sigma  \tau
   ^2+9U \sigma ^2 +9\frac{   U}{V}\tau ^2\,,\\
       \mathcal{G}^\text{free}_{2233}&=
1+3U\sigma^2+3\frac{U}{V}\tau^2+6\sqrt{U}\sigma+
    6\sqrt{\frac{U}{V}}\tau+12\frac{U}{\sqrt{V}}\sigma\tau\,,\\
       \mathcal{G}^\text{free}_{2323}&=
3 {U}^{\frac14}+ U^{\frac54}\sigma ^2+3\frac{ U^{\frac54}}{V}\tau ^2+6    U^{\frac34}\sigma +12\frac{    U^{\frac34}}{\sqrt{V}} \tau+6\frac{ U^{\frac54}}{\sqrt{V}} \sigma  \tau \,.\\
}
By comparing this to the superconformal block expansion \eqref{SBDecomp}, we can read off the OPE coefficients listed in Table~\ref{Avalues}, where the scaling dimensions of the long multiplets that appear in each correlator are
\es{superFree}{
\langle{2222}\rangle^\text{free},\langle{2233}\rangle^\text{free}&:\qquad {(A,0)^{[0000]}_{\Delta,\ell}}\quad\qquad\qquad\quad\,\text{for}\quad\Delta=\ell+1,\ell+2n\,,\qquad\;\;\; \\
\langle{3333}\rangle^\text{free}&:\qquad {(A,0)^{[0000]}_{\Delta,\ell}}\quad\qquad\qquad\quad\,\text{for}\quad\Delta=\ell+n\,,\qquad\qquad\qquad \\\
&\quad\qquad {(A,0)^{[0100]}_{\Delta,\ell}}\,, {(A,0)^{[0020]}_{\Delta,\ell}}\quad\text{for}\quad\Delta=\ell+2\,,\ell+2n+1\,,\quad \\
\langle{2323}\rangle^\text{free}&:\qquad {(A,0)^{[0010]}_{\Delta,\ell}}\quad\qquad\qquad\quad\,\text{for}\quad\Delta=\ell+\frac32,\ell+2n+\frac12\,,\;\;\, \\
}
for $n=1,2,3,\dots$. Note that $ {(A,0)^{[0000]}_{\ell+1,\ell}}$ for $\ell>0$ are conserved currents that only exist in a free theory, while ${(A,0)^{[0100]}_{\ell+2,\ell}}$, $ {(A,0)^{[0020]}_{\ell+2,\ell}}$, and $ {(A,0)^{[0010]}_{\ell+\frac32,\ell}}$ are equivalent to ${(A,2)^{[0100]}_{\ell+2,\ell}}$, ${(A,+)^{[0020]}_{\ell+2,\ell}}$, and  ${(A,+)^{[0010]}_{\ell+\frac32,\ell}}$, respectively. We can use the free theory values of $\lambda_{22(B,+)^{[0020]}_{1,0}}$, $\lambda_{33(B,+)^{[0020]}_{1,0}}$, $\lambda_{23(B,+)^{[0030]}_{\frac32,0}}$ and $c_T$ along with the general relations \eqref{stressOps} to find
\es{opetoCT}{
\lambda_{22(B,+)^{[0020]}_{1,0}}=\frac{16}{\sqrt{c_T}}\,,\qquad \lambda_{33(B,+)^{[0020]}_{1,0}}= \frac{24}{\sqrt{c_T}}\,,\qquad \lambda_{23(B,+)^{[0030]}_{\frac32,0}}=\frac{16\sqrt{3}}{\sqrt{c_T}}\,.
}

Another limit in which we can compute the correlators \eqref{4points} explicitly is in the generalized free field theory (GFFT) where the operators $ S_2$ and $S_3$ are treated as generalized free fields with two-point functions $\langle S_2(\vec{x}, Y_1)S_2(0, Y_2) \rangle = \frac{(Y_1 \cdot Y_2)^2}{\abs{x}^2}$, $\langle S_3(\vec{x}, Y_1)S_3(0, Y_2) \rangle = \frac{(Y_1 \cdot Y_2)^3}{\abs{x}^{3}}$ and $\langle S_2(\vec{x}, Y_1)S_3(0, Y_2) \rangle =0$.  The GFFT describes the $c_T\to\infty$ limit of $\mathcal{N}=8$ theories.  Performing the Wick contractions, we then find that \eqref{FourPoint2} equals:
\es{mean4}{
\mathcal{G}^\text{GFFT}_{2222}&=
1+U\sigma^2+\frac{U}{V}\tau^2\,,\qquad
    \mathcal{G}^\text{GFFT}_{3333}=
1+ U^{\frac32}\sigma ^3+\frac{ U^{\frac32}}{V^{\frac32}}\tau ^3\,,\\
       \mathcal{G}^\text{GFFT}_{2233}&=
1\,,\qquad\qquad\qquad\quad\;\;\;
       \mathcal{G}^\text{GFFT}_{2323}=
 U^{\frac54}\sigma ^2 \,.\\
}
By comparing this to the superconformal block expansion, we can read off the OPE coefficients listed in Table \ref{Avalues}, where the scaling dimensions of the long multiplets that appear in each correlator are
\es{superMean}{
\langle{2222}\rangle^\text{GFFT},\langle{2233}\rangle^\text{GFFT}&:\qquad {(A,0)^{[0000]}_{\Delta,\ell}}\quad\qquad\qquad\quad\,\text{for}\quad \Delta=\ell+2n\,,\qquad\;\;\; \\
\langle{3333}\rangle^\text{GFFT}&:\qquad {(A,0)^{[0000]}_{\Delta,\ell}}\quad\qquad\qquad\quad\,\text{for}\quad\Delta=\ell+2n+1\,,\quad \\
&\quad\qquad {(A,0)^{[0100]}_{\Delta,\ell}}\,, {(A,0)^{[0020]}_{\Delta,\ell}}\quad\text{for}\quad\Delta=\ell+2n+1\,,\quad \\
\langle{2323}\rangle^\text{GFFT}&:\qquad {(A,0)^{[0010]}_{\Delta,\ell}}\quad\qquad\qquad\quad\,\text{for}\quad \Delta=\ell+2n+\frac12\,,\;\;\, \\
}
for $n=1,2,3,\dots$.

 \begin{table}[htpbp]
\begin{center}
\begin{tabular}{|l|c|c|c|}
\hline
 \multicolumn{1}{|c|}{Type $\cal M$} & Correlator&Free theory $|\lambda_{\cal M}|$  & Generalized free field theory $|\lambda_{\cal M}|$ \\
   \hline
$(B,+)_{\frac12,0}^{[0010]}$&$\langle2323\rangle$& $\sqrt{6}$ & $0$\\
  \hline
  $(B,+)_{1,0}^{[0020]}$ &$\langle2222\rangle$& $4$ & $0$\\
  &$\langle3333\rangle$& $6$ & $0$\\
  \hline
$(B,+)_{\frac32,0}^{[0030]}$&$\langle2323\rangle$& $4\sqrt{3}$ & $0$\\
  \hline
   $(B,+)_{2,0}^{[0040]}$ &$\langle2222\rangle$& $4$ & $\frac{4}{\sqrt{3}}$\\
  &$\langle3333\rangle$& $12$ & $0$\\
  \hline
  $(B,+)_{\frac52,0}^{[0050]}$&$\langle2323\rangle$& $4\sqrt{2}$ & $\frac{4}{\sqrt{5}}$\\
  \hline
  $(B,+)_{3,0}^{[0060]}$&$\langle3333\rangle$& $8$ & $4\sqrt{\frac{2}{{5}}}$\\
  \hline
  $(B,2)_{\frac32,0}^{[0110]}$ &$\langle2323\rangle$& $0$ & $0$\\
  \hline
  $(B,2)_{2,0}^{[0200]}$ &$\langle2222\rangle$& $0$ & $4\sqrt{\frac23}$\\
  &$\langle3333\rangle$& $0$ & $0$\\
  \hline
  $(B,2)_{\frac52,0}^{[0210]}$ &$\langle2323\rangle$& $0$ & $4$\\
  \hline
  $(B,2)_{\frac52,0}^{[0130]}$ &$\langle2323\rangle$& $0$ & $\frac{8}{\sqrt{5}}$\\
  \hline
   $(B,2)_{3,0}^{[0220]}$ &$\langle3333\rangle$& $0$ & $12\sqrt{\frac25}$\\
  \hline
    $(A,+)_{2,0}^{[0020]}$ &$\langle2222\rangle$& $4\sqrt{\frac23}$ & $\frac83$\\
  \hline
   $(A,+)_{4,2}^{[0020]}$ &$\langle2222\rangle$& $ 16 \frac{\sqrt{82}}{35}$ & $\frac{128}{35}$\\
  \hline
  $(A,+)_{\frac52,0}^{[0030]}$ &$\langle2323\rangle$& $\frac{8}{\sqrt{7}}$ & $\frac{16}{3}\sqrt{\frac{5}{21}}$\\
  \hline
   $(A,+)_{\frac72,1}^{[0030]}$ &$\langle2323\rangle$& $\frac{16}{3\sqrt{39}}$ & $\frac{32}{3}\sqrt{\frac{14}{143}}$\\
  \hline
   $(A,+)_{3,0}^{[0040]}$ &$\langle3333\rangle$& $6\sqrt{2}$ & $2\sqrt{5}$\\
  \hline
   $(A,2)_{3,1}^{[0100]}$ &$\langle2222\rangle$& $8\sqrt{\frac{2}{21}}$ & $\frac{32}{\sqrt{105}}$\\
  \hline
   $(A,2)_{\frac52,0}^{[0110]}$ &$\langle2323\rangle$& $0$ & $\frac{8}{3}\sqrt{\frac{5}{3}}$\\
  \hline
   $(A,2)_{3,0}^{[0200]}$ &$\langle3333\rangle$& $0$ & $4\sqrt{2}$\\
  \hline
   $(A,2)_{4,1}^{[0120]}$ &$\langle3333\rangle$& $0$ & $\frac{20}{3}\sqrt{\frac{7}{3}}$\\
  \hline
  $(A,0)_{1,0}^{[0000]}$ &$\langle2222\rangle$& $2$ & $0$\\
  &$\langle3333\rangle$& $3$ & $0$\\
  \hline
  $(A,0)_{\frac32,0}^{[0010]}$ &$\langle2323\rangle$& $4\sqrt{\frac{6}{5}}$ & $0$\\
  \hline
  $(A,0)_{3,1}^{[0100]}$ &$\langle3333\rangle$& $8\sqrt{\frac{6}{7}}$ & $0$\\
  \hline
  $(A,0)_{2,0}^{[0020]}$ &$\langle3333\rangle$& $4\sqrt{{6}}$ & $0$\\
  \hline
\end{tabular}
\end{center}
\caption{Values of OPE coefficients of the superconformal primaries of short multiplets and the lowest semishort and long multiplets of each type in correlators of $S_2$ and $S_3$ in the $s$-channel of the free and generalized free field theory limits.}\label{Avalues}
\end{table}

\section{Protected 1d topological sector}
\label{1d}

All ${\cal N} = 8$ SCFTs contain 1d topological sectors \cite{Beem:2013sza,Chester:2014mea,Beem:2016cbd}, which, for theories with Lagrangian descriptions, can be computed using supersymmetric localization \cite{Dedushenko:2016jxl,Dedushenko:2017avn,Dedushenko:2018icp}.  The supersymmetric localization setup of \cite{Dedushenko:2016jxl,Dedushenko:2017avn,Dedushenko:2018icp} applies to any ${\cal N} \geq 4$ theory constructed from ${\cal N} = 4$ vector multiplets and hypermultiplets.  The theory we are interested in, ABJM$_{N, 1}$, is, however, not of this type due to the presence of the Chern-Simons interactions.  However, as shown in \cite{Bashkirov:2011pt}, ABJM$_{N, 1}$ does have a dual description as an ${\cal N} = 4$ $U(N)$ gauge theory with a fundamental hyper and an adjoint hyper, for which the results of \cite{Dedushenko:2016jxl,Dedushenko:2017avn,Dedushenko:2018icp} do apply.  This is the description we use in this section.

A generic ${\cal N} = 4$ theory has two distinct topological sectors:  one corresponding to the Higgs branch and one corresponding to the Coulomb branch.  Abstractly, these two sectors can be described as follows.  The R-symmetry of an ${\cal N} = 4$ SCFT is $\mathfrak{so}(4) \cong \mathfrak{su}(2)_H \oplus \mathfrak{su}(2)_C$.  The Higgs branch sector can be constructed from 3d half-BPS scalar operators with scaling dimension equal to the $\mathfrak{su}(2)_H$ spin, $\Delta = j_H$, that are invariant under $\mathfrak{su}(2)_C$.  Such an operator, ${\cal O}_{(a_1 \cdots a_{2j_H})}$, where $a_i = 1, 2$ are $\mathfrak{su}(2)_H$ fundamental indices, placed on a line and contracted with specific position-dependent polarization vectors, defines a 1d operator 
  \es{TopOpHiggs}{
   {\cal O}(x) \equiv u^{a_1}(x) \cdots u^{a_{2j_H}}(x){\cal O}_{(a_1 \cdots a_{2j_H})}(0, 0, x)
    \,, \qquad
     u^a(x) = \left(1, \frac{x}{2r} \right) 
  }
in the Higgs branch topological sector.  Here, $r$ is a dimensionful parameter with units of length.  The Coulomb branch sector can be constructed similarly from half-BPS scalar operators ${\cal O}_{(\dot a_1 \cdots \dot a_{2j_C})}$ with $\Delta = j_C$ and $j_H = 0$:
 \es{TopOpCoulomb}{
  {\cal O}(x) \equiv v^{\dot a_1}(x) \cdots v^{\dot a_{2j_C}}(x){\cal O}_{(\dot a_1 \cdots \dot a_{2j_C})}(0, 0, x)
    \,, \qquad
     v^{\dot a}(x) = \left(1, \frac{x}{2r} \right)  \,.
 }
The operators \eqref{TopOpHiggs} (or \eqref{TopOpCoulomb}) can be argued to have topological correlation functions because they belong to the cohomology of a nilpotent supercharge of $Q + r S$ type \cite{Beem:2013sza,Chester:2014mea, Beem:2016cbd}.  

While for a generic ${\cal N} = 4$ SCFT, the two sectors described above are unrelated, for the particular case of ${\cal N} = 8$ SCFTs it can be argued that they must be isomorphic. Indeed, it can be shown that the ${\cal N} = 4$ operators with $\Delta = j_H$ and $j_C = 0$ and those with $\Delta = j_C$ and $j_H = 0$ can only originate in pairs from ${\cal N} = 8$ operators that are at least quarter-BPS, and ${\cal N} = 8$ SUSY relates their correlation functions.  Since the description of the Higgs branch sector coming from supersymmetric localization \cite{Dedushenko:2016jxl}  is simpler than that of the Coulomb branch sector \cite{Dedushenko:2017avn,Dedushenko:2018icp}, we will thus focus only on the Higgs branch one.  

\subsection{General features}

Before restricting our attention to ABJM$_{N,1}$, we first derive some general constraints valid for the 1d sector of any 3d CFT with ${\cal N}=8$ SUSY\@. In addition to topological invariance, these theories have an $\mathfrak{su}(2)_F$ flavor symmetry, which from the 3d perspective originates as a subalgebra of the $\mathfrak{so}(8)$ R-symmetry that commutes with the ${\cal N} = 4$ R-symmetry $\mathfrak{su}(2)_H \oplus \mathfrak{su}(2)_C$.\footnote{What commutes with $\mathfrak{su}(2)_H \oplus \mathfrak{su}(2)_C$ inside $\mathfrak{so}(8)$ is an $\mathfrak{so}(4)$ algebra, so more generally, the 1d topological sector could have $\mathfrak{so}(4) \cong \mathfrak{su}(2)_F \oplus \mathfrak{su}(2)_F'$ flavor symmetry.  The 1d theories we focus on here will be invariant under $\mathfrak{su}(2)_F'$, so we will only consider $\mathfrak{su}(2)_F$ as a flavor symmetry.} We can therefore consider operators $\cO_{(a_1 \cdots a_{2j_F})}(\vphi)$, $a_i = 1, 2$, transforming in the spin-$j_F$ irrep of $\mathfrak{su}(2)_F$. Their indices can be contracted with polarization vectors $y^a$ to form homogeneous polynomials in $y$ of degree $2 j_F$:
\es{polyOp}{
\cO(\vphi, y) \equiv \cO_{(a_1 \cdots a_{2j_F})}(\vphi) y^{a_1} \cdots y^{2a_{j_F}} \,.
}

Correlation functions of such operators in any 1d theory are heavily constrained by symmetry and topological considerations. For convenience, we consider a basis of unnormalized orthogonal operators $\{ \cO_A \}$ in the 1d theory with definite $\mathfrak{su}(2)_F$ spin $j_A$ originating from 3d operators of scaling dimension $\Delta_A$. Their two-point functions are fixed to an overall constant
 \es{TwoPointGen}{
  \langle {\cO}_A(\varphi_1, y_1) {\cO}_B(\varphi_2, y_2) \rangle &= \delta_{AB} B_{\cO_A} \langle y_1, y_2 \rangle^{2j_A} (\sgn \varphi_{21})^{2 \Delta_A} \,,
 }
where $\langle y_1, y_2 \rangle = \epsilon_{ab} y_1^a y_2^b$ is the $\mathfrak{su}(2)_F$ invariant product of the polarization vectors. Similarly, three-point functions in the theory take the form
 \es{ThreePtGen}{
  \langle  {\cal O}_A(&\varphi_1, y_1) {\cal O}_B(\varphi_2, y_2)  {\cal O}_C(\varphi_3, y_3) \rangle 
   = C_{{\cal O}_A {\cal O}_B {\cal O}_C} \\
   &{}\times \langle y_1, y_2 \rangle^{j_{ABC}} \langle y_2, y_3 \rangle^{j_{BCA}} \langle y_3, y_1 \rangle^{j_{CAB}}
   (\sgn \varphi_{21})^{\Delta_{ABC}}  (\sgn \varphi_{32})^{\Delta_{BCA}}  (\sgn \varphi_{13})^{\Delta_{CAB}} \,,
 }
where $\Delta_{ABC} \equiv \Delta_A + \Delta_B - \Delta_C$, $j_{ABC} = j_A + j_B - j_C$, and $\{j_A,j_B,j_C\}$ obey the triangle inequality. For triplets of operators for which this is not the case, the right hand side of \eqref{ThreePtGen} vanishes.  Note that \eqref{TwoPointGen} and \eqref{ThreePtGen} depend only on the overall ordering of the operators on $S^1$, as expected of any topological theory.

The four-point structures can be used to relate observables in the 1d theory to CFT data in the 3d theory. From \eqref{TwoPointGen} and \eqref{ThreePtGen} we can first extract the OPE of a generic 1d theory. Assuming that $\varphi_1, \varphi_2 < \varphi$, operator products decompose as
 \es{OPE}{
  {\cal O}_A(\varphi_1, y_1) {\cal O}_B(\varphi_2, y_2) 
   &= \sum_{\cal O} \frac{C_{{\cal O}_A {\cal O}_B {\cal O}}}{B_{{\cal O}}}
   (-1)^{j + j_A - j_B}   \langle y_1, y_2 \rangle^{j_A + j_B - j}   \\
    &{}\times
       (\sgn \varphi_{21})^{\Delta_A + \Delta_B - \Delta}   (-1)^{\Delta  + \Delta_A - \Delta_B}
      {\cal O}(\varphi; y_1,y_2) \,,
 }
where we have introduced the symmetrized operator
\es{symPoly}{
\cO(\vphi; y_1, y_2) \equiv \frac{1}{(2j)!}\sum_{\sigma \in S_{2j}} \cO_{a_{\sigma(1)} \cdots a_{\sigma(2j)}}(\vphi) y_1^{a_1} \cdots y_1^{a_{j+j_1-j_2}} y_2^{a_{j+j_2-j_2+1}} \cdots y_2^{a_{2j}} \,.
}
The four-point structures are determined by applying this OPE \eqref{OPE} twice. Taking $\varphi_{1}, \varphi_{2} < \varphi_{3}, \varphi_{4}$ gives the s-channel expansion,
 \es{FourPoint}{
  &\langle {\cal O}_A(\varphi_1, y_1) {\cal O}_B(\varphi_2, y_2)  {\cal O}_C(\varphi_3, y_3) {\cal O}_D(\varphi_4, y_4)  \rangle
   = \sum_{\cal O} \frac{C_{{\cal O}_A {\cal O}_B {\cal O}} C_{{\cal O}_C {\cal O}_D {\cal O}}}{B_{{\cal O}}}
    (-1)^{2j +  j_A - j_B+ j_C - j_D}  \\
   &\times  (-1)^{2 \Delta + \Delta_A - \Delta_B + \Delta_C - \Delta_D}  
   (\sgn \varphi_{21})^{\Delta_A + \Delta_B - \Delta} 
    (\sgn \varphi_{43})^{\Delta_C + \Delta_D - \Delta} 
    \langle y_1, y_2 \rangle^{j_A + j_B - j}
    \langle y_3, y_4 \rangle^{j_C + j_D - j}
    \langle y_{1, 2}, y_{3, 4} \rangle^{2j} 
    \,.
 }
Here, the quantity $\langle y_{1, 2}, y_{3, 4} \rangle^{2j}$ is an $\mathfrak{su}(2)_F$ invariant quantity constructed from $(j - j_A + j_B)$ copies of $y_1$, $(j - j_B + j_A)$ copies of $y_2$, $(j - j_C + j_D)$ copies of $y_3$, and $(j - j_D + j_C)$ copies of $y_4$ that reduces to $\langle y, y' \rangle^{2j}$ if we were to set $y_1 = y_2 = y$ and $y_3 = y_4 = y'$.
 
Introducing the $\mathfrak{su}(2)_F$-invariant cross-ratio
 \es{wDef}{
  w \equiv \frac{\langle y_1, y_2 \rangle \langle y_3, y_4 \rangle }{\langle y_1, y_3 \rangle  \langle y_2, y_4 \rangle} \,, 
 }
the four-point function \eqref{FourPoint} simplifies to
 \es{FourPointSimp}{
  &\langle {\cal O}_A(\varphi_1, y_1) {\cal O}_B(\varphi_2, y_2)  {\cal O}_C(\varphi_3, y_3) {\cal O}_D(\varphi_4, y_4)  \rangle
   =\sum_{\cal O} \frac{C_{{\cal O}_A {\cal O}_B {\cal O}} C_{{\cal O}_C {\cal O}_D {\cal O}}}{B_{{\cal O}}} g_{\Delta,j}
    \,,
 }
where
\es{1dBlock}{
g_{\Delta,j}(w) &\equiv (-1)^{2 \Delta + \Delta_A - \Delta_B + \Delta_C - \Delta_D}  
   (\sgn \varphi_{21})^{\Delta_A + \Delta_B - \Delta} 
    (\sgn \varphi_{43})^{\Delta_C + \Delta_D - \Delta}    \langle y_1, y_2 \rangle^{j_A+j_B} \langle y_3, y_4 \rangle^{j_C+j_D}  \\ &\times\left( \frac{\langle y_1, y_4 \rangle}{\langle y_2, y_4 \rangle} \right)^{j_A - j_B} \left( \frac{\langle y_1, y_3 \rangle}{\langle y_1, y_4 \rangle} \right)^{j_C - j_D} w^{j_C-j_D}
    P_{j+j_C-j_D}^{(j_A - j_B - j_C + j_D ,j_B-j_A-j_C+j_D)}\left( \frac{2}{w} - 1 \right) \,.
} 
The $P_n^{(a,b)}(x)$ are Jacobi polynomials normalized such that
\es{JacobiPNorm}{
P_n^{(a,b)}\left( \frac{2}{w} - 1 \right) = w^{-n+a+b}(1+O(w)) \,.
}

 With the four-point structures in hand, we can read off the 3d OPE coefficients.  We first project the four-point function \eqref{FourPoint2} to 1d by placing the operators on a line and implementing the twist in \eqref{TopOpHiggs} (for more details, see \cite{Chester:2014mea}). Comparing this with the superconformal block expansion \eqref{blockExp}, a tedious calculation yields
 \es{1dTo3dBlock}{
\lambda_{\cal M} = 2^\Delta  \lim_{w \to \infty} \sqrt{\left| P_{j+j_C-j_D}^{(j_A - j_B - j_C + j_D ,j_B-j_A-j_C+j_D)}\left( \frac{2}{w} - 1 \right) \right| } \frac{ C_{\cO_A \cO_B \cO}  }{B^{1/2}_{\cO_A} B^{1/2}_{\cO_B} B^{1/2}_{\cO} }.
 }
(Here, we restrict to the cases $\langle S_2 S_2 S_2 S_2 \rangle$, $\langle S_2 S_3 S_2 S_3 \rangle$, $\langle S_3 S_3 S_3 S_3 \rangle$, so that in the conformal block decomposition \eqref{blockExp} we always have squares of OPE coefficients.)

Using the dictionary \eqref{1dTo3dBlock}, we can take advantage of the 1d crossing symmetry to derive a set of constraints involving only a  finite number of 3d OPE coefficients. Denoting the OPE between two operators as $\wick{\c1 \cO_A \c1 \cO_B}$, implementing 1d crossing yields
 \es{1dCrossing}{
 &\underline{\langle \wick{\c1 \cO_2 \c1 \cO_2} \wick{\c1 \cO_2 \c1 \cO_2} \rangle =   \langle \wick{\c1 \cO_2 \c2 \cO_2 \c1 \cO_2 \c2 \cO_2}\rangle} \qquad \Longrightarrow
  \\ &  4\lambda^2_{22(B,+)_{1,0}^{[0020]}} - 5 \lambda^2_{22(B,+)_{2,0}^{[0040]}} + \lambda^2_{22(B,2)^{[0200]}_{2,0}} + 16 = 0 \,. \\ \\
&\underline{\langle \wick{\c1 \cO_3 \c1 \cO_3} \wick{\c1 \cO_3 \c1 \cO_3} \rangle  =   \langle \wick{\c1 \cO_3 \c2 \cO_3 \c1 \cO_3 \c2 \cO_3}\rangle}  \qquad \Longrightarrow \\
& \lambda^2_{33(B,2)_{3,0}^{[0220]}}+16\lambda^2_{33(B,+)_{1,0}^{[0020]}}-9\lambda^2_{33(B,+)_{3,0}^{[0060]}}= 0 \,, \\
&\lambda^2_{33(B,2)^{[0200]}_{2,0}}+16+\lambda^2_{33(B,+)_{2,0}^{[0040]}}-\frac52\lambda^2_{33(B,+)_{3,0}^{[0060]}} = 0 \,. \\ \\
&\underline{\langle \wick{\c1 \cO_2 \c1 \cO_3} \wick{\c1 \cO_2 \c1 \cO_3} \rangle  =   \langle \wick{\c1 \cO_2 \c2 \cO_3 \c2 \cO_2 \c1 \cO_3}\rangle} \qquad \Longrightarrow \\
&-9\lambda^2_{23(B,+)_{\frac52,0}^{[0050]}} + 16 \lambda^2_{23(B,+)_{\frac12,0}^{[0010]}} +4 \lambda^2_{23(B,2)_{\frac32,0}^{[0110]}}+4 \lambda^2_{23(B,+)_{\frac32,0}^{[0030]}}  + \lambda^2_{23(B,2)_{\frac52,0}^{[0210]}}  + \lambda^2_{23(B,2)_{\frac52,0}^{[0130]}}  = 0 \,. \\ \\
&\underline{ \langle \wick{\c1 \cO_2 \c1 \cO_3} \wick{\c1 \cO_2 \c1 \cO_3} \rangle  =   \langle \wick{\c1 \cO_2 \c2 \cO_3 \c1 \cO_2 \c2 \cO_3}\rangle} \qquad \Longrightarrow \\
&-\lambda^2_{23(B,2)_{\frac52,0}^{[0210]}} - 16\lambda^2_{23(B,+)_{\frac12,0}^{[0010]}} + 4\lambda^2_{23(B,2)_{\frac32,0}^{[0110]}} + 16 + \bar{\lambda}^2_{(B,2)_{2,0}^{[0200]}} - \frac{20}{3} \bar{\lambda}^2_{(B,+)_{1,0}^{[0020]}} + 5 \bar{\lambda}^2_{(B,+)_{2,0}^{[0040]}}  = 0 \,, \\
&15 \lambda^2_{23(B,2)_{\frac52,0}^{[0130]}} - 60\lambda^2_{23(B,+)_{\frac32,0}^{[0030]}} - 192 + 32 \bar{\lambda}^2_{(B,+)_{1,0}^{[0020]}} - 12 \bar{\lambda}^2_{(B,2)_{2,0}^{[0200]}} + 48\bar{\lambda}^2_{(B,+)_{2,0}^{[0040]}} = 0 \,, \\
&-5 \lambda^2_{23(B,+)_{\frac52,0}^{[0050]}} + 16 + \bar{\lambda}^2_{(B,2)_{2,0}^{[0200]}} + 4 \bar{\lambda}^2_{(B,+)_{1,0}^{[0020]}} + \bar{\lambda}^2_{(B,+)_{2,0}^{[0040]}} = 0 \,.
  }
The quantity $\bar{\lambda}_\cM^2$ is defined as the contribution of a multiplet $\cM \in S_2 \times S_2 \cap S_3 \times S_3$. More precisely, an orthogonal set of 1d operators $\cO_i$, all descending from the 3d multiplet $\cM_i = \cM$, contribute as
\es{}{
\bar{\lambda}_\cM^2 = \sum_i \lambda_{22 \cM_i} \lambda_{33 \cM_i} \,.
}
Note that the relations \eqref{1dCrossing} are derived solely based on representation theory arguments, so they do not rely on the specific details of ABJM theory and of its 1d sector.

\subsection{1d sector of ABJM$_{N,1}$}

We now specialize to the 1d sector of ABJM$_{N,1}$, in the dual presentation where this theory is an ${\cal N} = 4$ $U(N)$ gauge theory with a fundamental hypermultiplet and an adjoint hypermultiplet.   To obtain a Lagrangian description of the 1d Higgs branch sector of a general ${\cal N} = 4$ gauge theory with hypermultiplet matter fields, Ref~\cite{Dedushenko:2016jxl} first used the stereographic projection to map the 3d theory from flat space to a round $S^3$ whose radius of curvature $r$ is taken to be the same as the parameter $r$ appearing in \eqref{TopOpHiggs}--\eqref{TopOpCoulomb}.  Since under this mapping a straight line passing through the origin of $\R^3$ gets mapped to a great circle on $S^3$, the 1d sector of the $S^3$ theory will be defined on a great circle.  An explicit description was then obtained in \cite{Dedushenko:2016jxl} using supersymmetric localization.  The 1d partition function takes the form of a Gaussian theory coupled to a matrix model, where the Gaussian fields come from the hypermultiplet matter and the matrix degrees of freedom come from one of the scalars in the ${\cal N} = 4$ vector multiplet.  For the particular case of a $U(N)$ gauge theory with an adjoint hyper and a fundamental hyper, the 1d theory is \cite{Dedushenko:2016jxl}
\es{ABJM1d}{
Z_{\text{ABJM}_{N,1}} &= \frac{1}{N!} \int d^N\sigma  \prod_{\alpha<\beta} 4 \sinh^2(\pi \sigma_{\alpha\beta})
\int {\cal D}Q {\cal D}\widetilde{Q} {\cal D}X {\cal D}\widetilde{X} e^{-S_Q - S_X} \,, 
}
with
\es{ABJMact}{
S_Q &= -4\pi r\int_{-\pi}^{\pi} d\varphi \left[\widetilde{Q}_\alpha \dot{Q}^\alpha + \sigma_\alpha \widetilde{Q}_\alpha Q^\alpha \right] \,, \\
 S_X &=  -4\pi r\int_{-\pi}^{\pi} d\varphi \left[\widetilde{X}^{\:\: \beta}_{\alpha} \dot{X}^{\alpha}_{\:\: \beta}
+  \sum_{\alpha< \beta} \sigma_{\alpha \beta}(\widetilde{X}^{\:\: \beta}_{\alpha}X^{\alpha}_{\:\: \beta} - \widetilde{X}^{\:\: \alpha}_{\beta}X^{\beta}_{\:\: \alpha}) \right]  \,,
}
where $\alpha,\beta = 1,\ldots, N$, and $\sigma_{\alpha \beta} \equiv \sigma_\alpha - \sigma_\beta$.  The 1d fields $X^{\alpha}_{\:\: \beta}$ and $\widetilde{X}^{\:\: \beta}_{\alpha}$ correspond to the adjoint hypermultiplet, $Q^\alpha$ and $\widetilde{Q}_\alpha$ correspond to the fundamental hypermultiplet, and $\sigma_\alpha$ are the matrix degrees of freedom in the Cartan of $U(N)$. The 1d operators can be constructed as gauge-invariant products of $X$, $\widetilde{X}$, $Q$, and $\widetilde{Q}$.  However, the D-term relations of the 3d theory imply that one can always trade the product of $Q$ and $\widetilde{Q}$ for products of $X$ and $\widetilde{X}$, so without loss of generality we can construct all 1d operators in terms of $X$ and $\widetilde{X}$.  After integrating out $Q$ and $\widetilde{Q}$, the correlation functions become
\es{1dCorrelators}{
\langle \cO_1(\varphi) \cdots \cO_n(\varphi)\rangle =  \frac{1}{Z_{\text{ABJM}_{N,1}}} \frac{1}{N!} \int d^N\sigma  \, \prod_{\alpha<\beta} 4 \sinh^2(\pi \sigma_{\alpha\beta}) \, Z_\sigma
\langle \cO_1(\varphi) \cdots \cO_n(\varphi)\rangle_\sigma \,,
}
where 
 \es{ZSigma}{
  Z_\sigma \equiv \frac{1}
 {\prod_{\alpha, \beta=1}^N (2 \cosh (\pi \sigma_{\alpha\beta})) \prod_{\alpha=1}^N (2 \cosh (\pi \sigma_\alpha))}
 }
is the partition function of the Gaussian theory at fixed $\sigma$, and the correlation functions in this theory are given by
\es{1dGaussianCorrelator}{
\langle \cO_1(\varphi) \cdots \cO_n(\varphi)\rangle_\sigma = \frac{1}{Z_\sigma} \int {\cal D}X {\cal D}\widetilde{X} e^{-S_X} \cO_1(\varphi) \cdots \cO_n(\varphi) \,.
}
The Gaussian correlators with fixed $\sigma$ can be computed systematically by performing Wick contractions using the 1d propagator \cite{Dedushenko:2016jxl}
\es{Xprop}{
\braket{X^\alpha_{\:\: \beta}(\varphi_1)\widetilde{X}_\gamma^{\:\: \delta} (\varphi_2)}_\sigma = -\delta^\alpha_{\:\:\gamma} \delta^\delta_{\:\:\beta}\frac{\text{sgn} \: \varphi_{12} + \tanh(\pi \sigma_{\alpha\beta})}{2 \ell} e^{- \sigma_{\alpha\beta} \varphi_{12}}\,, \qquad \ell \equiv 4\pi r \,.
}

Note that the factorization of ABJM$_{N,1}$ into ABJM$^\text{int}_{N,1}$ and a free theory, as described in the Introduction, is manifest for this presentation of the 1d sector. Indeed, under the identification of the free fields $X_\text{free},\widetilde{X}_\text{free}$ as the traces of $X,\widetilde{X}$ and the interacting fields $X_\text{int}, \tX_\text{int}$ as the traceless parts,
 \es{TraceTraceless}{
   X_\text{free} \equiv \tr X \,, \qquad \tX_\text{free} \equiv  \tr \tilde X  \,, \qquad
    X_\text{int} \equiv X - \frac{{\bf 1}}{N} \tr X \,, \qquad
     \tX_\text{int} \equiv \tX - \frac{{\bf 1}}{N} \tr \tX \,,
 }
where ${\bf 1}$ is the $N \times N$ identity matrix,  the action \eqref{ABJMact} decomposes into
\es{1dSplitS}{
S_X[X,\widetilde{X},\sigma] = S_\text{free}[X_\text{free}, \widetilde{X}_\text{free}] + S_\text{int}[X_\text{int}, \widetilde{X}_\text{int},\sigma] \,,
}
and the partition function immediately factorizes as
\es{1dSplitZ}{
Z_{\text{ABJM}_{N,1}} &=  \left( \int \cD X_\text{free} \cD \widetilde{X}_\text{free} e^{-S_\text{free}} \right) \left(  \int d^N\sigma  \frac{\prod_{\alpha<\beta} 4 \sinh^2(\pi \sigma_{\alpha\beta})}
 {N!\prod_\alpha (2 \cosh \pi \sigma_\alpha)} \int \cD X_\text{int} \cD \widetilde{X}_\text{int} e^{-S_\text{int}} \right)
\,.
}

\subsection{Operator content}

We seek the operator content of the 1d sector of ABJM$_{N,1}^\text{int}$. Under the $\mathfrak{su}(2)_F$ symmetry, $\widetilde{X}$ and $X^T$ transform as a doublet.  To simplify notation, we group $\widetilde{X}$ and $X^T$ into a quantity 
\es{compositeA}{
{\cal X} (\varphi,y) = y^1 \widetilde{X}(\varphi) + y^2 X^T(\varphi)
}
that depends on an additional polarization variable $y^a = (y^1, y^2)$. The observables of the theory are then given by gauge-invariant products of $2 j_F$ ${\cal X}$'s.

As explained in \cite{Chester:2014mea}, the ${\cal N} = 8$ multiplets of type $(B, 2)$, $(B, 3)$, $(B, +)$, and $(B, -)$ whose superconformal primary transforms in the $[0a_2a_3a_4]$ irrep of $\mathfrak{so}(8)$ contain ${\cal N} = 4$ Higgs branch operators with $\Delta = j_H  = a_2 + \frac{a_3 + a_4}{2}$ and $j_C = 0$ and $j_F = \frac{a_3}{2}$.  From these, we can construct 1d operators with $j_F = \frac{a_3}{2}$:
 \es{so8Tosu2}{
  (B, 2), (B, 3), (B, \pm) \text{ in } [0 a_2 a_3 a_4] \  \to \ 
   \text{1d op with $j_F = \frac{a_3}{2}$} 
 }
by taking a gauge-invariant product of precisely $2a_2 + a_3 + a_4$ ${\cal X}$'s.   No other ${\cal N} = 8$ multiplets yield operators in the 1d theory.

Based on the analysis of the previous section, we are looking to find the 1d operators that arise from the following ${\cal N} = 8$ multiplets.  The external operators $S_2$ and $S_3$ are part of the multiplets
 \es{ExternalMult}{
  (B, +)^{[0020]}_{1,0}  \,, \qquad
   (B, +)^{[0030]}_{\frac 32,0} \,,
 }
respectively, while the internal operators that appear in the OPEs of the operators \eqref{ExternalMult} and have representatives in the 1d theory can be part of any of the following multiplets: 
 \es{InternalMult}{
  &(B, +)^{[00k0]}_{\frac k2, 0} \,, \qquad \text{for } k = 1, 2, \ldots, 6 \,, \\
  &(B, 2)^{[0010]}_{\frac 32, 0} \,, \qquad 
   (B, 2)^{[0200]}_{2, 0} \qquad (B, 2)^{[0210]}_{\frac 52, 0} \,, \qquad
   (B, 2)^{[0130]}_{\frac 52, 0} \,, \qquad (B, 2)^{[0220]}_{3, 0} \,.
 }

Let us now construct these 1d operators.  The most general 1d operators one can construct are products $\tr {\cal X}^{n_1}  \tr {\cal X}^{n_2} \tr {\cal X}^{n_3} \cdots$. The operator $\tr {\cal X}$ and its various powers are the 1d operators that come from the free sector.  The operators in the interacting sector are then constructed from the traceless part of ${\cal X}$, namely
 \es{Traceless}{
  \widehat {\cal X} = {\cal X} - \frac {\mathbf 1} N \tr {\cal X} =  y^1 \widetilde{X}_\text{int} (\varphi) + y^2 X_\text{int}^T(\varphi)\,,
 }
where as in \eqref{TraceTraceless}, $\mathbf 1$ is the $N \times N$ identity matrix.   We will only focus on 1d operators which descend from superconformal primaries in the interacting sector. 

For the external multiplets \eqref{ExternalMult}, the 1d operators constructed out of $\widehat {\cal X}$ are unique and single trace:
 \es{O23Def}{
  (B,+)^{[0020]}_{1,0} \to j_F = 1: \qquad {\cal O}_2 &= 
   \tr \widehat {\cal X}^2 = \tr {\cal X}^2 - \frac{1}{N}  (\tr {\cal X})^2 \,, \\
  (B,+)^{[0030]}_{\frac32,0} \to j_F = \frac32: \qquad {\cal O}_3 &=  
   \tr \widehat {\cal X}^3  =  \tr {\cal X}^3 - \frac{3}{N} \tr {\cal X}^2 \tr {\cal X} + \frac{2}{N^2} (\tr {\cal X})^3 \,.
 }
The 1d operators corresponding to the internal multiplets \eqref{InternalMult} can be single trace or double trace.  To determine them, we make use of the fact that in 1d the operator products are well-defined and require no regularization.  Any operator $\cO \in \cO_A \times \cO_B$ arising from a 3d operator with $\Delta = \Delta_{\cO_A} + \Delta_{\cO_B}$ and $\mathfrak{su}(2)_F$ spin $j_F$ can be expressed as the projection of $(\cO_A \cO_B)_{a_1 \cdots a_{2(j_A+j_B)}}$ to the spin $j_F$ component.

We first focus on internal operators $\cO_k$ with $j_F  = k/2$ coming from the $(B, +)$ multiplets in \eqref{InternalMult} with $\Delta = k/2$.   For $k=1$, it is impossible to construct a 1d operator out of $\tr {\cal X}$.  For $k=2, 3$ these operators are just those in \eqref{O23Def}.  For $k = 4, 5, 6$ there are several linearly-independent operators one can construct in principle, but not all of them appear in the OPE of ${\cal O}_2$ and ${\cal O}_3$.  Indeed, operators with $k=5, 6$ can only appear in the ${\cal O}_2 \times {\cal O}_3$ and ${\cal O}_3 \times {\cal O}_3$ OPEs, respectively, so we can simply define ${\cal O}_5$ and ${\cal O}_6$ from the OPEs themselves.  They are $\cO_2 \cO_3$ and $\cO_3 \cO_3$, respectively.  Operators with $k=4$ appear both in $\cO_2 \times \cO_2$ and $\cO_3 \times \cO_3$, so in general there are two such operators we should consider, one of them being $\cO_2 \cO_2$ and the other being a linear combination of it and $\tr \widehat {\cal X}^4$. In summary:
\es{Internal}{
  (B,+)^{[0020]}_{1,0} \to j_F = 1: \qquad &{\cal O}_2 \,, \\
  (B,+)^{[0030]}_{\frac32, 0} \to j_F = \frac32: \qquad &{\cal O}_3  \,, \\
   (B,+)^{[0040]}_{2,0} \to j_F = 2: \qquad &\cO_{4,1} = (\cO_2)^2 \,, \\
 &\cO_{4,2} =  \tr \widehat {\cal X}^4  + B_{4,4} {\cal O}_{4, 1}\\
  &\quad \ \ \, =  \tr {\cal X}^4 - \frac{4}{N} \tr {\cal X} \tr {\cal X}^3+ B_{4,4} {\cal O}_{4, 1} \,, \\
    (B,+)^{[0050]}_{\frac52, 0} \to j_F = \frac52: \qquad &\cO_5 = \cO_3 \cO_2 \,, \\
  (B,+)^{[0060]}_{3,0} \to j_F = 3: \qquad &\cO_6 = (\cO_3)^2  \,.
}
Here, the coefficient $B_{4,4}$ appearing in the definition of the second operator with $k=2$ is fixed by requiring $\langle \cO_{4,1} \cO_{4,2} \rangle = 0$.\footnote{In the $U(3)$ theory, orthogonality of $\cO_{4,1}$ and $\cO_{4,2}$ implies $\cO_{4,2}$ vanishes identically, i.e.~there is only a single operator with $(\Delta, j_F) = (2,2)$ in the $\cO_3 \times \cO_3$ OPE.}

Next we consider operators which descend from the quarter-BPS superconformal primaries in the second line of \eqref{InternalMult}.  For a given dimension $\Delta$ of the 3d operator, the two possibilities for the 1d operators under consideration are operators $\cO'_{2\Delta}$ with $j_F = \Delta - 1$ and operators $\cO''_{2\Delta}$ with $j_F = \Delta-2$.   All these operators are the unique operators appearing in the OPEs of $\cO_2$ and $\cO_3$, so (up to mixing with lower dimension operators) they can simply be written as products of $\cO_2$ and $\cO_3$, projected onto the spin $j_F$ sector:  
 \es{Internal2}{
  (B,2)^{[0130]}_{\frac52, 0} \to j_F = \frac32: \qquad &{\cal O}_5' =\epsilon^{ac} ({\cal O}_2)_{ab}({\cal O}_3)_{cde} y^b y^d y^e \,, \\
  (B,2)^{[0200]}_{2,0} \to  j_F = 0: \qquad  &{\cal O}_4'' =\epsilon^{ac} \epsilon^{bd} ({\cal O}_2)_{ab}({\cal O}_2)_{cd} + B''_{4,0} \mathds{1}  \,, \\
  (B,2)^{[0210]}_{\frac52,0} \to  j_F = \frac 12: \qquad  &{\cal O}_5'' =\epsilon^{ac} \epsilon^{bd} ({\cal O}_2)_{ab}({\cal O}_3)_{cde} y^e \,, \\
  (B,2)^{[0220]}_{3,0} \to  j_F = 1: \qquad &{\cal O}_6'' =\epsilon^{ad} \epsilon^{be} ({\cal O}_3)_{abc}({\cal O}_3)_{def} y^c y^f + B''_{6,2} {\cal O}_2 \,.
 }
Here, $\mathds{1}$ is identity operator. The coefficients $B''_{4,0}, B''_{6,2}$ are fixed by requiring different operators of the same isospin be orthogonal, i.e.~$\langle \cO_4'' \rangle = 0$ and $\langle \cO''_6 \cO_2 \rangle = 0$, respectively.

Notably absent from the interacting sector are operators that descend from 3d operators in the $[0010]$ and $[0110]$ irreps of $\mathfrak{so}(8)_R$.   While symmetry considerations do not forbid these operators from appearing in the relevant OPEs, they cannot be constructed in the 1d theory corresponding to the interacting sector of the ABJM$_{N, 1}$ theory.  Indeed, had they existed, these operators would have $j_F = \frac12$ and satisfy $j_F = \Delta$ and $j_F = \Delta -1$, respectively. The only operator satisfying the first of these relations (namely $\Delta = j_F = 1/2$) is $\tr {\cal X}$ itself, which is an operator in the free sector.  Similarly, operators obeying $j_F = 1/2$ and $\Delta = 3/2$ cannot be constructed in the interacting sector either.

\subsection{Correlation functions}

As mentioned previously, we can compute correlation functions of arbitrary operator insertions through Wick contractions and subsequently performing the matrix integral over $\sigma$ using \eqref{1dCorrelators}. For those under consideration, the number of fundamental fields that enter into the computation is large, rendering the combinatorics unwieldy. That being said, with some effort the fixed $\sigma$ correlators can be computed exactly as a function of $N$. For instance, the two-point functions of the external operators are:
\es{sampleABJMCorr}{
\langle \cO_2(\vphi_1,y_1) \cO_2(\vphi_2,y_2) \rangle_\sigma &= \left( \frac{N^2-1}{2 \ell ^2} - \frac{1}{2\ell^2} \sum _{\alpha _1=1}^N \sum _{\alpha _2=1}^N
   \tanh^2 (\pi \sigma_{\alpha _1,\alpha _2}) \right) \langle y_1, y_2 \rangle^2 (\sgn \vphi_{21})^2  \,, \\
\langle \cO_3(\vphi_1,y_1) \cO_3(\vphi_2,y_2) \rangle_\sigma &= \left( \frac{3
   \left(N^4-5 N^2+4\right)}{8 N \ell ^3} + \frac{9}{4 N \ell ^3} \sum _{\alpha _1=1}^N \sum _{\alpha _2=1}^N 
   \tanh^2 ( \pi \sigma_{\alpha _1,\alpha _2})  \right. \\ &\left. -\frac{9}{8 \ell ^3}\sum
   _{\alpha _1=1}^N \sum _{\alpha _2=1}^N \sum
   _{\alpha _3=1}^N  \tanh (\pi \sigma_{\alpha _1 \alpha _2})
   \tanh (\pi \sigma_{\alpha _1\alpha _3}) \right)\langle y_1, y_2 \rangle^3 (\sgn \vphi_{21})^3  \,.
}
The expressions for the other correlators are quite complicated, so we relegate them to Appendix \ref{1dGaussianCorrelators}.
 
Although we can determine the mixed correlators in the Gaussian theory for arbitrary $N$, it is difficult to compute the full observables in the 1d sector when the rank of the gauge group is large; indeed, numerical integration over large-dimensional spaces is infeasible when a high degree of precision is required.\footnote{Monte Caro and related numerical integration methods are typically the only option for large-dimensional integrals, but these techniques have infamously slow square-root convergence and so are unsuitable for our purposes.}
Consequently, we restrict ourselves to the computation of the 1d data in $\text{ABJM}^\text{int}_{N,1}$ with $N=3,4$ and list the results in Tables \ref{2ptCoeff} and \ref{3ptCoeff}.\footnote{The full correlators can be shown to obey extra relations for all $N$, which we list in Appendix~\ref{extraRel}.}

\begin{table}[htpbp]
\begin{center}
\begin{tabular}{|l|c|c|}
\hline
 \multicolumn{1}{|c|}{$B_{\cO}$} & $\text{ABJM}^\text{int}_{3,1}$ & $\text{ABJM}^\text{int}_{4,1}$ \\
   \hline
$B_{\cO_2}$ & $\frac{10 \pi -31}{2 (\pi -3) \ell ^2}$ & $2.387353007033 \ell^{-2}$ \\ \hline
$B_{\cO_3}$ & $\frac{35 \pi -111}{8 (\pi -3) \ell ^3}$ & $2.376727867088 \ell^{-3}$ \\ \hline
$B_{\cO_{4,1}}$ & $\frac{840 \pi -2629}{10 (\pi -3) \ell ^4}$
   & $15.75431552556 \ell^{-4}$ \\ \hline
$B_{\cO_{4,2}}$ & $0$ & $1.232222330097 \ell^{-4}$ \\ \hline
$B_{\cO_5}$ & $\frac{1540 \pi -4849}{20 (\pi -3) \ell
   ^5}$ & $12.06393636645 \ell^{-5}$ \\ \hline
$B_{\cO_6}$ & $\frac{33845 \pi -106182}{160 (\pi -3) \ell
   ^6}$ & $30.17627414811 \ell^{-6}$ \\ \hline
$B_{\cO'_5}$ & $\frac{3745 \pi -11772}{32 (\pi -3) \ell
   ^5}$ & $5.646860418180 \ell^{-5}$ \\ \hline
$B_{\cO''_4}$ & $\frac{3 (3888+\pi  (420 \pi -2557))}{4
   (\pi -3)^2 \ell ^4}$ & $23.88992289342 \ell^{-4}$ \\ \hline
$B_{\cO''_5}$ & $0$ & $4.067464511258 \ell^{-5}$ \\ \hline
$B_{\cO''_6}$ & $\frac{2675592+5 \pi  (55440 \pi
   -344503)}{144 (\pi -3) (10 \pi -31) \ell ^6}$ & $7.238637771987 \ell^{-6}$\\
\hline
\end{tabular}
\end{center}
\caption{Values of unnormalized two-point coefficients in the 1d theory corresponding to superconformal primaries in ABJM$^\text{int}_{N,1}$ with $N=3,4$.}\label{2ptCoeff}
\end{table}

 \begin{table}[htpbp]
\begin{center}
\begin{tabular}{|l|c|c|}
\hline
 \multicolumn{1}{|c|}{$C_{{\cal O}_A {\cal O}_B {\cal O}_C}$} & $\text{ABJM}^\text{int}_{3,1}$ & $\text{ABJM}^\text{int}_{4,1}$ \\
   \hline
$C_{\cO_2 \cO_2 \cO_2}$ & $\frac{31-10 \pi }{(\pi -3) \ell ^3}$ & $4.774706014065 \ell^{-3}$ \\ \hline
$C_{\cO_2 \cO_2 \cO_4}$ & $\frac{840 \pi -2629}{10 (\pi -3) \ell ^4}$
   & $15.75431552556 \ell^{-4}$ \\ \hline
$C_{\cO_2 \cO_2 \cO''_4}$ & $\frac{3888+\pi  (420 \pi -2557)}{4 (\pi
   -3)^2 \ell ^4}$ & $7.963307631139 \ell^{-4}$ \\ \hline
$C_{\cO_3 \cO_3 \cO_2}$ & $\frac{3 (111-35 \pi )}{8 (\pi -3) \ell ^4}$ &
  $7.130183601265 \ell^{-4}$ \\ \hline
$C_{\cO_3 \cO_3 \cO_{4,1}}$ & $\frac{3 (840 \pi -2629)}{40 (\pi -3) \ell
   ^5}$ & $13.47542823883 \ell^{-5}$ \\ \hline
$C_{\cO_3 \cO_3 \cO_{4,2}}$ & $0$ & $5.545000485437 \ell^{-5}$\\ \hline
$C_{\cO_3 \cO_3 \cO''_4}$ & $-\frac{3 (3888+\pi  (420 \pi -2557))}{8
   (\pi -3)^2 \ell ^5}$ & $-8.958721085031 \ell^{-5}$ \\ \hline
$C_{\cO_3 \cO_3 \cO_6}$ & $\frac{33845 \pi -106182}{160 (\pi -3) \ell
   ^6}$ & $30.17627414811 \ell^{-6}$ \\ \hline
$C_{\cO_3 \cO_3 \cO''_6}$ & $\frac{2675592+5 \pi  (55440 \pi
   -344503)}{160 (\pi -3) (10 \pi -31) \ell ^6}$ & $6.514773994788 \ell^{-6}$ \\ \hline
$C_{\cO_2 \cO_3 \cO_3}$ & $\frac{3 (111-35 \pi)}{8 (\pi -3) \ell
   ^4}$ & $7.130183601265 \ell^{-4}$ \\ \hline
$C_{\cO_2 \cO_3 \cO_5}$ & $\frac{4849 - 1540\pi}{20 (\pi -3) \ell ^5}$
  & $12.06393636645 \ell^{-5}$ \\ \hline
$C_{\cO_2 \cO_3 \cO'_5}$ & $\frac{3 (11772-3745 \pi)}{80 (\pi -3)
   \ell ^5}$  & $6.776232501815 \ell^{-5}$ \\ \hline
$C_{\cO_2 \cO_3 \cO''_5}$ & $0$ & $2.033732255629 \ell^{-5}$ \\
  \hline
\end{tabular}
\end{center}
\caption{Values of unnormalized three-point coefficients in the 1d theory corresponding to superconformal primaries in ABJM$^\text{int}_{N,1}$ with $N=3,4$.}\label{3ptCoeff}
\end{table}

\subsection{OPE coefficients of ABJM$_{N,1}^\text{int}$}

Using our results for the two- and three-point functions in Tables \ref{2ptCoeff} and \ref{3ptCoeff}, we can use the dictionary \eqref{1dTo3dBlock} to extract the 3d OPE coefficients.  We list our results for ABJM$^\text{int}_{N,1}$ with $N=3,4$ in Table \ref{ABJMvalues}. 
As a consistency check, it is straightforward to verify that these coefficients obey the 1d crossing relations in \eqref{1dCrossing}.

\begin{table}[htpbp]
\begin{center}
\begin{tabular}{|l|c|c|c|}
\hline
 \multicolumn{1}{|c|}{Type $\cal M$} & OPE & ABJM$_{3,1}$ $\lambda_{\cal M}$  & ABJM$_{4,1}$ $\lambda_{\cal M}$ \\
   \hline
$(B,+)_{\frac12,0}^{[0010]}$&$S_2 \times S_3$& $0$ & $0$\\
  \hline
  $(B,+)_{1,0}^{[0020]}$ &$S_2 \times S_2$& $4 \sqrt{\frac{\pi -3}{10 \pi -31}}$ & $1.83057140076$\\
  &$S_3 \times S_3$& $6 \sqrt{\frac{\pi -3}{10 \pi -31}}$ & $2.74585710114$\\
  \hline
$(B,+)_{\frac32,0}^{[0030]}$&$S_2 \times S_3$& $4 \sqrt{\frac{3 (\pi -3)}{10 \pi -31}}$ & $3.17064267300$\\
  \hline
   $(B,+)_{2,0}^{[0040],1}$ &$S_2 \times S_2$& $\frac{4 \sqrt{\frac{1}{15} (\pi -3) (840 \pi -2629)}}{10 \pi -31}$ & $2.71498537721$\\
  &$S_3 \times S_3$& $\frac{12 \sqrt{\frac{2}{5} (\pi -3) (840 \pi -2629)}}{111-35 \pi }$ & $2.33263997442$\\
   $(B,+)_{2,0}^{[0040],2}$ &$S_3 \times S_3$& $0$ & $3.43211600867$\\
  \hline
  $(B,+)_{\frac52,0}^{[0050]}$&$S_2 \times S_3$& $\frac{8}{5} \sqrt{\frac{(4849-1540 \pi ) (\pi -3)}{(111-35 \pi ) (10 \pi -31)}}$ & $2.60838358490$\\
  \hline
  $(B,+)_{3,0}^{[0060]}$&$S_3 \times S_3$& $\frac{4 \sqrt{2 (\pi -3) (33845 \pi -106182)}}{5 (111-35 \pi )}$ & $4.13455086483$\\
  \hline
  $(B,2)_{\frac32,0}^{[0110]}$ &$S_2 \times S_3$& $0$ & $0$\\
  \hline
  $(B,2)_{2,0}^{[0200]}$ &$S_2 \times S_2$& $\frac{4 \sqrt{\frac{1}{3} (3888+\pi  (420 \pi -2557))}}{10 \pi -31}$ & $2.72979145342$\\
  &$S_3 \times S_3$& $-\frac{4 \sqrt{\frac{3}{5} (\pi -3) (840 \pi -2629)}}{111-35 \pi }$ & $-3.08474433097$\\
  \hline
  $(B,2)_{\frac52,0}^{[0210]}$ &$S_2 \times S_3$& $0$ & $2.39474254580$\\
  \hline
  $(B,2)_{\frac52,0}^{[0130]}$ &$S_2 \times S_3$& $\frac{8}{5} \sqrt{\frac{3 (11772-3745 \pi ) (\pi -3)}{(111-35 \pi ) (10 \pi -31)}}$ & $3.90976887215$\\
  \hline
   $(B,2)_{3,0}^{[0220]}$ &$S_3 \times S_3$& $\frac{12 (2675592-5 (344503-55440 \pi ) \pi ) \sqrt{\frac{2 (\pi -3)}{(10 \pi -31) \left(8094992-5159515 \pi +822200 \pi ^2\right)}}}{5 (111-35 \pi )}$ & $5.76323677864$\\
  \hline
\end{tabular}
\end{center}
\caption{Values of signed OPE coefficients of the superconformal primaries of short multiplets in correlators of $S_2$ and $S_3$ in ABJM$^\text{int}_{3,1}$ and ABJM$^\text{int}_{4,1}$. By construction, the multiplet $(B,+)_{2,0}^{[0040],1}$ appears in both the $S_2 \times S_2$ and $S_3 \times S_3$ OPEs, while $(B,+)_{2,0}^{[0040],2}$ appears only in $S_3 \times S_3$. All of the coefficients except $\lambda_{33 (B,2)^{[0200]}_{2,0}}$ are non-negative.} \label{ABJMvalues}
\end{table}

\section{Numerical bootstrap}
\label{numerics}

We will now use the results of the previous sections to derive the mixed correlator crossing equations for interacting 3d $\mathcal{N}=8$ SCFTs, taking into account some important redundancies due to supersymmetry. We then use these crossing equations to perform the numerical bootstrap for general $N$, i.e. $c_T$, as well as $N=3,4$, using the values of the protected OPE coefficients derived in Section \ref{1d}. We will also show results of the single correlator $\langle2222\rangle$ bootstrap for $N\geq2$ using the OPE coefficients from Section \ref{1d}, as well as the all orders in $1/N$ expression in \cite{Agmon:2017xes} for the OPE coefficients that appear in the $S_2\times S_2$ OPE.

\subsection{Crossing equations}
\label{secCrossing}

To derive the crossing equations for the four point functions $\langle 2222\rangle$, $\langle 3333\rangle$, and $\langle 2233\rangle$, we equate all the independent channels:
\es{crossings}{
&\langle S_2(\vec x_1,Y_1)S_2(\vec x_2,Y_2)S_2(\vec x_3,Y_3)S_2(\vec x_4,Y_4) \rangle=\langle S_2(\vec x_3,Y_3)S_2(\vec x_2,Y_2)S_2(\vec x_1,Y_1)S_2(\vec x_4,Y_4) \rangle\,,\\
&\langle S_3(\vec x_1,Y_1)S_3(\vec x_2,Y_2)S_3(\vec x_3,Y_3)S_3(\vec x_4,Y_4) \rangle=\langle S_3(\vec x_3,Y_3)S_3(\vec x_2,Y_2)S_3(\vec x_1,Y_1)S_3(\vec x_4,Y_4) \rangle\,,\\
&\langle S_2(\vec x_1,Y_1)S_3(\vec x_2,Y_2)S_2(\vec x_3,Y_3)S_3(\vec x_4,Y_4) \rangle=\langle S_2(\vec x_3,Y_3)S_3(\vec x_2,Y_2)S_2(\vec x_1,Y_1)S_3(\vec x_4,Y_4) \rangle\,,\\
&\langle S_2(\vec x_1,Y_1)S_2(\vec x_2,Y_2)S_3(\vec x_3,Y_3)S_3(\vec x_4,Y_4) \rangle=\langle S_3(\vec x_3,Y_3)S_2(\vec x_2,Y_2)S_2(\vec x_1,Y_1)S_3(\vec x_4,Y_4) \rangle\,.\\
}
In terms of the $A^{k_{12},k_{34}}_{nm}(U,V)$ basis in \eqref{Ybasis}, we can use the explicit expressions for the $\mathfrak{so}(8)_R$ structures $Y^{0,0}_{nm}(\sigma,\tau)$, $Y^{-1,-1}_{nm}(\sigma,\tau)$, and $Y^{1,1}_{nm}(\sigma,\tau)$ in Appendix \ref{Ys} to write these crossing relations as a 34-dimensional vector $\vec V(U,V)$ whose entries are linear combinations of
\es{Fs}{
F_{\pm,nm}^{k_{1}k_2k_{3}k_4}(U,V)\equiv V^{\frac{k_2+k_3}{2}}A_{nm}^{k_{12},k_{34}}(U,V)\pm U^{\frac{k_2+k_3}{2}}A_{nm}^{k_{12},k_{34}}(V,U)
}
for the irreps $[0\,n-m\, 2m\,0]$ that appear in $[0\,0\,2\,0]\times [0\,0\,2\,0]$, $[0\,0\,2\,0]\times [0\,0\,3\,0]$, and $[0\,0\,3\,0]\times [0\,0\,3\,0]$. The explicit expression for $\vec V(U,V)$ can be found in Appendix \ref{Vs}.

Combining the crossing equations with the superconformal block decomposition in \eqref{FourPoint2}, \eqref{SBDecomp}, \eqref{GExpansion}, we can then define a $\vec V_\mathcal{M}$ for each supermultiplet $\mathcal{M}$ listed in Tables \ref{opemult4} and \ref{opemult5} by replacing each $A_{nm}^{k_{12},k_{34}}$ in $\vec V$ by the linear combination of conformal blocks that appears for the superblock $\mathfrak{G}^{k_{12},k_{34}}_{\mathcal{M}}$.  The quantity $\vec{V}_{\cal M}$ is matrix-valued and denoted by $\vec{\bf V}_{\cal M}$ in the case where ${\cal M}$ appears in several correlation functions.   For instance, we use the explicit expression for the stress tensor multiplet $(B,+)_{1,0}^{[0020]}$ superblock in \eqref{stressBlock} to write $\vec {\bf V}_{(B,+)_{1,0}^{[0020]}}$ in \eqref{stressV}. The crossing equations in terms of these $\vec V_\mathcal{M}$ are
\es{crossing}{
0=\sum_{\mathcal{M}\in S_2\times S_2} \begin{pmatrix} \lambda_{22\mathcal{M} } &  \lambda_{33\mathcal{M}} \end{pmatrix}\vec {\bf V}_\mathcal{M} \begin{pmatrix} \lambda_{22\mathcal{M}}  \\  \lambda_{33\mathcal{M}} \end{pmatrix}+\sum_{\mathcal{M}\in S_2\times S_3}\lambda^2_{23\mathcal{M}}\vec V_\mathcal{M}+\sum_{\substack{\mathcal{M}\in S_3\times S_3\\\text{s.t.}\notin S_2\times S_2}}\lambda^2_{33\mathcal{M}}\vec V_\mathcal{M}\,.
}
We can simplify these crossing equations in four steps. First we restrict to interacting SCFTs by not including the free multiplet $(B,+)_{\frac12,0}^{[0010]}$. Second, we normalize the unit operator OPE coefficient as $\lambda_{k_ik_i \text{Id}}=1$. Third, we relate OPE coefficients with two stress tensor multiplets to $1/\sqrt{c_T}$ as in \eqref{opetoCT}. Lastly, we use the 1d crossing relations \eqref{1dCrossing} as well as the relations \eqref{opetoCT} to set $\lambda_{(B,2)_{\frac32,0}^{0110}}=0$ and write all the remaining pairs of short OPE coefficients in terms of $c_T$, $\lambda^2_{kk(B,2)_{2,0}^{[0200]}}$, $\lambda^2_{(B,+)_{3,0}^{[0060]}}$, $\lambda^2_{(B,+)_{\frac52,0}^{[0050]}}$, $\lambda^2_{(B,2)_{\frac52,0}^{[0130]}}$. The resulting crossing equations are
\es{crossing2}{
0=&\vec V_\text{Id}+ \frac{1}{c_T}\vec V_{(B,+)_{1,0}^{[0020]}}+ \vec V^{2222}_{(B,2)_{2,0}^{[0200]}}\lambda^2_{22(B,2)_{2,0}^{[0200]}}+ \vec V^{3333}_{(B,2)_{2,0}^{[0200]}}\lambda^2_{33(B,2)_{2,0}^{[0200]}}\\
&+ \vec V'_{(B,+)_{3,0}^{[0060]}}\lambda^2_{33(B,+)_{3,0}^{[0060]}}+ \vec V'_{(B,+)_{\frac52,0}^{[0050]}}\lambda^2_{23 (B,+)_{\frac52,0}^{[0050]} }+ \vec V'_{(B,2)_{\frac52,0}^{[0130]}} \lambda^2_{23 (B,2)_{\frac52,0}^{[0130]} }\\
&+\hspace{-.2cm}\sum_{\mathcal{M}\in S_2\times S_2\vert_A} \begin{pmatrix} \lambda_{22\mathcal{M} } &  \lambda_{33\mathcal{M}} \end{pmatrix}\vec {\bf V}_\mathcal{M} \begin{pmatrix} \lambda_{22\mathcal{M}}  \\  \lambda_{33\mathcal{M}} \end{pmatrix}+\hspace{-.2cm}\sum_{\mathcal{M}\in S_2\times S_3\vert_A}\lambda^2_{23\mathcal{M}}\vec V_\mathcal{M}+\hspace{-.2cm}\sum_{\substack{\mathcal{M}\in S_3\times S_3\vert_A\\\text{s.t.}\notin S_2\times S_2\vert_A}}\hspace{-.2cm}\lambda^2_{33\mathcal{M}}\vec V_\mathcal{M}\,,
}
where the $\vec V_\mathcal{M}$ in the first couple lines are written as linear combinations of $\vec V_\mathcal{M}$ for other short multiplets, whose explicit form are given in Appendix \ref{Vs}, and $\vert_A$ in the last line denotes that only the semishort and long $A$-type multiplets are considered here.

\subsection{Redundant equations}
\label{secRedundant}

As was first noted for the $\langle2222\rangle$ crossing equations in \cite{Chester:2014fya}, $\mathcal{N}=8$ superconformal symmetry makes many crossing equations linearly dependent. If the redundancies in the 34 crossing equations in $\vec V_{M}$ are not removed, then they will cause numerical instabilities in the bootstrap algorithm. To analyze most of these redundancies, we do not need to consider explicit conformal blocks, and can instead consider the general crossing relations $\vec V$ in \eqref{V} that is written in terms of crossed functions $F_{\pm,nm}^{k_{1}k_2k_{3}k_4}(U,V)$ \eqref{Fs} of generic functions $A_{nm}^{k_{12},k_{34}}(U,V)$ of the conformal cross ratios. For the numerical bootstrap we consider the expansion of these functions around the crossing symmetric point $z=\bar z=\frac12$ :
\es{crossDer}{
F_{+,nm}^{k_{1}k_2k_{3}k_4}(U,V)&=\sum_{\substack{p+q=\text{even}\\\text{s.t. }p\leq q}}\frac{2}{p!q!}\left(z-\frac12\right)^p\left(\bar z-\frac12\right)^q\partial^p_z\partial_{\bar z}^q F_{+,nm}^{k_{1}k_2k_{3}k_4}(U,V)\vert_{z=\bar z=\frac12}\,,\\
F_{-,nm}^{k_{1}k_2k_{3}k_4}(U,V)&=\sum_{\substack{p+q=\text{odd}\\\text{s.t. }p<q}}\frac{2}{p!q!}\left(z-\frac12\right)^p\left(\bar z-\frac12\right)^q\partial^p_z\partial_{\bar z}^q F_{-,nm}^{k_{1}k_2k_{3}k_4}(U,V)\vert_{z=\bar z=\frac12}\,,
}
where $z,\bar z$ are written in terms of $U,V$ in \eqref{UVtozzbar}, and in the sums we only consider terms are nonzero and independent according to the definition \eqref{Fs}. We can then truncate these sums to a finite number of terms by imposing that
\es{nmax}{
p+q\leq\Lambda\,,
}
and then consider the finite dimensional matrix $\widetilde V_i^{(p,q)}$ whose rows as labeled by $i=1,\dots34$ are those of $\vec V$, and whose columns as labeled by ${(p,q)}$ are the coefficients of the $\partial^p_z\partial^q_{\bar z} A_{nm}^{k_{12},k_{34}}(U,V)\vert_{z=\bar z=\frac12}$ that appear in each entry of $\vec V$ after expanding like \eqref{crossDer} using the definition \eqref{Fs} of $F_{\pm,nm}^{k_{1}k_2k_{3}k_4}(U,V)$ in terms of $A_{nm}^{k_{12},k_{34}}(U,V)$. The superconformal Ward identities \eqref{ward} as expressed in the basis \eqref{Ybasis} impose relations between the $\partial^p_z\partial^q_{\bar z} A_{nm}^{k_{12},k_{34}}(U,V)\vert_{z=\bar z=\frac12}$, so we can define the analogous matrix $ V_i^{(p,q)}$ using these reduced coefficients. Unlike $\widetilde V_i^{(p,q)}$, the matrix $V_i^{(p,q)}$ is degenerate, and we find that a linearly independent subspace is given by the entries:
\es{indies2}{
\{
V_{1}^{(p,0)}\,,
V_{2}^{(p,q)}\,,
V_{7}^{(p,q)}\,,
V_{8}^{(p,q)}\,,
V_{9}^{(p,0)}\,,
V_{10}^{(p,0)}\,,
V_{11}^{(p,q)}\,,
V_{17}^{(p,q)}\,,
V_{18}^{(p,0)}\,,
V_{23}^{(p,q)}\,,
V_{24}^{(p,q)}\,,
V_{25}^{(p,0)}\,,
V_{26}^{(p,0)}\,,
V_{27}^{(p,0)}\,,
V_{28}^{(0,0)}\}\,,
}
where $V_i^{(p,0)}$ means that we only consider derivatives in $z$, and $V_i^{(0,0)}$ means that we only consider the term with no derivatives.

Even more redundancies occur once we impose the 1d crossing equations, as we did to get \eqref{crossing2}. To find these redundancies, we used the explicit expressions for the crossing equations in \eqref{crossing2} in terms of superblocks, expanded in $z,\bar z$ derivatives as in \eqref{crossDer} where now $A_{nm}^{k_{12},k_{34}}(U,V)$ is a linear combination of conformal blocks for each supermultiplet, restricted to the $V_i^{(p,q)}$ in \eqref{indies2}, and then checked numerically to see which further crossing equations were linearly independent. We found that the crossing equations in \eqref{indies2} remained linearly independent if we removed the $V_i^{(0,0)}$ component of each of the $V_{i}^{(p,0)}$ for $i=1,9,10,18,25,26,27$.

\subsection{Bootstrap algorithms}
\label{secAlg}

We now have all the ingredients to perform the numerical bootstrap using the crossing equations \eqref{crossing2} where the 1d crossing relations have been imposed, the free multiplet has been excluded, and we restrict to the linearly independent set of crossing equations in \eqref{indies2}. We will describe the bootstrap algorithms we use following \cite{Chester:2014fya,Kos:2014bka}.

Consider a linear functional $\alpha_i^{(p,q)}$ acting on the crossing equations \eqref{crossing2} expanded in $\partial^p_z,\partial^q_{\bar z}$ derivatives around the crossing symmetric point \eqref{crossDer}. The index $i$ refers to the 34 crossing equations in $ V_i$, which we have restricted to the linearly independent set in \eqref{indies2}.  In the following we will suppress the indices on $\alpha$ for simplicity.

To find upper/lower bounds on a given OPE coefficient of a short or semishort multiplet ${\cal M}^*$ that appears in \eqref{crossing2} with a scalar constraint, we consider linear functionals $\alpha$ satisfying 
 \es{CondOPE}{
  &\alpha(\vec{V}_{\cal M^*}) = s \,, \qquad\qquad \qquad\qquad\qquad \text{$s=1$ for upper bounds, $s=-1$ for lower bounds} \,,  \\
  &\alpha(\vec{V}_{\cal M}) \geq 0 \quad \text{and} \quad \alpha(\vec{{\bf V}}_{\cal M}) \succeq 0
    \,, \;\;\quad \text{for all short and semi-short ${\cal M} \notin \{ \text{Id}, (B,+)_{1,0}^{[0020]}, \cal M^* \}$} \,, \\
  &\alpha(\vec{V}_{\cal M}) \geq 0 \quad \text{and} \quad \alpha(\vec{{\bf V}}_{\cal M}) \succeq 0 \,, \quad\;\; \text{for all long $\mathcal{M}$ with $\Delta_{\mathcal{M}}\geq \Delta'_{\mathcal{M}}$}\,,
 }
where the lower bounds $\Delta'_{\mathcal{M}}$ can be taken to be the unitarity bounds $\Delta'_{(A,0)_{\Delta,\ell}^{[0000]}}=\ell+1$, $\Delta'_{(A,0)_{\Delta,\ell}^{[0100]}}=\ell+2$, $\Delta'_{(A,0)_{\Delta,\ell}^{[0020]}}=\ell+2$, and $\Delta'_{(A,0)_{\Delta,\ell}^{[0010]}}=\ell+\frac12$ in Tables \ref{opemult4} and \ref{opemult5}. If such a functional $\alpha$ exists, then this $\alpha$ applied to \eqref{crossing2} along with the reality of $\lambda_{k_1 k_2 {\cal M}}$ except, possibly, for that of $\lambda_{k_1k_2\mathcal{M}^*}$, implies that
 \es{UpperOPE}{
  &\text{if $s=1$, then}\qquad\;\;\;\lambda_{k_1k_2{\cal M}^*}^2 \leq - \alpha (\vec{V}_\text{Id})  - \frac{1}{c_T}\alpha( \vec{V}_{(B,+)_{1,0}^{[0020]}} ) \,,\\
    &\text{if $s=-1$, then}\qquad \lambda_{k_1k_2{\cal M}^*}^2 \geq  \alpha (\vec{V}_\text{Id})  + \frac{1}{c_T}\alpha( \vec{V}_{(B,+)_{1,0}^{[0020]}} ) \,,\\
 }
 for the associated $k_1,k_2$. If we are bounding the OPE coefficient squared of $(B,+)_{1,0}^{[0020]}$ itself, which is proportional to $\frac{1}{c_T}$, then we should remove the second term from the RHS of \eqref{UpperOPE}. To obtain the most stringent upper/lower bound on $\lambda_{k_1k_2{\cal M}^*}^2$, one should then minimize/maximize the RHS of \eqref{UpperOPE} under the constraints \eqref{CondOPE}. Note that for ${\cal M}^*$ that appear in \eqref{crossing2} with a matrix constraint, e.g. $(A,+)_{\ell+2,\ell}^{[0020]}$ or $(A,2)_{\ell+2,\ell}^{[0100]}$, we cannot obtain any constraints using this algorithm since the LHS of \eqref{UpperOPE} must be a scalar. We can also find islands in the space of OPE coefficients $(\lambda_{\mathcal{M}^*},\lambda_{\mathcal{M}^{**}})$ for a given $c_T$, by removing $\mathcal{M}^{**}$ from the second line of \eqref{CondOPE} and replacing the RHS of \eqref{UpperOPE} by
 \es{islandAlg}{
  \alpha (\vec{V}_\text{Id})  + \frac{1}{c_T}\alpha( \vec{V}_{(B,+)_{1,0}^{[0020]}} )+\lambda^2_{k_1k_2\mathcal{M}}\alpha(\vec{V}_{  \mathcal{M}^{**}} )\,,
 }
 where we input the value of $\lambda^2_{k_1k_2\mathcal{M}}$ by hand for the appropriate $k_1,k_2$.
 
 To find upper bounds on the scaling dimension $\Delta^*$ of the lowest dimension operator in a long supermultiplet with spin $\ell^*$ that appears in \eqref{crossing2}, we consider linear functionals $\alpha$ satisfying
  \es{CondLong}{
  &\alpha(\vec{V}_\text{Id})+ \frac{1}{c_T}\alpha( \vec{V}_{(B,+)_{1,0}^{[0020]}} ) = 1 \,,   \\
  &\alpha(\vec{V}_{\cal M}) \geq 0 \quad \text{and} \quad \alpha(\vec{{\bf V}}_{\cal M}) \succeq 0
    \,, \;\;\quad \text{for all short and semi-short ${\cal M} \notin \{ \text{Id}, (B,+)_{1,0}^{[0020]} \}$} \,, \\
  &\alpha(\vec{V}_{\cal M}) \geq 0 \quad \text{and} \quad \alpha(\vec{{\bf V}}_{\cal M}) \succeq 0 \,, \quad\;\; \text{for all long $\mathcal{M}$ with $\Delta_{\mathcal{M}}\geq \Delta'_{\mathcal{M}}$}\,,  
 }
 where we set the lower bounds $\Delta'_{\mathcal{M}}$ to their unitarity bounds except for $\Delta'_{\mathcal{M}*}$. If such a functional $\alpha$ exists, then this $\alpha$ applied to \eqref{crossing2} along with the reality of $\lambda_{k_1k_2\mathcal{M}}$ would lead to a contradiction. By running this algorithm for many values of $(c_T,\Delta'_{\mathcal{M}*})$ we can find an upper bound on $\Delta'_{\mathcal{M}*}$ in this plane.

In the above algorithms, we fixed the SCFT by inputting the value of $c_T$. We can further fix the theory by also putting in the values of all the short OPE coefficients, that can be fixed using the 1d theory in Section \ref{1d}. We should then replace the RHS of \eqref{UpperOPE} and the first line of \eqref{CondLong} by
\es{1dfix}{
&\alpha(\vec{V}_\text{Id})+ \frac{1}{c_T}\alpha( \vec{V}_{(B,+)_{1,0}^{[0020]}} )+\lambda^2_{22(B,2)_{2,0}^{[0200]}}\alpha( \vec V^{2222}_{(B,2)_{2,0}^{[0200]}}) +\lambda^2_{33(B,2)_{2,0}^{[0200]}}\alpha( \vec V^{3333}_{(B,2)_{2,0}^{[0200]}})\\
&+\lambda^2_{33(B,+)_{3,0}^{[0060]}}\alpha( \vec V'_{(B,+)_{3,0}^{[0060]}}) + \lambda^2_{23 (B,+)_{\frac52,0}^{[0050]} } \alpha(\vec V'_{(B,+)_{\frac52,0}^{[0050]}} )+ \lambda^2_{23 (B,2)_{\frac52,0}^{[0130]} } \alpha(\vec V'_{(B,2)_{\frac52,0}^{[0130]}})\,.
}
For island plots, we should furthermore add $\lambda^2_{k_1k_2\mathcal{M}}\alpha(\vec{V}_{  \mathcal{M}^{**}} )$ as in \eqref{islandAlg}.

The numerical implementation of the algorithms described above requires two truncations: one in the number of derivatives $\Lambda$ defined in \eqref{nmax} that are used to construct $\alpha$, and one in the spins $\ell_\text{max}$ that can appear for the long multiplets. For the OPE coefficient extremization we use spins in $\{0,\dots,44\}\cup\{47,48,51,52,55,56,59,60,63,64,67,68\}$ and derivatives parameter $\Lambda=35$, while for the scaling dimension upper bounds we used spins in $\{0,\dots,26\}\cup\{29,30,33,34,37,38,41,42,45,46,49,50\}$ and derivatives parameter $\Lambda=27$. For the single correlator bounds we used spins in $\{0,\dots,64\}\cup\{67,68,71,72,75,76,79,80,83,84,87,88\}$ and $\Lambda=43$. The truncated  problem can now be rephrased as a semidefinite programing problem using the method developed in \cite{Kos:2014bka}. This problem can be solved efficiently using {\tt SDPB} \cite{Simmons-Duffin:2015qma}.

\subsection{Single correlator bounds for ABJM$_{N,1}^\text{int}$ with $N\geq2$}
\label{singleBounds}

We begin by presenting bootstrap results derived from just the single correlator $\langle2222\rangle$. The bootstrap algorithm is identical to the mixed correlator algorithms presented before, except we only use the first two crossing equations $V_1^{(p,0)}$ and $V_2^{(p,q)}$ in \eqref{indies2}, and only the operators shown in black in Table \ref{opemult4} appear. Note that only three short operators appear: the stress tensor $(B,+)_{1,0}^{[0020]}$, $(B,+)_{2,0}^{[0040]}$, and $(B,2)_{2,0}^{[0200]}$. We computed these expressions for ABJM$^\text{int}_{N,1}$ with $N=3,4$ in Section \ref{1d} (the $N=2$ values were already known from \cite{Chester:2014mea}), and to all orders in $1/N$ in \cite{Agmon:2017xes}, which was found to be very accurate even down to $N=2$. We will restrict our bootstrap results to ABJM$^\text{int}_{N,1}$ by imposing the values of these short OPE coefficients. As discussed in subsection \ref{secRedundant}, we must remove the $V_1^{(0,0)}$ component to eliminate redundancies that occur when these short operators have been imposed.

 \begin{table}[htpbtp]
\begin{center}
\begin{tabular}{|l|c|c|c|c|}
\hline
 $N=$     & $\lambda^2_{(A,+)_{2,0}^{[0020]}}$ Bounds    & $\lambda^2_{(A,+)_{2,0}^{[0020]}}$ Error    & $\lambda^2_{(A,+)_{4,2}^{[0020]}}$ Bounds    & $\lambda^2_{(A,+)_{4,2}^{[0020]}}$ Error  \\
 \hline 
  2      & $9.53052-9.54437$ & $0.036\%$  &   $15.91101-15.93224$ & $0.033\%$\\
 3      & $8.19080-8.19991$ &   $0.028\%$&  $14.50650-14.51986$ & $0.023\%$ \\
4       & $7.76839-7.77502$ & $0.021\%$  & $14.06399-14.07340$ & $0.017\%$  \\
 20     & $7.16636-7.16710$ & $0.0026\%$  & $13.43291-13.43376$ & $0.0016\%$  \\
 50     & $7.11611-7.11628$ &   $0.00088\%$&  $13.38957-13.38981$ & $0.00046\%$ \\
  100       & $7.12520-7.12545$ & $0.00060\%$  & $13.37998-13.38014$ & $0.00029\%$  \\
 \hline
\end{tabular}
\end{center}
\caption{Upper and lower bounds on OPE coefficients squared for semishort $(A,+)_{\ell+2,\ell}^{[0020]}$, computed using the single correlator $\langle2222\rangle$ with $\Lambda = 43$ and the short OPE coefficients fixed to their values for ABJM$_{N,1}^\text{int}$ using the 1d theory in Section \ref{1d} for $N=2,3,4$ ($N=2$ was already reported in \cite{Chester:2014mea}), and the all orders in $1/N$ results from \cite{Agmon:2017xes} for $N=20,50,100$.  }
\label{UpLowS}
\end{table}

In Table \ref{UpLowS}, we show upper and lower bounds on the semishort $(A,+)_{\ell+2,\ell}^{[0020]}$ that appear for all even spins $\ell$ in the $S_2\times S_2$ OPE\@. We compute the percent error as the difference between these lower bounds normalized by the average:
\es{error}{
\lambda^2_{\mathcal{M}}{}^\text{\% error}=100\times\frac{\lambda^2_{\mathcal{M}}{}^\text{upper}-\lambda^2_{\mathcal{M}}{}^\text{lower}}{\left[\lambda^2_{\mathcal{M}}{}^\text{upper}+\lambda^2_{\mathcal{M}}{}^\text{lower}\right]/2}\,.
}
In general, the error is extremely small and seems to decrease with increasing spin and $N$. 

\begin{figure}[]
\begin{center}
   \includegraphics[width=0.49\textwidth]{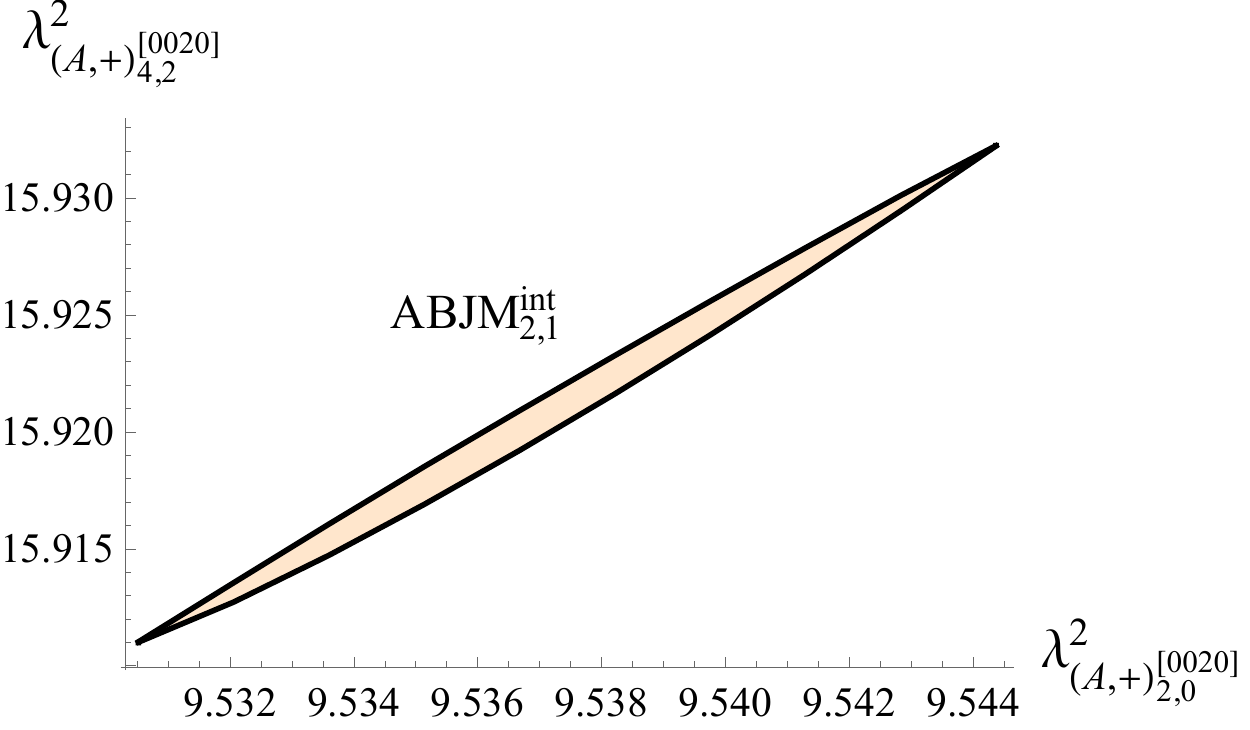}
    \includegraphics[width=0.49\textwidth]{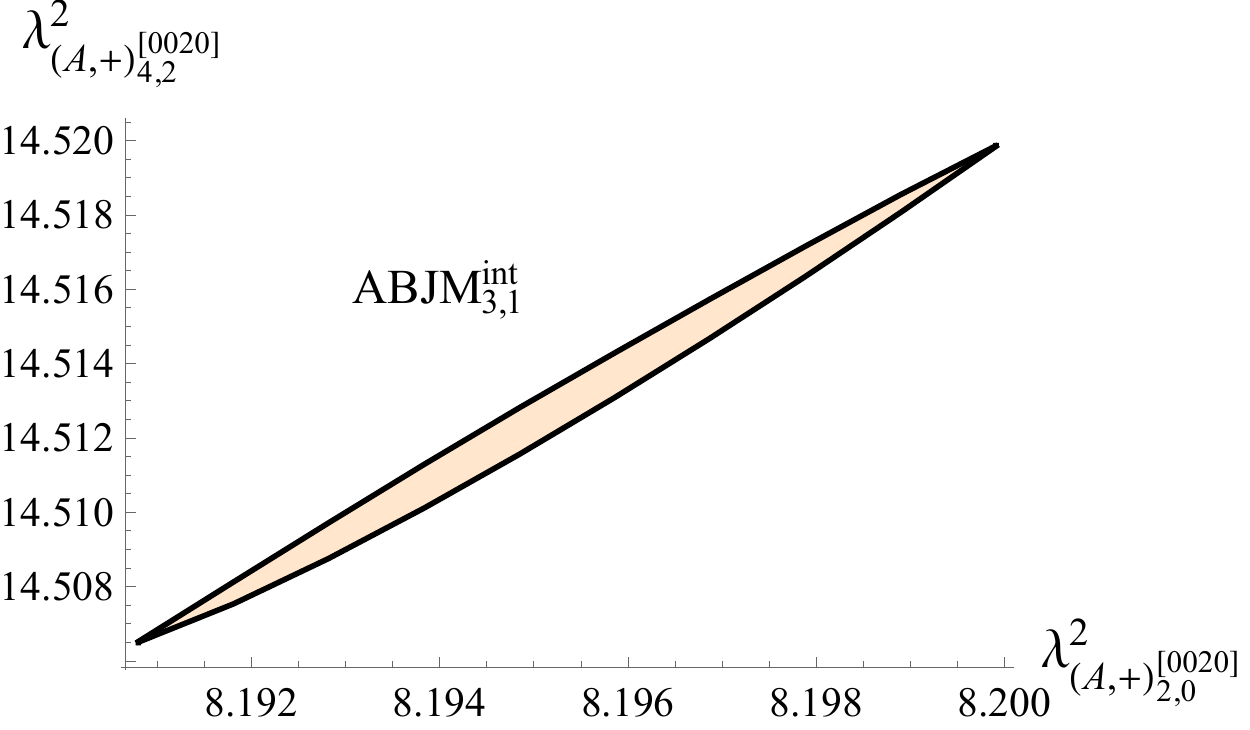}
       \includegraphics[width=0.49\textwidth]{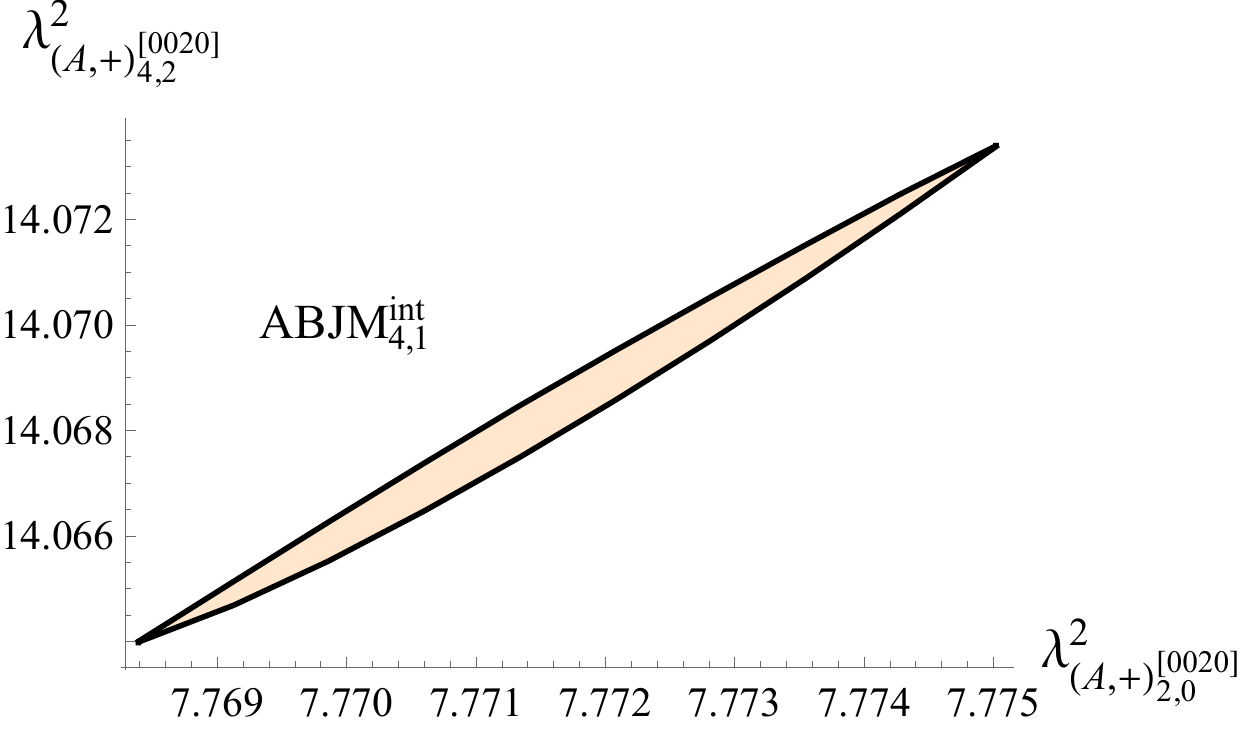}
    \includegraphics[width=0.49\textwidth]{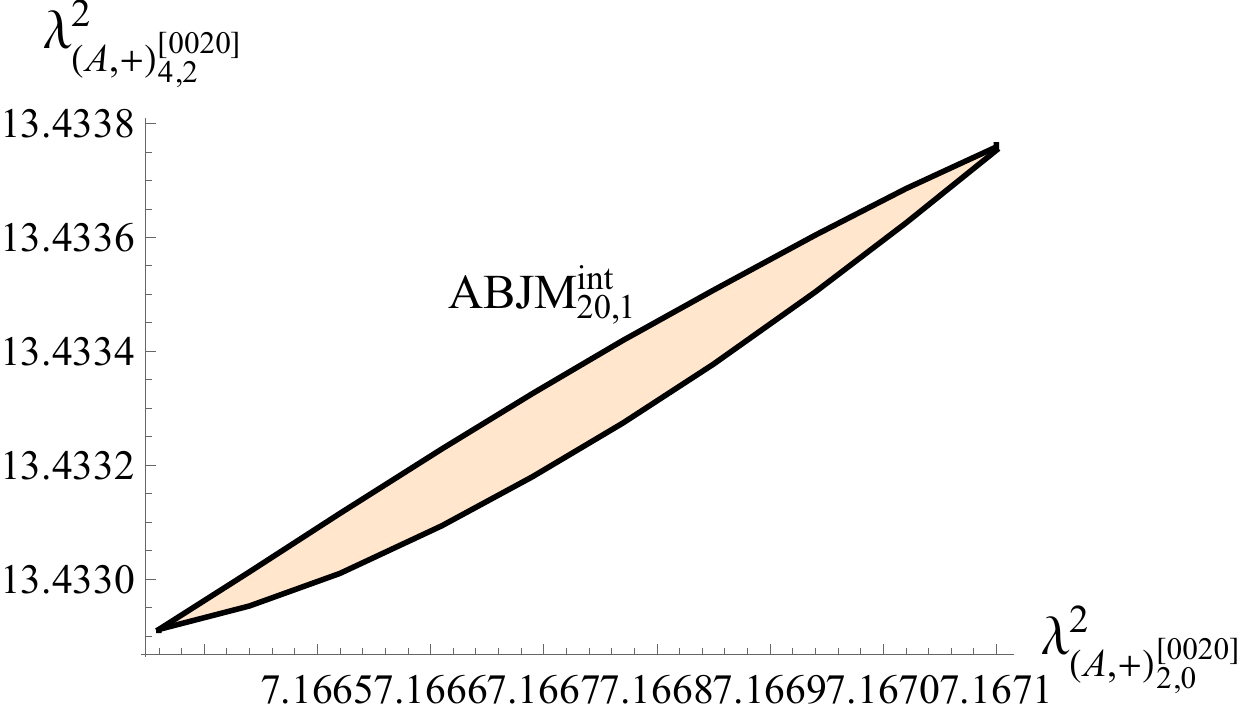}
       \includegraphics[width=0.49\textwidth]{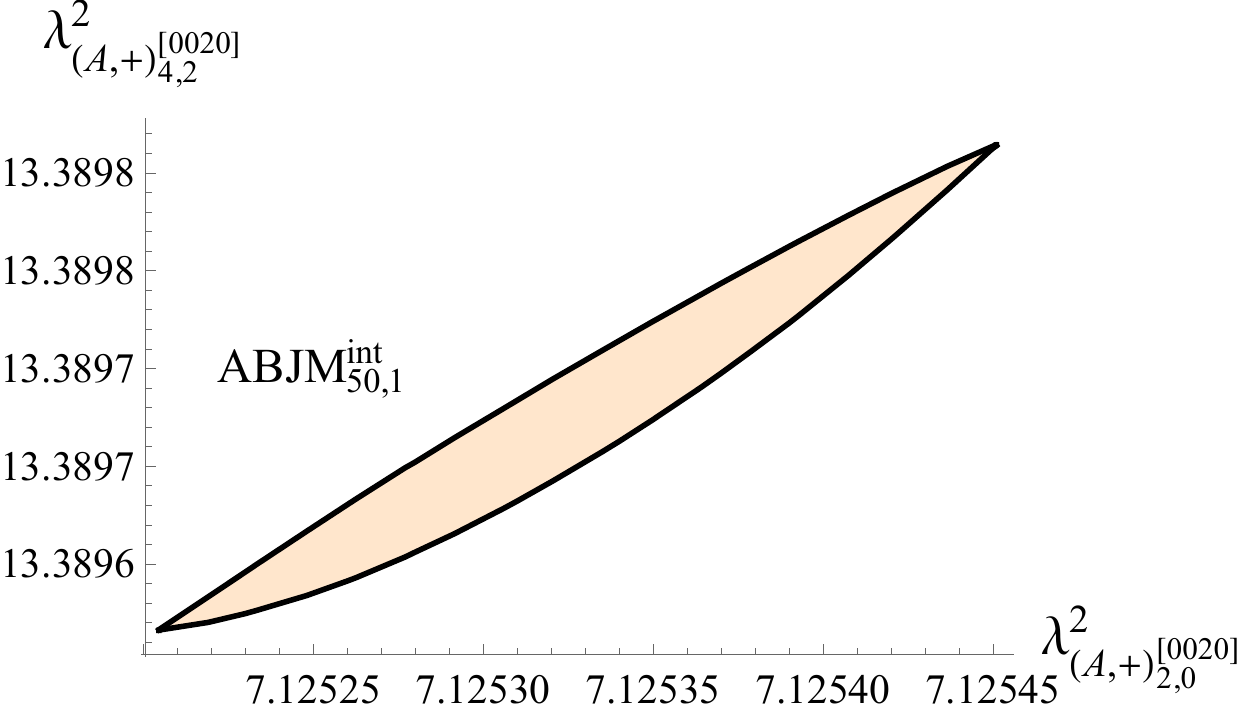}
    \includegraphics[width=0.49\textwidth]{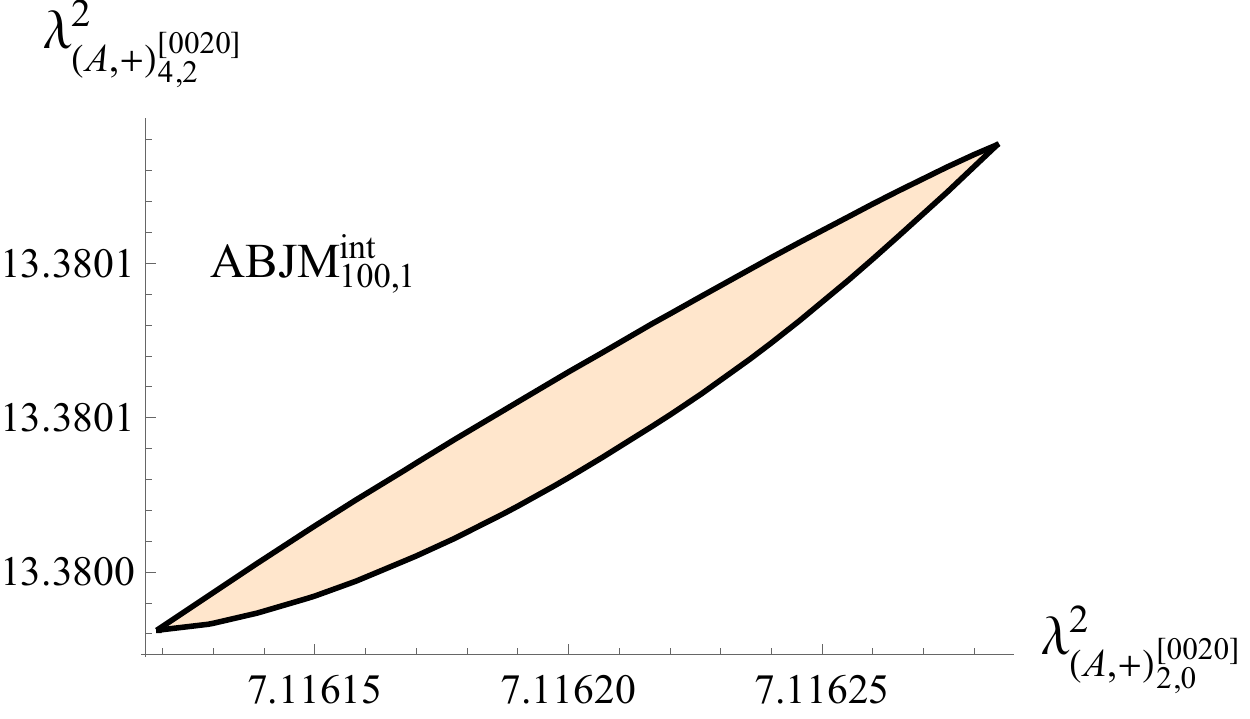}
\caption{Islands in the space of OPE coefficients $\lambda_{(A, +)_{2,0}^{[0020]}}^2$, $\lambda_{(A, +)_{4,2}^{[0020]}}^2$ for ABJM$_{N,1}^\text{int}$ for $N=2,3,4,20,50,100$, where orange is allowed.   These bounds are derived from the single correlator $\langle2222\rangle$ with $\Lambda = 43$ and the short OPE coefficients fixed to their ABJM$_{N,1}^\text{int}$ values using the 1d theory in Section \ref{1d} and \cite{Chester:2014mea} for $N=2,3,4$, and from the all orders in $1/N$ formulae in \cite{Agmon:2017xes} for $N=20,50,100$. }
\label{islandsS}
\end{center}
\end{figure}  

In Figures~\ref{islandsCombined} and~\ref{islandsS} we show islands in the space of $((A,+)_{2,0}^{[0020]},(A,+)_{4,2}^{[0020]})$ for $N=2,3,4,20,50,100$. For $N=2$ we use the exact results reported in \cite{Chester:2014mea}, for $N=3,4$ we use the exact results from Section \ref{1d}, while for $N=20,50,100$ we use the all orders in $1/N$ expression from \cite{Agmon:2017xes}. These extremely small islands interpolate between the upper and lower bounds for these values in Table \ref{UpLowS}. As noted above, these islands get smaller with increasing~$N$.

\subsection{Mixed correlator bounds for general $N$}
\label{genN}

\begin{figure}[t!]
\begin{center}
   \includegraphics[width=0.49\textwidth]{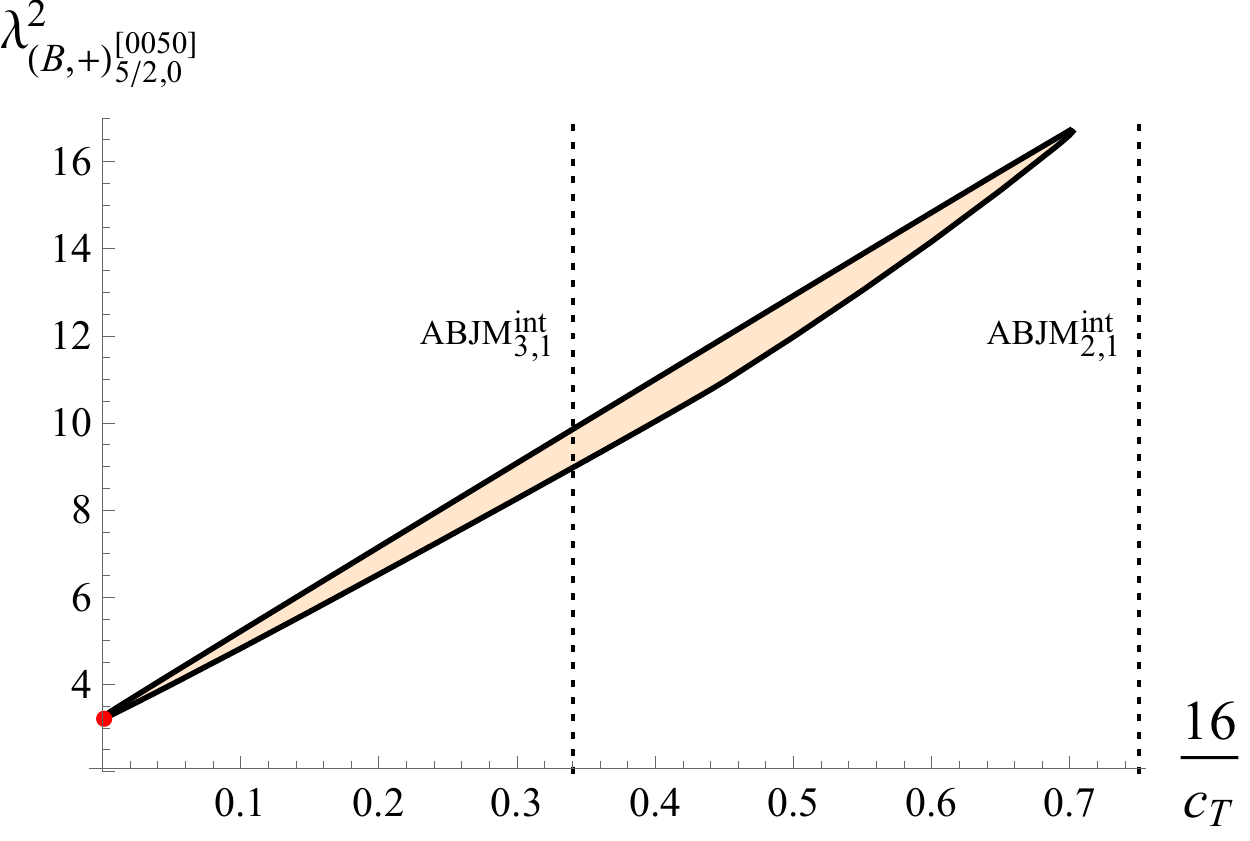}
    \includegraphics[width=0.49\textwidth]{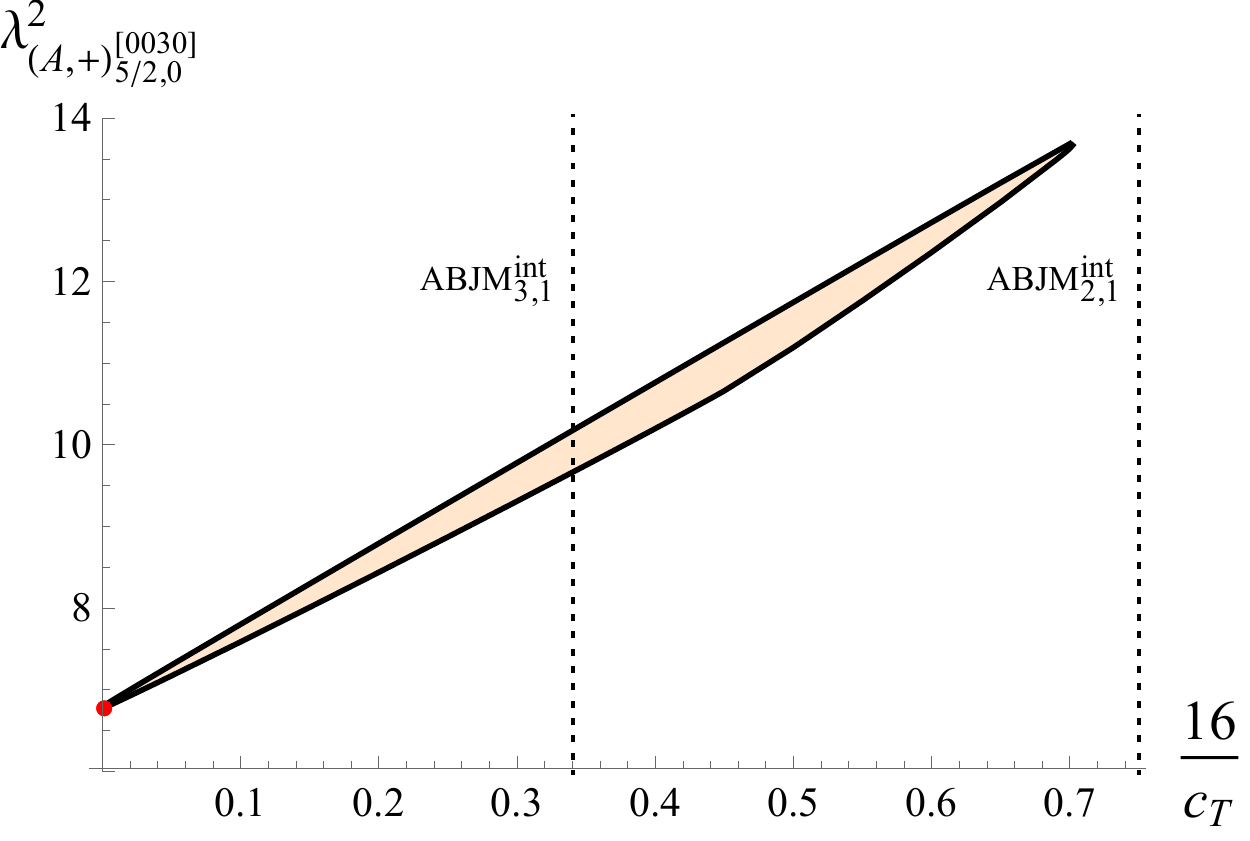}
\caption{Upper and lower bounds on $\lambda_{(B, +)_{\frac52,0}^{[0050]}}^2$ (left) and $\lambda_{(A, +)_{\frac52,0}^{[0030]}}^2$ (right) OPE coefficients, where the orange shaded regions are allowed. These bounds are computed using mixed correlators with $\Lambda = 35$, the 1d crossing relations imposed, and the OPE coefficient of the free multiplet set to zero. The red dot corresponds to the known GFFT values in Table \ref{Avalues}. The black dotted vertical lines correspond to ABJM$_{3,1}^\text{int}$ with $\frac{16}{c_T}=\frac{\pi -3}{10 \pi -31}\sim.34$ and ABJM$_{2,1}^\text{int}$ with $\frac{16}{c_T}=\frac34$.}
\label{ApBp}
\end{center}
\end{figure}  

We now move one to the full mixed correlator bootstrap, and present the results of the OPE coefficient extremization \eqref{CondOPE} and scaling dimension upper bound \eqref{CondLong} algorithms for general $N$, i.e.~$c_T$. Since many supermultiplets appear in \eqref{crossing2}, we will only show results for a representative sample of these multiplets.

We start with the upper bound on the OPE coefficient squared of the stress tensor multiplet $(B,+)_{1,0}^{[0020]}$, which is equivalent to a lower bound on $c_T$. Using $\Lambda = 35$ derivatives, we find
\es{cTbound}{
c_T\geq22.7354\qquad\Leftrightarrow\qquad \frac{16}{c_T}\leq.703749\,.
}
As expected, this excludes the free theory with $c_T^\text{free}=16$, as well as the theory with next lowest $c_T$, ABJM$_{2,1}^\text{int}\cong \text{ABJ}_1$ with $c_T^\text{ABJM$_{2,1}^\text{int}$}=21.3333$, which does not contain an $S_3$ operator and so cannot appear in this mixed bootstrap study. Curiously, this bound on $c_T$ occurs at the exact same point where a kink was observed in the $\langle2222\rangle$ study of \cite{Chester:2014fya,Chester:2014mea}, which was also where $\lambda^2_{(B,2)_{2,0}^{[0200]}}$ was observed to vanish. This kink had been conjectured to correspond to ABJM$_{2,1}^\text{int}\cong \text{ABJ}_1$ up to numerical error, which is the highest known $c_T$ theory with vanishing $\lambda^2_{(B,2)_{2,0}^{[0200]}}$, but our new upper bound suggests there must be a different explanation.

Next, in Figure \ref{ApBp} we show upper/lower bounds on the OPE coefficients squared of the short $(B,+)_{\frac52,0}^{[0050]}$ and semishort $(A,+)_{2,0}^{[0030]}$ supermultiplets as a function of $\frac{16}{c_T}$, where the orange is the allowed region. The red dot denotes the GFFT values in Table \ref{Avalues} at $c_T\to\infty$, while the dotted lines show the values of $c_T$ for various known SCFTs. As expected, both plots terminate at the upper bound on $\frac{16}{c_T}$ in \eqref{cTbound}.

\begin{figure}[t!]
\begin{center}
   \includegraphics[width=0.7\textwidth]{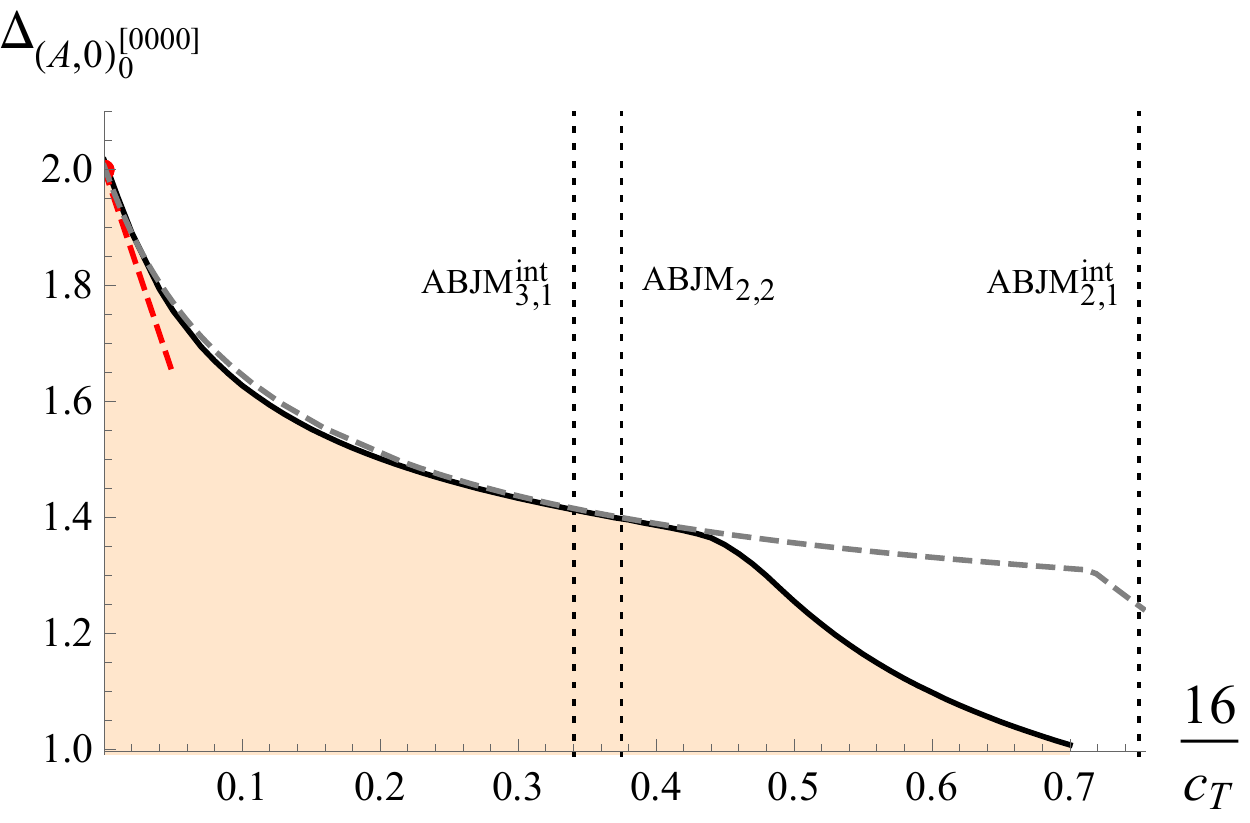}
\caption{Upper bounds on the scaling dimension $\Delta_{(A,0)_{\Delta,0}^{[0000]}}$ of the lowest dimension scalar long multiplet $(A,0)_{\Delta,0}^{[0000]}$. The black line was computed using the mixed correlators introduced in this work with $\Lambda = 27$, the 1d crossing relations imposed, and the OPE coefficient of the free multiplet set to zero. The gray dotted line corresponds upper bounds computed using just the $\langle2222\rangle$ correlator in \cite{Chester:2014fya} with $\Lambda=43$, and without any 1d crossing relations or conditions on the free theory imposed. The red dotted line shows the $\frac{1}{c_T}$ correction computed in \cite{Zhou:2017zaw}. The black dotted vertical lines correspond to ABJM$_{3,1}^\text{int}$ with $\frac{16}{c_T}=\frac{\pi -3}{10 \pi -31}\sim.34$ and ABJM$_{2,1}^\text{int}\cong \text{ABJ}_1$ with $\frac{16}{c_T}=\frac34$.}
\label{A0Mix}
\end{center}
\end{figure}  

Lastly, in Figure \ref{A0Mix}, we show upper bounds on the scaling dimension of the lowest dimension long $(A,0)_{\Delta,0}^{[0000]}$ multiplet as a function of $\frac{16}{c_T}$, where the orange is the allowed region. The red dot denotes the GFFT value at $c_T\to\infty$, the red dotted line shows the $\frac{1}{c_T}$ correction computed in \cite{Zhou:2017zaw}, and the dotted lines show the values of $c_T$ for various known SCFTs. The gray dashed line shows the upper bounds for the analogous quantity computed using only the $\langle2222\rangle$ crossing equations in \cite{Chester:2014fya}, which is slightly less constraining in general. Curiously, the bounds start becoming very different around $\frac{16}{c_T}\sim.4$, and the new bound goes to the free theory value at the upper bound on $\frac{16}{c_T}$ in \eqref{cTbound}.

\subsection{Mixed correlator bounds for ABJM$_{N,1}^\text{int}$ with $N=3,4$}
\label{N34}

 \begin{table}[htpbtp]
\begin{center}
\begin{tabular}{|l|c|c|c|c|}
\hline
$\lambda^2$: Type     & $N=3$ Bounds    & $N=3$ Error    & $N=4$ Bounds    & $N=4$ Error  \\
 \hline 
  $(A,+)^{[0030]}_{{5/2,0}}$       & $9.7383-9.7675$ & $0.300\%$  &   $8.5718-8.5935$ & $0.253\%$\\[3pt]
 $(A,+)^{[0030]}_{{7/2,1}}$       & $13.100-13.115$ &   $0.113\%$&  $12.337-12.347$ & $0.087\%$ \\[3pt]
 $(A,+)^{[0040]}_{{3,0}}$       & $35.851-36.030$ & $0.497\%$  & $29.589-29.736$ & $0.496\%$  \\[3pt]
 $(A,+)^{[0040]}_{{5,2}}$       & $87.344-87.782$ &  $0.500\%$ &  $76.569-76.921$ & $0.459\%$ \\[3pt]
  $(A,2)^{[0110]}_{{5/2,0}}$       & $10.135-10.275$ & $1.370\%$  & $10.747-10.842$ & $0.875\%$  \\[3pt]
 $(A,2)^{[0110]}_{{7/2,1}}$       & $13.146-13.528$ &   $2.858\%$&  $15.913-16.191$ & $1.735\%$ \\[3pt]
  $(A,2)^{[0120]}_{{4,1}}$       & $80.551-83.278$ & $3.329\%$  & $87.870-90.058$ & $2.459\%$  \\[3pt]
 $(A,2)^{[0120]}_{{6,3}}$       & $220.20-225.51$ & $2.385\%$  & $231.57-235.71$ & $1.770\%$  \\[3pt]
  $(A,2)^{[0200]}_{{3,0}}$       & $13.341-16.143$ &  $19.009\%$ & $20.022-22.166$ & $10.161\%$ \\[3pt]
 $(A,2)^{[0200]}_{{5,2}}$       & $69.327-74.529$ & $7.232\%$  & $84.647-88.502$& $4.453\%$  \\
 \hline
\end{tabular}
\end{center}
\caption{Upper and lower bounds on OPE coefficients squared for semishort operators, computed with $\Lambda = 35$ using mixed correlators and the short OPE coefficients fixed to their values for ABJM$_{3,1}^\text{int}$ and ABJM$_{4,1}^\text{int}$ using the 1d theory in Section \ref{1d}.  }
\label{UpLow}
\end{table}

We now restrict to the ABJM$^\text{int}_{N,1}$ theories for $N=3,4$ by imposing the values of the short OPE coefficients computed in these cases in Section \ref{1d}. In Table \ref{UpLow}, we show upper and lower bounds on various semishort multiplets that appear in \eqref{crossing2}. We compute the percent error as in \eqref{error}. In general, for a given multiplet the error seems to decrease with increasing spin and $N$. While the error is bigger than the single correlator OPE coefficients in Table \ref{UpLowS}, it is still quite small.

\begin{figure}[t!]
\begin{center}
   \includegraphics[width=0.49\textwidth]{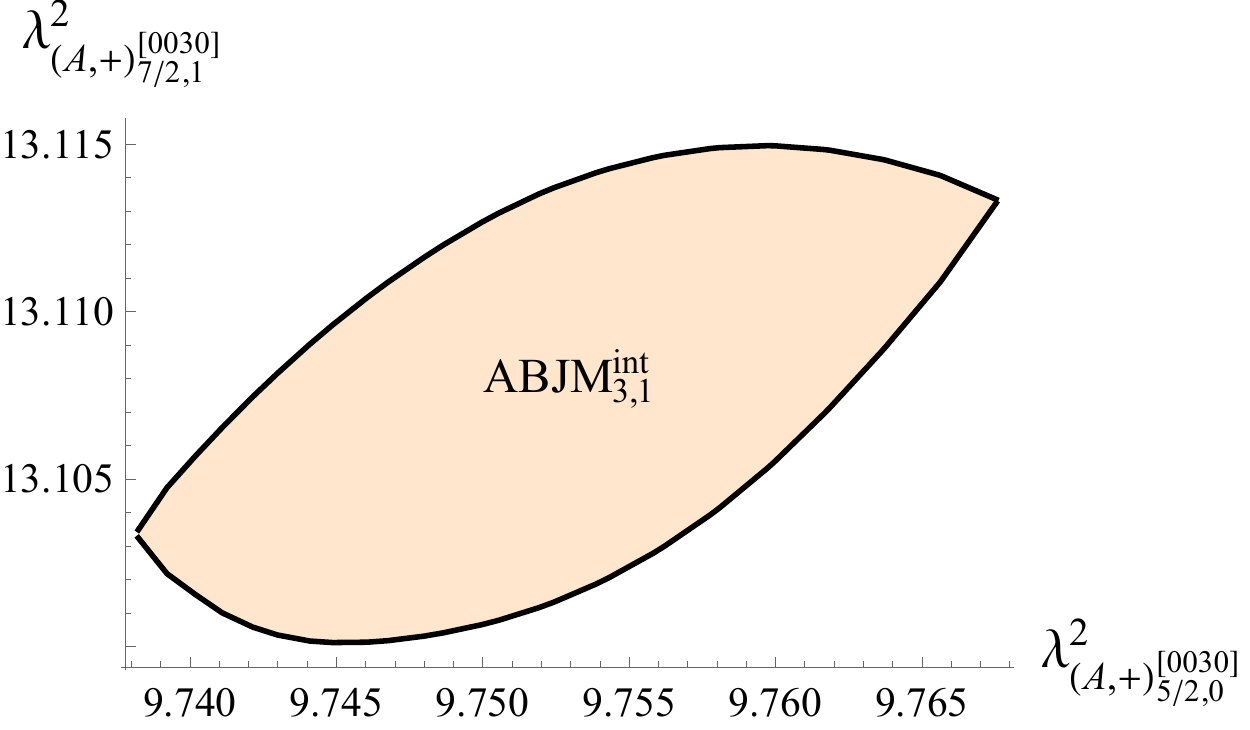}
    \includegraphics[width=0.49\textwidth]{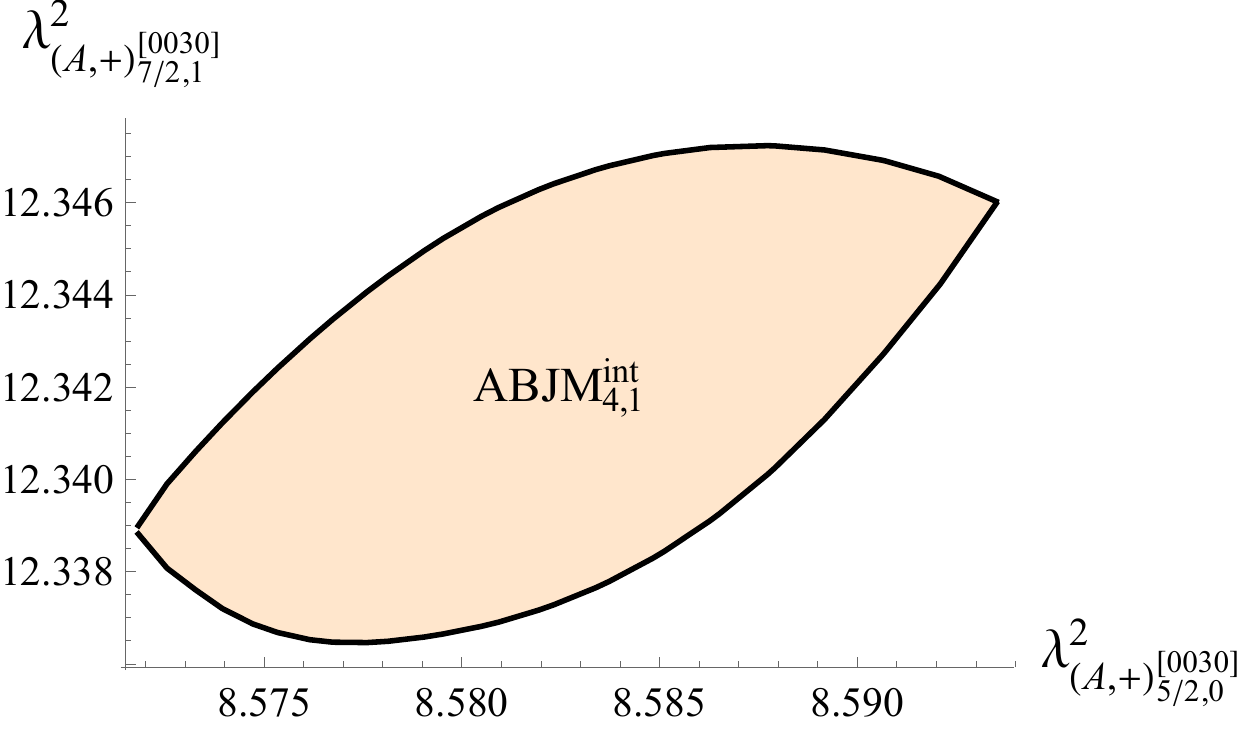}
        \includegraphics[width=0.7\textwidth]{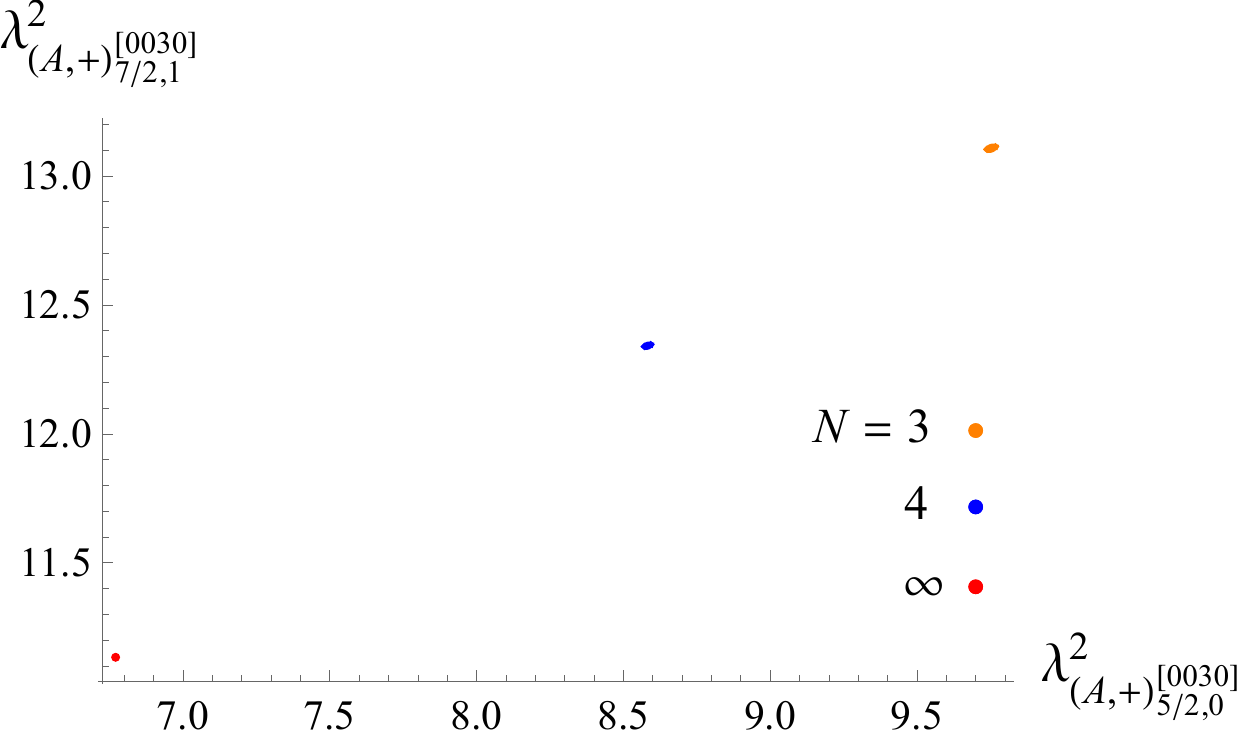}
\caption{Islands in the space of OPE coefficients $\lambda_{(A, +)_{\frac52,0}^{[0030]}}^2$ and $\lambda_{(A, +)_{\frac72,1}^{[0030]}}^2$ for ABJM$_{N,1}^\text{int}$ for $N=3,4$, where the orange shaded regions are allowed. These bounds are computed using mixed correlators with $\Lambda = 35$ and the short OPE coefficients fixed to their values for ABJM$_{3,1}^\text{int}$ and ABJM$_{4,1}^\text{int}$ using the 1d theory in Section \ref{1d}. The bottom plot compares these islands to the $N\to\infty$ GFFT values in Table \ref{Avalues}.}
\label{islands}
\end{center}
\end{figure}

In Figure \ref{islands} we show islands in the space of $((A,+)_{\frac52,0}^{[0030]},(A,+)_{\frac72,1}^{[0030]})$, which were the two most accurately known OPE coefficients in Table \ref{UpLow}, for $N=3,4$. These extremely small islands interpolate between the upper and lower bounds for these values in Table \ref{UpLow}. As noted above, these islands get smaller with increasing $N$.

\section{Discussion}
\label{disc}

In this paper we bootstrapped a mixed correlator system of $S_2$ and $S_3$ operators in 3d $\mathcal{N}=8$ SCFTs, generalizing the single correlator bootstrap of $\langle S_2S_2S_2S_2\rangle$ \cite{Chester:2014fya,Chester:2014mea,Agmon:2017xes}. This bootstrap study automatically excludes all known 3d $\mathcal{N}=8$ SCFTs except those that are dual to ABJM$_{N,1}^\text{int}$ with $N\geq3$ or the free theory, since these other theories do not contain an $S_3$ operator. After further restricting to interacting theories by setting the free multiplet OPE coefficients to zero, we computed the lower bound $c_T\gtrsim22.8$, which is curiously the same value were a kink was observed in all plots of the previous single correlator bootstrap studies. We then computed upper and lower bounds on protected operators and an upper bound on the lowest dimension unprotected scaling dimension as a function of $c_T$ in general interacting 3d $\mathcal{N}=8$ theories. This scaling dimension bound is stronger than the same bound in the single correlator boostrap studies, and so implies that the latter bound, which was conjectured to be saturated by one of the ABJ(M) theories, must in fact be saturated by one of the $k=2$ ABJ(M) theories, which is likely the $U(N+1)_2\times U(N)_{-2}$ ABJ theory. Lastly, we imposed the values of short scalar operator OPE coefficients as computed from the 1d theory for ABJM$_{N,1}$ with $N=3,4$, and derived precise islands in the space of semishort operator OPE coefficients. Since these semishort operators appear for an infinite number of possible spins, infinite many islands of this type can be computed.

It is instructive to compare these islands against those computed previously in the literature for the $O(N)$ critical models \cite{Kos:2014bka,Simmons-Duffin:2015qma,Kos:2015mba,Kos:2016ysd}, with $N=1$ being the Ising model. For each $N$, the islands could be computed in the space of the finite number of relevant operators and their OPE coefficients. At small values of $N$, these islands were observed to grow in size fairly rapidly as $N$ increased. For instance, the lowest dimension scaling dimension computed with $\Lambda=35$ and scanning over OPE coefficients in \cite{Kos:2016ysd} had percent error 0.0004\% for $N=1$, 0.12\% for $N=2$, and 0.24\% for $N=3$. In this work, for a given theory we can compute islands in the space of several families of semishort operator OPE coefficients that exist for infinitely many spins. As shown in Tables~\ref{UpLowS} and \ref{UpLow}, these island have comparable precision as the $O(N)$ islands for $N>1$ and always shrink with both increasing spin and $N$.

The main technical challenge of this paper was computing the superconformal blocks, and finding a linearly independent set of crossing equations. The expressions for the blocks are quite complicated, and are included in an attached \texttt{Mathematica} file. Supersymmetry makes many of the original 34 mixed correlator crossing equations redundant when they are expanded in terms of derivatives of $z$ and $\bar z$, as was originally observed for the single correlator case in \cite{Chester:2014fya}. After taking into account these redundancies using both analytical and numerical methods, we are left with seven full crossing equations, seven crossing equations in which only derivatives of $z$ are independent, and lastly one crossing equation in which only the leading term in the derivative expansion is independent.  

Looking ahead, there are several ways we could improve both the analytical and numerical calculations in this work. In the single correlator bootstrap study \cite{Agmon:2017xes}, the scalar short operator OPE coefficients were computed to all orders in $1/N$, which was found to be very accurate even for $N=2$. If similar analytic expressions could be derived for all the short operators in the mixed correlator system, then we could use them to derive precise islands for all values of $N$ in the mixed correlator system (as we did for the single correlator system), not just $N=3,4$. We could also check if the numerics saturate some of these analytic expressions, in which case we could read off all the CFT data that appears in the relevant OPEs, as was done for the single correlator case in \cite{Agmon:2017xes}. Since the mixed correlator system automatically excludes the $k=2$ ABJ(M) theories,\footnote{Except for ABJM$_{3,2}$, which is dual to ABJM$_{4,1}^\text{int}$.} there would be no ambiguities as to which theory was saturating the bounds, as was the case in the single correlator study. A more careful study of the spectrum as a function of $N$ could perhaps shed light on the mysterious point with $c_T\sim22.8$, which does not correspond to any known theory.

From the numerical perspective, the islands derived in this work come from the single and mixed correlator systems that only have around one and seven crossing equations each, respectively, which makes them almost as simple as the Ising model case, which has one and five crossing equations for the single and mixed correlator systems \cite{Kos:2014bka}, respectively. These islands also appear to still be shrinking as the derivatives parameter $\Lambda$ is increased. It should thus be possible to use the rapidly developing \texttt{SDPB} program to drastically increase $\Lambda$, so that these islands become small enough that they can be used to constrain the $1/c_T$ expansion of the respective four-point functions, which in the flat space limit can be used to constrain the small momentum expansion of the M-theory S-matrix \cite{Chester:2018aca,Binder:2018yvd,Binder:2019mpb}.

Finally, there are more constraints from localization that could be imposed on the numerical bootstrap. In particular, the squashed sphere free energy $F(b,m_i)$ in ABJ(M) theory can be computed using localization as a function of three masses $m_i$ as well as a squashing parameter $b$. The integrated four point function of $S_2$ can be related to taking four derivatives of these parameters and setting $m_i=0$ and $b=1$. Indeed, the quantity $\partial_{m_i}F\vert_{m_i=0}$ was already used in \cite{Agmon:2017xes} to compute the short OPE coefficient, while $\partial_{m_i}\partial_{m_j}F\vert_{m_i=m_j=0}$ for $i\neq j$ was used in \cite{Binder:2018yvd,Binder:2019mpb} to constrain $\langle S_2S_2S_2S_2\rangle$ in the large $c_T$ limit. Imposing new constraints of this type on the numerical bootstrap could further constrain the precise islands presented in this work.

\section*{Acknowledgments} 

We thank David Poland for useful discussions.  We also thank the organizers of ``Bootstrap 2019'' and Perimeter Institute for Theoretical Physics for its hospitality during the course of this work.  NBA thanks the organizers of and participants at the TASI 2019 summer school, and SMC those at The 36th Advanced School in Physics in Jerusalem, where part of this work was completed.   SMC is supported by the Zuckerman STEM Leadership Fellowship.  The work of SSP is supported in part by the Simons Foundation Grant No.~488653 and in part by an Alfred P.~Sloan Research Fellowship.

\appendix

\section{$\mathfrak{so}(8)$ harmonics}
\label{Ys}

We are interested in the $\mathfrak{so}(8)$ harmonics $Y^{k_{12},k_{34}}_{nm}(\sigma,\tau)$ for irreps $[0\,n-m\, 2m\,0]$ that appear in the tensor products $[00k_10]\otimes[00k_20]$ \eqref{tens0} for $k_i=2,3$. These can be computed as in Appendix D of \cite{Nirschl:2004pa}, with the following results:

\es{3333poly}{
Y^{0,0}_{00}&= 1, \\
Y^{0,0}_{10}&= \sigma - \tau, \\
Y^{0,0}_{11}& = \sigma + \tau - \frac14 , \\
Y^{0,0}_{20}&= \sigma^2 + \tau^2 -2\sigma \tau - \frac13 (\sigma + \tau) + \frac{1}{21}, \\
Y^{0,0}_{21}&= \sigma^2 - \tau^2 - \frac25 (\sigma - \tau), \\
Y^{0,0}_{22}& = \sigma^2 + \tau^2 + 4\sigma \tau - \frac23 (\sigma + \tau) + \frac{1}{15}, \\
Y^{0,0}_{30} &= \sigma^3 - \tau^3 + 3(\sigma \tau^2 - \sigma^2 \tau) + \frac34 (\tau^2 - \sigma^2)
 + \frac16 (\sigma - \tau), \\
Y^{0,0}_{31} &= \sigma^3 + \tau^3 -(\sigma^2 \tau + \sigma \tau^2) - \frac79 (\tau^2 + \sigma^2) -\frac29 \sigma \tau + \frac{13}{81}(\sigma + \tau) - \frac{1}{81}, \\
Y^{0,0}_{32} &= \sigma^3 - \tau^3 +3(\sigma^2 \tau - \sigma \tau^2) + \frac67 (\tau^2 - \sigma^2) + \frac17 (\sigma - \tau), \\
Y^{0,0}_{33} &= \sigma^3 + \tau^3 + 9(\sigma^2 \tau + \sigma \tau^2) - \frac98 (\sigma^2 + \tau^2) -\frac92 \sigma \tau + \frac{9}{28}(\sigma + \tau) - \frac{1}{56}, \\
}
and
\es{2323poly}{
Y^{-1,-1}_{\frac12\frac12} &= 1, \\
Y^{-1,-1}_{\frac32\frac12} &= \sigma - \tau-\frac17, \\
Y^{-1,-1}_{\frac32\frac32} &= \sigma + 2\tau - \frac25 , \\
Y^{-1,-1}_{\frac52\frac12}&= \sigma ^2-2 \sigma  \tau -\frac{14 \sigma }{27}+\tau ^2-\frac{4 \tau }{27}+\frac{2}{27}, \\
Y^{-1,-1}_{\frac52\frac32} &=\sigma ^2+\sigma  \tau -\frac{16 \sigma }{27}-2 \tau ^2+\frac{13 \tau }{27}+\frac{1}{27}, \\
Y^{-1,-1}_{\frac52\frac52} &=\sigma ^2+6 \sigma  \tau -\frac{6 \sigma }{7}+3 \tau ^2-\frac{12 \tau }{7}+\frac{1}{7}, 
}
and
\es{3223poly}{
Y^{1,-1}_{\frac12\frac12} &= 1, \\
Y^{1,-1}_{\frac32\frac12} &= \tau - \sigma-\frac17, \\
Y^{1,-1}_{\frac32\frac32} &= \tau + 2\sigma - \frac25 , \\
Y^{1,-1}_{\frac52\frac12}&= \tau ^2-2   \sigma\tau -\frac{14 \tau }{27}+\sigma ^2-\frac{4 \sigma }{27}+\frac{2}{27}, \\
Y^{1,-1}_{\frac52\frac32} &=\tau ^2+\sigma  \tau -\frac{16 \tau }{27}-2 \sigma ^2+\frac{13 \sigma }{27}+\frac{1}{27}, \\
Y^{1,-1}_{\frac52\frac52} &=\tau ^2+6 \sigma  \tau -\frac{6 \tau }{7}+3 \sigma ^2-\frac{12 \sigma }{7}+\frac{1}{7}.
}

\section{Supermultiplets}
\label{superMults}
 
 The conformal primaries that contribute to a given supermultiplet can be found by decomposing the characters of $\mathfrak{osp}(8|4)$ into its maximal bosonic subgroup $\mathfrak{so}(5)\oplus\mathfrak{so}(8)_R$. This can be done using the formulae in Appendix B of \cite{Chester:2014fya}, based on \cite{Dolan:2008vc}. In the tables below we give the results for the supermultiplets that appear in the $S_2\times S_2$, $S_2\times S_3$, and $S_3\times S_3$ OPEs. We denote in red those primaries that are allowed by symmetry, but vanish once we compute the explicit superblocks. As explained in the main text, these primaries have the opposite parity as the superconformal primary.

\begin{table}[htpb]
\centering
\begin{tabular}{|l||c|c|c|}
\hline 
$(B,+)^{[0010]}_{\frac12,0}$ spin: &{\text{dimension}}  \\
\hline
$\mathfrak{so}(8)_R$ irrep&1/2\\         
\hline\hline
$[0010]=\bold{8}_c$      & ${0}$           \\
\hline
\end{tabular}
\caption{All possible conformal primaries in $S_2\times S_3$ corresponding to the $(B,+)^{[0010]}_{\frac12,0}$ superconformal multiplet.}\label{Bp0010S2S3}
\end{table}

\begin{table}[htpb]
\centering
\begin{tabular}{|l||c|c|c|}
\hline 
$(B,+)^{[0030]}_{\frac32,0}$ spin: & \multicolumn{3}{c|}{\text{dimension}}  \\
\hline
$\mathfrak{so}(8)_R$ irrep&3/2&5/2&7/2\\         
\hline\hline
$[0010]=\bold{8}_c$     & --         & --          & ${2}$           \\
$[0110]=\bold{160}_c$     & --         & ${1}$          & --         \\
$[0030]=\bold{112}_c$     & ${0}$       & --          & --      \\
\hline
\end{tabular}
\caption{All possible conformal primaries in $S_2\times S_3$ corresponding to the $(B,+)^{[0030]}_{\frac32,0}$ superconformal multiplet.}\label{Bp0030}
\end{table}

\begin{table}[htpb]
\centering
\begin{tabular}{|l||c|c|c|c|c|}
\hline 
$(B,+)^{[0050]}_{\frac52,0}$ spin: & \multicolumn{5}{c|}{\text{dimension}}  \\
\hline
$\mathfrak{so}(8)_R$ irrep&5/2&7/2&9/2&11/2&13/2\\         
\hline\hline
$[0010]=\bold{8}_c$     & --         & --          & --          & --          & {0}   \\
$[0110]=\bold{160}_c$     & --         & --          & --          & {1}          & --    \\
$[0030]=\bold{112}_c$     & --         & --          & {2}        & --          & --    \\
$[0210]=\bold{1400}_c$     & --         & --        &$ {0} $         & --          & --    \\
$[0130]=\bold{1568}_c$     & --         &$ {1} $         & --          & --          & --    \\
$[0050]=\bold{672}^\prime_c$     &$ {0}$      & --          & --          & --          & --    \\
\hline
\end{tabular}
\caption{All possible conformal primaries in $S_2\times S_3$ corresponding to the $(B,+)^{[0050]}_{\frac52,0}$ superconformal multiplet.}\label{Bp0050}
\end{table}

\begin{table}[htpb]
\centering
\begin{tabular}{|l||c|c|c|c|}
\hline 
$(B,2)^{[0110]}_{\frac32,0}$ spin: & \multicolumn{4}{c|}{\text{dimension}}  \\
\hline
$\mathfrak{so}(8)_R$ irrep&3/2&5/2&7/2&9/2\\         
\hline\hline
$[0010]=\bold{8}_c$   &--  & ${1}$        & ${\red1,2}$         & ${\red2,3}$      \\
$[0110]=\bold{160}_c$    &${0}$ & ${\red0,1}$         & ${\red1,2}$        & -- \\
$[0030]=\bold{112}_c$     & --         & ${1}$ & --  & --   \\
\hline
\end{tabular}
\caption{All possible conformal primaries in $S_2\times S_3$ corresponding to the $(B,2)^{[0110]}_{\frac32,0}$ superconformal multiplet.}\label{B20110}
\end{table}

\begin{table}[htpb]
\centering
\begin{tabular}{|l||c|c|c|c|c|c|}
\hline 
$(B,2)^{[0130]}_{\frac52,0}$ spin: & \multicolumn{6}{c|}{\text{dimension}}  \\
\hline
$\mathfrak{so}(8)_R$ irrep&5/2&7/2&9/2&11/2&13/2&15/2\\         
\hline\hline
$[0010]=\bold{8}_c$     & --         & --          & --          & ${1}$     & ${0,\red1}$ &${1}$  \\
$[0110]=\bold{160}_c$     & --         & --          & ${0,\red1,2}$   & {\red0,1,\red2}     & ${0,\red1,2}$&--    \\
$[0030]=\bold{112}_c$     & --         & ${1}$  & {\red1,2}   & ${1,\red2,3}$    & -- &--   \\
$[0210]=\bold{1400}_c$     & --         & ${1}$      &$ {0,\red1} $         & ${1}$      & --   &-- \\
$[0130]=\bold{1568}_c$     & $0$     &$ {\red0,1} $         & ${0,\red1,2}$       & --          & -- &--   \\
$[0050]=\bold{672}^\prime_c$     &--     & ${1}$          & --          & --          & -- &--   \\
\hline
\end{tabular}
\caption{All possible conformal primaries in $S_2\times S_3$ corresponding to the $(B,2)^{[0130]}_{\frac52,0}$ superconformal multiplet.}\label{B20130}
\end{table}

\begin{table}[htpb]
\centering
\begin{tabular}{|l||c|c|c|c|c|c|c|}
\hline 
$(B,2)^{[0210]}_{\frac52,0}$ spin: & \multicolumn{7}{c|}{\text{dimension}}  \\
\hline
$\mathfrak{so}(8)_R$ irrep&5/2&7/2&9/2&11/2&13/2&15/2&17/2\\         
\hline\hline
$[0010]=\bold{8}_c$     & --         & --          & ${0}$     & ${\red0,1}$     & ${0,\red1,2}$ &${\red0,1}$&${0}$  \\
$[0110]=\bold{160}_c$     & --         & ${1}$        & ${0,\red1,2}$   & {\red0,1,\red2,3}     & ${0,\red1,2}$&${1}$ &--   \\
$[0030]=\bold{112}_c$     & --         & --  & {0,\red1,2}   & ${1,\red2}$    & ${2}$ &--  &-- \\
$[0210]=\bold{1400}_c$     & ${0}$        & ${\red0,1}$      &$ {0,\red1,2} $         & ${\red0,1}$      & ${0}$   &--&-- \\
$[0130]=\bold{1568}_c$     & --    &$ {1} $         & ${0,\red1}$       & ${1}$          & -- &--  &-- \\
$[0050]=\bold{672}^\prime_c$     &--     & --         & ${0}$        & --          & -- &--&--   \\
\hline
\end{tabular}
\caption{All possible conformal primaries in $S_2\times S_3$ corresponding to the $(B,2)^{[0210]}_{\frac52,0}$ superconformal multiplet.}\label{B20210}
\end{table}

\begin{table}[htpb]
\centering
\begin{tabular}{|l||c|c|c|c|c|c|c|}
\hline 
$(A,+)^{[0030]}_{\ell+\frac52,\ell}$ spin: & \multicolumn{7}{c|}{\text{dimension}}  \\
\hline
$\mathfrak{so}(8)_R$ irrep&$\ell+\frac52$&$\ell+\frac72$&$\ell+\frac92$&$\ell+\frac{11}{2}$&$\ell+\frac{13}{2}$&$\ell+\frac{15}{2}$&$\ell+\frac{17}{2}$\\         
\hline\hline
\multirow{2}{*}{$[0010]=\bold{8}_c$   }  & --         & --          & ${\ell,\red {\ell\pm1},}$     & ${\red \ell,\ell\pm1,}$     & ${ \ell,\red{\ell+1},}$ &${\ell+1},\red{\ell+2}$&${\ell+2}$  \\
&         &           & ${\ell\pm2}$     & $\red{\ell+2}$     & ${\ell+2}$ & &  \\
\hline
\multirow{2}{*}{$[0110]=\bold{160}_c$ }    & --         & ${\red \ell,{\ell\pm1}}$        & ${ \ell,\red{\ell\pm1},}$   & ${\red \ell,{\ell\pm1},}$     & ${ \ell,\red{\ell+1},}$&${\ell+1},\red{\ell+2},$ &--   \\
&        &         &  ${\ell+2}$  & $\red{\ell+2},{\ell+3}$     & ${\ell+2},\red{\ell+3}$&${\ell+3}$ &   \\
\hline
\multirow{2}{*}{$[0030]=\bold{112}_c$}     & ${\ell}$         & ${\red \ell,\ell+1}$ &$ {\ell,\red{\ell+1},}  $ & ${\red \ell,\ell+1,}$    & ${\ell,\red{\ell+1},\ell+2,}$ &--  &-- \\
&    &  & ${\ell+2}$& $\red{\ell+2},\ell+3$    & $\red{\ell+3},\ell+4$ &--  &-- \\
\hline
\multirow{2}{*}{$[0210]=\bold{1400}_c$   }  &--       & --    &$ {\ell,\red{\ell+1},} $         & ${\ell+1,}$      & ${\ell+2}$   &--&-- \\
&     &    &$ {\ell+2} $         & $\red{\ell+2}$      &    & & \\
\hline
\multirow{2}{*}{$[0130]=\bold{1568}_c$  }   & --    &$ {\ell+1} $         & $\red{\ell+1},{\ell+2}$       & ${\ell+1},\red{\ell+2},$          & -- &--  &-- \\
&     &        &     & ${\ell+3}$          &  & & \\
\hline
$[0050]=\bold{672}^\prime_c$     &--     & --         & ${\ell+2}$        & --          & -- &--&--   \\
\hline
\end{tabular}
\caption{All possible conformal primaries in $S_2\times S_3$ corresponding to the $(A,+)^{[0030]}_{\ell+\frac52,\ell}$ superconformal multiplet. For $\ell=0$, the conformal primaries with negative spin should be omitted.}\label{Ap0030}
\end{table}

\begin{table}[htpb]
\centering
\begin{tabular}{|l||c|c|c|c|c|c|c|c|}
\hline 
$(A,2)^{[0110]}_{\ell+\frac52,\ell}$ spin: & \multicolumn{8}{c|}{\text{dimension}}  \\
\hline
$\mathfrak{so}(8)_R$ irrep&$\ell+\frac52$&$\ell+\frac72$&$\ell+\frac92$&$\ell+\frac{11}{2}$&$\ell+\frac{13}{2}$&$\ell+\frac{15}{2}$&$\ell+\frac{17}{2}$&$\ell+\frac{19}{2}$\\         
\hline\hline
\multirow{3}{*}{$[0010]=\bold{8}_c$   }  & --         & $\red \ell,{\ell\pm1}$         & ${ \ell,\red{\ell\pm1},}$     & ${\red \ell,{\ell\pm1},}$     & ${ \ell,\red{\ell\pm1},}$ &${\red \ell,{\ell\pm1}}$&${ \ell,\red{\ell+1}}$  &${\ell+1}$\\
&         &           & $\red{\ell\pm2}$     & $\red{\ell\pm2,}$     & ${\ell\pm2}$ & $\red{\ell+2}$&  ${\ell+2}$&\\
&         &           &                 & ${\ell\pm3}$     & $\red{\ell+3}$ &${\ell+3}$ & & \\
\hline
\multirow{4}{*}{$[0110]=\bold{160}_c$ }    & $\ell$        & ${\red \ell,\ell\pm1}$        & ${\ell,\red{\ell\pm1},}$   & ${\red \ell,\ell\pm1,}$     & ${\ell,\red{\ell\pm1},}$&${\red \ell,\ell\pm1,}$ &${\ell,\red{\ell+1},}$&--  \\
&        &         &  ${\ell\pm2}$  & $\red{\ell\pm2}$     & ${\ell\pm2,}$&$\red{\ell+2,}$ &  ${\ell+2}$& \\
&        &         &                   & ${\ell+3}$     & $\red{\ell+3,}$&${\ell+3}$ & &  \\
&        &         &                   &                   & ${\ell+4}$& & &  \\
\hline
\multirow{3}{*}{$[0030]=\bold{112}_c$}     &     --   & ${\red \ell,\ell\pm1}$ &$ {\ell,\red{\ell\pm1},}  $ & ${\red \ell,\ell\pm1,}$    & ${\ell,\red{\ell\pm1},}$ &$\red \ell,\ell\pm1$  &--& --\\
&    &  & ${\ell+2}$& $\red{\ell+2}$    & ${\ell+2,}$ &$\red{\ell+2}$  && \\
&    &  &            & ${\ell+3}$    & $\red{\ell+3,}$ &$\ell+3$  && \\
\hline
\multirow{3}{*}{$[0210]=\bold{1400}_c$   }  &--       & $\ell+1$   &$ {\ell,\red{\ell+1},} $         & ${\red \ell,\ell\pm1,}$      & ${\ell,\red{\ell+1}}$   &${\ell+1}$&-- &--\\
&     &    &   $ {\ell+2} $                    & $\red{\ell+2,}$      &  ${\ell+2}$  & && \\
&     &    &        & ${\ell+3}$      &    & && \\
\hline
\multirow{2}{*}{$[0130]=\bold{1568}_c$  }   & --    &--         & ${\ell,\red{\ell+1},}$       & ${\red \ell,\ell+1,}$          & ${\ell,\red{\ell+1}}$ &--  &-- &--\\
&     &        &  ${\ell+2}$   & $\red{\ell+2}$          & ${\ell+2}$ & && \\
\hline
$[0050]=\bold{672}^\prime_c$     &--     & --         & --     & $\ell+1$          & -- &--&-- &--  \\
\hline
\end{tabular}
\caption{All possible conformal primaries in $S_2\times S_3$ corresponding to the $(A,2)^{[0110]}_{\ell+\frac52,\ell}$ superconformal multiplet. For $\ell=0,1,2$, the conformal primaries with negative spin should be omitted.}\label{A20110}
\end{table}

\begin{table}[htpb]
\centering
\begin{tabular}{|l||c|c|c|c|c|c|c|c|c|}
\hline 
$(A,0)^{[0010]}_{\Delta,\ell}$ spin: & \multicolumn{9}{c|}{\text{dimension}}  \\
\hline
$\mathfrak{so}(8)_R$ irrep&$\Delta$&$\Delta+1$&$\Delta+2$&$\Delta+3$&$\Delta+4$&$\Delta+5$&$\Delta+6$&$\Delta+7$&$\Delta+8$\\         
\hline\hline
\multirow{4}{*}{$[0010]=\bold{8}_c$   }  & $\ell$         & ${\red \ell,\ell\pm1}$         & ${\ell,\red{\ell\pm1},}$     & ${\red \ell,\ell\pm1,}$     & ${\ell,\red{\ell\pm1},}$ &${\red \ell,\ell\pm1}$&${\ell,\red{\ell\pm1}}$  &${\red \ell,\ell\pm1}$&${\ell}$\\
&         &           & ${\ell\pm2}$     & ${\red{\ell\pm2},}$     & ${\ell\pm2}$ & $\red{\ell\pm2}$&  ${\ell\pm2}$&&\\
&         &           &                 & ${\ell\pm3}$     & $\red{\ell\pm3}$ &${\ell\pm3}$ & && \\
&         &           &                 &                     & ${\ell\pm4}$ &                & && \\
\hline
\multirow{3}{*}{$[0110]=\bold{160}_c$ }    &   --    & ${\red \ell,\ell\pm1}$        & ${\ell,\red{\ell\pm1},}$   & ${\red \ell,\ell\pm1,}$     & ${\ell,\red{\ell\pm1},}$&${\red \ell,\ell\pm1,}$ &${\ell,\red{\ell\pm1},}$&${\red \ell,\ell\pm1}$&--  \\
&        &         &  ${\ell\pm2}$  & $\red{\ell\pm2}$     & ${\ell\pm2,}$&$\red{\ell\pm2,}$ &  ${\ell\pm2}$&& \\
&        &         &                   & ${\ell\pm3}$     & $\red{\ell\pm3,}$&${\ell\pm3}$ & &&  \\
\hline
\multirow{2}{*}{$[0030]=\bold{112}_c$}     &   --       &--   &$ {\ell,\red{\ell\pm1},}  $ & ${\red \ell,\ell\pm1,}$    & ${\ell,\red{\ell\pm1},}$ &$\red \ell,\ell\pm1$  &${\ell,\red{\ell\pm1}}$&--&-- \\
&    &  & ${\ell\pm2}$& $\red{\ell\pm2}$    & ${\ell\pm2,}$ &$\red{\ell\pm2}$  &${\ell\pm2}$& &\\
\hline
\multirow{2}{*}{$[0210]=\bold{1400}_c$   }  &--       & -- &$ {\ell} $         & ${\red \ell,\ell\pm1,}$      & ${\ell,\red{\ell\pm1}}$   &${\red \ell,\ell\pm1}$&$\ell$ &--&--\\
&     &    &                      &       &  ${\ell\pm2}$  & && &\\
\hline
$[0130]=\bold{1568}_c$     & --    &--         & --       & ${\red \ell,\ell\pm1,}$          & ${\ell,\red{\ell\pm1}}$ &${\red \ell,\ell\pm1}$&-- &--&--\\
\hline
$[0050]=\bold{672}^\prime_c$     &--     & --         & --     &--        &  $\ell$   &--&-- &-- &-- \\
\hline
\end{tabular}
\caption{All possible conformal primaries in $S_2\times S_3$ corresponding to the $(A,0)^{[0010]}_{\Delta,\ell}$ superconformal multiplet. For $\ell=0,1,2,3$, the conformal primaries with negative spin should be omitted.}\label{A00010}
\end{table}
 
 \begin{table}[htpb]
\centering
\begin{tabular}{|l||c|c|c|c|}
\hline 
$(B,+)^{[0020]}_{1,0}$ spin: & \multicolumn{3}{c|}{\text{dimension}}  \\
\hline
$\mathfrak{so}(8)_R$ irrep& 1&2&3\\         
\hline\hline
{$[0000]=\bold{1}$}       & --          & --    &  $ {2}$   \\
{$[0100]=\bold{28}$}      & --          & ${1} $    & --\\
$[0020]=\bold{35}_c$   & {0}         & --       &--\\
\hline
\end{tabular}
\caption{All possible conformal primaries in $S_2\times S_2$ and $S_3\times S_3$ corresponding to the $(B,+)^{[0020]}_{1,0}$ superconformal multiplet. }\label{Bp0020}
\end{table}

\begin{table}[htpb]
\centering
\begin{tabular}{|l||c|c|c|c|c|}
\hline 
$(B,+)^{[0040]}_{2,0}$ spin: & \multicolumn{5}{c|}{\text{dimension}}  \\
\hline
$\mathfrak{so}(8)_R$ irrep& 2&3&4&5&6\\         
\hline\hline
{$[0000]=\bold{1}$  }     & --          & --          & --          & --          & ${0}$         \\
{$[0100]=\bold{28}$ }     & --          & --          & --          & ${1} $         & --      \\
$[0020]=\bold{35}_c$   & --          & --          & ${2}$         & --        & --      \\
{$[0200]=\bold{300}$  }   & --          & --         &${ 0}  $        & --          & --      \\
$[0120]=\bold{567}_c$     & --         &$ {1}$          & --          & --          & --    \\
$[0040]=\bold{294}_c$     & ${0}$         & --          & --          & --          & --    \\
\hline
\end{tabular}
\caption{All possible conformal primaries in $S_2\times S_2$ and $S_3\times S_3$ corresponding to the $(B,+)^{[0040]}_{2,0}$ superconformal multiplet. } \label{Bp0040}
\end{table}

\begin{table}[htpb]
\centering
\begin{tabular}{|l||c|c|c|c|c|}
\hline 
$(B,+)^{[0060]}_{3,0}$ spin: & \multicolumn{5}{c|}{\text{dimension}}  \\
\hline
$\mathfrak{so}(8)_R$ irrep&3&4&5&6&7\\         
\hline\hline
$[0020]=\bold{35}_c$     & --         & --          & --          & --          & {0}   \\
$[0120]=\bold{567}_c$     & --         & --          & --          & {1}          & --    \\
$[0040]=\bold{294}_c$     & --         & --          & {2}        & --          & --    \\
$[0220]=\bold{4312}_c$     & --         & --        &$ {0} $         & --          & --    \\
$[0140]=\bold{3696}_c$     & --         &$ {1} $         & --          & --          & --    \\
$[0060]=\bold{1386}_c$     &$ {0}$      & --          & --          & --          & --    \\
\hline
\end{tabular}
\caption{All possible conformal primaries in $S_3\times S_3$ corresponding to the $(B,+)^{[0060]}_{3,0}$ superconformal multiplet.}\label{Bp0060}
\end{table}

\begin{table}[htpb]
\centering
\begin{tabular}{|l||c|c|c|c|c|c|c|}
\hline 
$(B,2)^{[0200]}_{2,0}$ spin: & \multicolumn{7}{c|}{\text{dimension}}  \\
\hline
$\mathfrak{so}(8)_R$ irrep& 2&3&4&5&6&7&8\\         
\hline\hline
$[0000]=\bold{1}$       & --          & --          & 0          & \red{0}         &$ 0,2$&\red{0}&0         \\
$[0100]=\bold{28}$      & --          & 1         & \red{1}          & $1,3$          & \red{1}&1&--      \\
$[0020]=\bold{35}_c$   & --          & --          & $0,2$         & \red{2}        & 2&--&--      \\
$[0200]=\bold{300}$     & 0          & \red{0}         & $0,2$          & \red{0}          & 0&--&--      \\
$[0120]=\bold{567}_c$     & --         & 1          & \red{1}         & 1          & --&--&--    \\
$[0040]=\bold{294}_c$     & --         & --          & 0          & --          & -- &--&--   \\
\hline
\end{tabular}
\caption{All possible conformal primaries in $S_2\times S_2$ and $S_3\times S_3$ corresponding to the $(B,2)^{[0200]}_{2,0}$ superconformal multiplet.}\label{B20200}
\end{table}

\begin{table}[htpb]
\centering
\begin{tabular}{|l||c|c|c|c|c|c|c|}
\hline 
$(B,2)^{[0220]}_{3,0}$ spin: & \multicolumn{7}{c|}{\text{dimension}}  \\
\hline
$\mathfrak{so}(8)_R$ irrep&$3$&$4$&$5$&$6$&$7$&$8$&$9$\\         
\hline\hline
{$[0100]=\bold{28}$ }    &--  & --          & --          & $1$         & $\red1$         &$1$&--         \\
$[0020]=\bold{35}_c$     &--  & --          & $0$          & $\red0$         & $0,2$         &$\red0$&$0$         \\
{$[0200]=\bold{300}$  }    &-- & --          & ${0,2}$         & ${\red0,\red2}$         & ${0,2}$         &--&--         \\
$[0120]=\bold{567}_c$     &--  & 1          & $\red1$         & $1,3$         & $\red1$         &1&--         \\
$[0040]=\bold{294}_c$       &--& --          & $0,2$          & $\red2$         & 2         &--&--         \\
{$[0300]=\bold{1925}$  }     &--& $1$          & $\red1$          & $1$         & --         &--&--         \\
$[0220]=\bold{4312}_c$     &0  & $\red0$          & $0,2$          & $\red0$         & 0         &--&--         \\
$[0140]=\bold{3696}_c$   &--    & 1          & $\red1$         & 1         & --         &--&--         \\
$[0060]=\bold{1386}_c$     &--  & --          & 0          & --         & --         &--&--         \\
\hline
\end{tabular}
\caption{All possible conformal primaries in $S_3\times S_3$ corresponding to the $(B,2)^{[0220]}_{3,0}$ superconformal multiplet.  }\label{B20220}
\end{table}

\begin{table}[htpb]
\centering
\begin{tabular}{|l||c|c|c|c|c|c|c|}
\hline 
$(A,+)^{[0020]}_{\ell+2,\ell}$ spin: & \multicolumn{7}{c|}{\text{dimension}}  \\
\hline
$\mathfrak{so}(8)_R$ irrep&$\ell+2$&$\ell+3$&$\ell+4$&$\ell+5$&$\ell+6$&$\ell+7$&$\ell+8$\\         
\hline\hline
$[0000]=\bold{1}$      & --          & --         & $\ell\pm2,\ell$    & $\red{\ell+2}$   &$\ell+2$&$\red{\ell+2}$&$\ell+2$         \\
$[0100]=\bold{28}$      & --          &$\ell\pm1$         & $\red{\ell\pm1}$        & $\ell\pm1,\ell+3$         &$\red{\ell+1},\red{\ell+3}$&$\ell+1,\ell+3$ &--        \\
$[0020]=\bold{35}_c$      & $\ell$          & $\red \ell$        & $\ell,\ell+2$         & $\red \ell,\red{\ell+2}$         &$\ell,\ell+2,\ell+4$&-- &--        \\
$[0200]=\bold{300}$       & --          & --         & $\ell,\ell+2$         & $\red{\ell+2}$        &$\ell+2$&--   &--      \\
$[0120]=\bold{567}_c$       & --          & $\ell+1$          & $\red{\ell+1}$         & $\ell+1,\ell+3$         &--&--   &--      \\
$[0040]=\bold{294}_c$   & --        & --        & $\ell+2$       & --      &--&--  &--       \\
\hline
\end{tabular}
\caption{All possible conformal primaries in $S_2\times S_2$ and $S_3\times S_3$ corresponding to the $(A,+)^{[0020]}_{\ell+2,\ell}$ superconformal multiplet.}\label{Ap0020}
\end{table}

\begin{table}[htpb]
\centering
\begin{tabular}{|l||c|c|c|c|c|c|c|}
\hline 
$(A,+)^{[0040]}_{\ell+3,\ell}$ spin: & \multicolumn{7}{c|}{\text{dimension}}  \\
\hline
$\mathfrak{so}(8)_R$ irrep&$\ell+3$&$\ell+4$&$\ell+5$&$\ell+6$&$\ell+7$&$\ell+8$&$\ell+9$\\         
\hline\hline
{$[0000]=\bold{1}$ }     & --         &--      & --         & --         &$\ell$&-- &--  \\
\hline
{$[0100]=\bold{28}$ }     & --         &--      & --         & $\ell\pm1$         &$\red{\ell+1}$&$\ell+1$ &--  \\
\hline
{$[0020]=\bold{35}_c$ }     & --         &--      & $\ell,\ell\pm2$         & $\red \ell,\red{\ell+2}$         &$\ell,\ell+2$&$\red{\ell+2}$ &$\ell+2$  \\   
\hline
{$[0200]=\bold{300}$ }     & --         &--      & $\ell$         & $\red \ell$         &$\ell,\ell+2$& -- &--  \\   
\hline
{$[0120]=\bold{567}_c$  }     & --          & $\ell\pm1$          & $\red{\ell\pm1}$         & $\ell\pm1,\ell+3$         &$\red{\ell+1},\red{\ell+3}$&$\ell+1,\ell+3$   &--  \\   
\hline
{$[0040]=\bold{294}_c$ }  & $\ell$        & $\red \ell$       & $\ell,\ell+2$      & $\red \ell,\red{\ell+2}$      &$\ell,\ell+2,\ell+4$&--  &--  \\  
\hline
{$[0300]=\bold{1925}$ }     & --         &--      & --         & $\ell+1$         &--& --&--  \\   
\hline
{$[0220]=\bold{4312}_c$ }     & --         &--      & $\ell,\ell+2$         & $\red{\ell+2}$         &$\ell+2$& --& -- \\   
\hline
{$[0140]=\bold{3696}_c$ }     & --         &$\ell+1$      & $\red{\ell+1}$         & $\ell+1,\ell+3$         &--&-- &--  \\  
\hline
{$[0060]=\bold{1386}_c$ }     & --         &--      & $\ell+2$         &    --      &--& --&--  \\  
\hline
\end{tabular}
\caption{All possible conformal primaries in $S_3\times S_3$ corresponding to the $(A,+)^{[0040]}_{\ell+3,\ell}$ superconformal multiplet.}\label{Ap0040}
\end{table}

\begin{table}[htpb]
\centering
\begin{tabular}{|l||c|c|c|c|c|c|c|c|}
\hline 
$(A,2)^{[0100]}_{\ell+2,\ell}$ spin: & \multicolumn{8}{c|}{\text{dimension}}  \\
\hline
$\mathfrak{so}(8)_R$ irrep&$\ell+2$&$\ell+3$&$\ell+4$&$\ell+5$&$\ell+6$&$\ell+7$&$\ell+8$&$\ell+9$\\         
\hline\hline
$[0000]=\bold{1}$      & --          & $\ell\pm1$        & $\red{\ell+1}$    & $\ell\pm1,\ell\pm3$   &$\red{\ell+1},\red{\ell+3}$&$\ell+1,\ell+3$&$\red{\ell+1}$&$\ell+1$         \\
$[0100]=\bold{28}$      & $\ell$          &$\red \ell$         & $\ell,\ell\pm2$        & $\red \ell,\red{\ell\pm2}$         &$\ell,\ell\pm2,\ell+4$&$\red \ell,\red{\ell+2}$ &$\ell,\ell+2$&--        \\
$[0020]=\bold{35}_c$      & --          & $\ell\pm1$        & $\red{\ell\pm1}$         & $\ell\pm1,\ell+3$         &$\red{\ell\pm1},\red{\ell+3}$&$\ell\pm1,\ell+3$ &--&--        \\
$[0200]=\bold{300}$       & --          & $\ell+1$         & $\red{\ell+1}$         & $\ell\pm1,\ell+3$        &$\red{\ell+1}$&$\ell+7$   &--&--      \\
$[0120]=\bold{567}_c$       & $\ell$         & $\red \ell$          & $\ell,\ell\pm2$         & $\red \ell,\red{\ell\pm2}$         &$\ell,\ell\pm2,\ell+4$&$\red \ell,\red{\ell+2}$   &$\ell,\ell+2$&--      \\
$[0040]=\bold{294}_c$   & --        & --        &--       & $\ell+1$      &--&--  &--    &--   \\
\hline
\end{tabular}
\caption{All possible conformal primaries in $S_2\times S_2$ and $S_3\times S_3$ corresponding to the $(A,2)^{[0100]}_{\ell+2,\ell}$ superconformal multiplet.}\label{A20100}
\end{table}

\begin{table}[htpb]
\centering
\begin{tabular}{|l||c|c|c|c|c|c|c|c|}
\hline 
$(A,2)^{[0120]}_{\ell+3,\ell}$ spin: & \multicolumn{8}{c|}{\text{dimension}}  \\
\hline
$\mathfrak{so}(8)_R$ irrep&$\ell+3$&$\ell+4$&$\ell+5$&$\ell+6$&$\ell+7$&$\ell+8$&$\ell+9$&$\ell+10$\\         
\hline\hline
{$[0000]=\bold{1}$}    &--   &--        &--      & $\ell\pm1$         & $\red{\ell\pm1}$         &$\ell\pm1$&-- &--  \\     
\hline
{$[0100]=\bold{28}$}    &--   &--        &$\ell,\ell\pm2$      & $\red \ell,\red{\ell\pm2}$         & $\ell,\ell\pm2$         &$\red \ell,\red{\ell+2}$&$ \ell,{\ell+2}$ &--  \\     
\hline
{$[0020]=\bold{35}_c$}    &--   & ${\ell\pm1}$        &$\red{\ell\pm1}$      & ${\ell\pm1},{\ell\pm3}$         & $\red{\ell\pm1},\red{\ell+3}$         &${\ell\pm1},{\ell+3}$&$\red{\ell+1}$ &${\ell+1}$  \\     
\hline
{$[0200]=\bold{300}$}    &--   & $\ell\pm1$        &$\red{\ell\pm1}$      & $\ell\pm1,\ell+3$         & $\red{\ell\pm1},\red{\ell+3}$         &$\ell\pm1,\ell+3$&-- &--  \\     
\hline
{$[0120]=\bold{567}_c$ } &$\ell$  & $\red \ell$        & $\ell,\ell\pm2$        & $\red \ell,\red{\ell\pm2}$       & $\ell,\ell\pm2,\ell+4$      &$\red \ell,\red{\ell+2}$&$\ell,\ell+2$  &-- \\  
\hline
{$[0040]=\bold{294}_c$ } &--  & $\ell\pm1$        & $\red{\ell\pm1}$        & $\ell\pm1,\ell+3$       & $\red{\ell\pm1},\red{\ell+3}$      &$\ell\pm1,\ell+3$&--  &-- \\  
\hline
{$[0300]=\bold{1925}$  }&--  & --       & $ \ell,{\ell+2}$        & $\red \ell,\red{\ell+2}$       & $ \ell,{\ell+2}$      &--&--  &--   \\ 
\hline
{$[0220]=\bold{4312}_c$  }&--  & ${\ell+1}$        & $\red{\ell+1}$        & ${\ell\pm1},{\ell+3}$       & $\red{\ell+1}$      &${\ell+1}$&--  &--   \\ 
\hline
{$[0140]=\bold{3696}_c$}  &-- & --    & $ \ell,{\ell+2}$        & $\red \ell,\red{\ell+2}$       & $ \ell,{\ell+2}$      &--&--  &--  \\  
\hline
$[0060]=\bold{1386}_c$  &--  & --  & --     & $\ell+1$       & --   &--&--  &--       \\
\hline
\end{tabular}
\caption{All possible conformal primaries in $S_3\times S_3$ corresponding to the $(A,2)^{[0120]}_{\ell+3,3}$ superconformal multiplet.}\label{A20120}
\end{table}

\clearpage

\begin{table}[htpb]
\centering
\begin{tabular}{|l||c|c|c|c|c|c|c|c|}
\hline 
$(A,2)^{[0200]}_{\ell+3,\ell}$ spin: & \multicolumn{8}{c|}{\text{dimension}}  \\
\hline
$\mathfrak{so}(8)_R$ irrep&$\ell+3$&$\ell+4$&$\ell+5$&$\ell+6$&$\ell+7$&$\ell+8$&$\ell+9$&$\ell+10$\\         
\hline\hline
{$[0000]=\bold{1}$}    &--   &--        &$\ell$      & $\red \ell$         & $\ell\pm2$         &$\red \ell$&$\ell$ &--  \\     
\hline
{$[0100]=\bold{28}$}    &--   &$\ell\pm1$        &$\red{\ell\pm1}$      & $\ell\pm1,\ell\pm3$         & $\red{\ell\pm1},\red{\ell+3}$         &$\ell\pm1,\ell+3$&$\red{\ell+1}$ &$\ell+1$  \\     
\hline
{$[0020]=\bold{35}_c$}    &--   & --        &$\ell\pm2$      & $\red{\ell\pm2}$         & $\ell\pm2$         &$\red \ell,\red{\ell+2}$&$\ell,\ell+2$ &--  \\     
\hline
{$[0200]=\bold{300}$}    &$\ell$  & $\red \ell$        &$\ell,\ell\pm2$      & $\red \ell,\red{\ell\pm2}$         & $\ell,\ell\pm2,\ell+4$         &$\red \ell,\red{\ell+2}$&$\ell,\ell+2$ &--  \\     
\hline
{$[0120]=\bold{567}_c$ } &--  & $\ell\pm1$        & $\red{\ell\pm1}$        & $\ell\pm1,\ell+3$       & $\red{\ell\pm1},\red{\ell+3}$      &$\ell\pm1,\ell+3$&--  &-- \\  
\hline
{$[0040]=\bold{294}_c$ } &--  & --        & $\ell$        & $\red{\ell}$       & $\ell,\ell+2$      &--&--  &-- \\  
\hline
{$[0300]=\bold{1925}$  }&--  & $\ell+1$       & $\red{\ell+1}$        & $\ell\pm1,\ell+3$       & $\red{\ell+1}$      &$\ell+1$&--  &--   \\ 
\hline
{$[0220]=\bold{4312}_c$  }&--  & --        & $\ell,\ell+2$        & $\red \ell,\red{\ell+2}$       & $\ell,\ell+2$      &--&--  &--   \\ 
\hline
{$[0140]=\bold{3696}_c$}  &-- & --    & --       & $\ell+1$       & --     &--&--  &--  \\  
\hline
\end{tabular}
\caption{All possible conformal primaries in $S_3\times S_3$ corresponding to the $(A,2)^{[0200]}_{\ell+3,3}$ superconformal multiplet.}\label{A20200}
\end{table}

\begin{table}[htpb]
\centering
\begin{tabular}{|l||c|c|c|c|c|c|c|c|c|}
\hline 
$(A,0)^{[0000]}_{\Delta,\ell}$ spin: & \multicolumn{9}{c|}{\text{dimension}}  \\
\hline
$\mathfrak{so}(8)_R$ irrep&$\Delta$&$\Delta+1$&$\Delta+2$&$\Delta+3$&$\Delta+4$&$\Delta+5$&$\Delta+6$&$\Delta+7$&$\Delta+8$   \\         
\hline\hline
\multirow{2}{*}{$[0000]=\bold{1}$}      & $\ell$    & $\red \ell$   & $\ell$   & $\red \ell$   & $\ell,\ell\pm2$   & $\red \ell$   & $\ell$   & $\red \ell$   & $\ell$      \\  
&    &   &   &    & $\ell\pm4$   &    &    & &     \\  
\hline
\multirow{2}{*}{$[0100]=\bold{28}$}      & --   & $\ell\pm1$   & $\red{\ell\pm1}$   & $\ell\pm1,$   & $\red{\ell\pm1},$   & $\ell\pm1,$   & $\red{\ell\pm1}$   & $\ell\pm1$   & --      \\  
&    &   &   & $  \ell\pm3$ & $\red{\ell\pm3}$   &  $\ell\pm3$  &    & &     \\  
\hline
{$[0020]=\bold{35}_c$}      & -- & --   & $\ell,\ell\pm2$   & $\red \ell,\red{\ell\pm2}$   & $\ell,\ell\pm2$   & $\red \ell,\red{\ell\pm2}$   & $\ell,\ell\pm2$   & --   & --        \\  
\hline
{$[0200]=\bold{300}$}      & --    & --   & $\ell$   & $\red \ell$   & $\ell,\ell\pm2$   & $\red \ell$   & $\ell$   & --   & --      \\  
\hline
{$[0120]=\bold{567}_c$}      & --   & --  & --   & $\ell\pm1$   & $\red {\ell\pm1}$   & $\ell\pm1$   & --   & --   & --      \\  
\hline
{$[0040]=\bold{294}_c$}      & --    & --   & --  & --   & $\ell$   & --   & --   & --   & --    \\  
\hline
\end{tabular}
\caption{All possible conformal primaries in $S_2\times S_2$ and $S_3\times S_3$ corresponding to the $(A,0)^{[0000]}_{\Delta,\ell}$ superconformal multiplet.}\label{A00000}
\end{table}

\begin{table}[htpb]
\centering
\begin{tabular}{|l||c|c|c|c|c|c|c|c|c|}
\hline 
$(A,0)^{[0020]}_{\Delta,\ell}$ spin: & \multicolumn{9}{c|}{\text{dimension}}  \\
\hline
$\mathfrak{so}(8)_R$ irrep&$\Delta$&$\Delta+1$&$\Delta+2$&$\Delta+3$&$\Delta+4$&$\Delta+5$&$\Delta+6$&$\Delta+7$&$\Delta+8$\\         
\hline\hline
{{$[0000]=\bold{1}$ }} &-- &--  &$\ell,\ell\pm2$ & $\red \ell,\red{\ell\pm2}$          & $\ell,\ell\pm2$         &$\red \ell,\red{\ell\pm2}$   &$\ell,\ell\pm2$&-- &--       \\
\hline
{\multirow{2}{*}{$[0100]=\bold{28}$ }}  &-- &$\ell\pm1$ &$\red{\ell\pm1}$ & $\ell\pm1,$  & $\red{\ell\pm1},$         & $\ell\pm1,$   &$\red{\ell\pm1}$&$\ell\pm1$ &--       \\
&& & & $\ell\pm3$  & $\red{\ell\pm3}$         & $\ell\pm3$   && &--       \\
\hline
{$[0020]=\bold{35}_c$ }   &$\ell$&$\red \ell$ &$\ell,\ell\pm2$ & $\red \ell,\red{\ell\pm2}$          & $\ell,\ell\pm2,\ell\pm4$        & $\red \ell,\red{\ell\pm2}$   &$\ell,\ell\pm2$&$\red \ell$ &$\ell$        \\
\hline
{{$[0200]=\bold{300}$}   }&-- &-- &$\ell,\ell\pm2$ & $\red \ell,\red{\ell\pm2}$          & $\ell,\ell\pm2$         & $\red \ell,\red{\ell\pm2}$  &$\ell,\ell\pm2$&--&--       \\
 \hline
\multirow{2}{*}{$[0120]=\bold{567}_c$}    &-- &$\ell\pm1$&$\red{\ell\pm1}$        & $\ell\pm1,$         & $\red{\ell\pm1},$   &$\ell\pm1,$&$\red{\ell\pm1},$&$\ell\pm1$&--         \\
 & &&     & ${\ell\pm3}$         & \red{$\ell\pm3$}   &{$\ell\pm3$}&&&         \\
\hline
{$[0040]=\bold{294}_c$ }   &-- &-- & $\ell,\ell\pm2$    &$\red \ell,\red{\ell\pm2}$     & $\ell,\ell\pm2$   &$\red \ell,\red{\ell\pm2}$&$\ell,\ell\pm2$&--&--        \\
 \hline
{{$[0300]=\bold{1925}$   } } &-- &--        & --     & $\ell\pm1$   &$\red {\ell\pm1}$&$\ell\pm1$  &--&--&--       \\
\hline
{$[0220]=\bold{4312}_c$  }   &-- & --        & $\ell$         & $\red \ell$   &$\ell,\ell\pm2$&$\red \ell$  &$\ell$ &--&--      \\
\hline
{$[0140]=\bold{3696}_c$ }    &-- & --     & --        & $\ell\pm1$   &$\red{\ell\pm1}$&$\ell\pm1$ &--&--&--       \\
\hline
{$[0060]=\bold{1386}_c$ }    &-- & --     & --        & --  &$\ell$&-- &--&--&--       \\
\hline
\end{tabular}
\caption{All possible conformal primaries in $S_3\times S_3$ corresponding to the $(A,0)^{[0020]}_{\Delta,\ell}$ superconformal multiplet.}\label{A00020}
\end{table}

\begin{table}[htpb]
\centering
\begin{tabular}{|l||c|c|c|c|c|c|c|c|c|}
\hline 
$(\Delta,\ell)^{[0100]}_{(A,0)}$ spin: & \multicolumn{9}{c|}{\text{dimension}}  \\
\hline
$\mathfrak{so}(8)_R$ irrep&$\Delta$&$\Delta+1$&$\Delta+2$&$\Delta+3$&$\Delta+4$&$\Delta+5$&$\Delta+6$&$\Delta+7$&$\Delta+8$\\         
\hline\hline
\multirow{2}{*}{$[0000]=\bold{1}$ } &-- &$\ell\pm1$  &$\red{\ell\pm1}$ & $\ell\pm1,$          & $\red{\ell\pm1},$         &$\ell\pm1$   &$\red{\ell\pm1}$&$\ell\pm1$ &--       \\
& & & & $\ell\pm3$          & $\red{\ell\pm3}$         &$\ell\pm3$   && &       \\
\hline
\multirow{2}{*}{$[0100]=\bold{28}$ }  &$\ell$&$\red \ell$ &$\ell,\ell\pm2$ & $\red \ell,\red{\ell\pm2}$  & $\ell,\ell\pm2,$         & $\red \ell,\red{\ell\pm2}$   &$\ell,\ell\pm2$&$\red \ell$ &$\ell$       \\
& & &&   & $\ell\pm4$         &   && &       \\
\hline
\multirow{2}{*}{$[0020]=\bold{35}_c$ }   &&$\ell\pm1$ &$\red {\ell\pm1}$ & $\ell\pm1,$          & $\red{\ell\pm1},$        & $\ell\pm1,$   &$\red{\ell\pm1}$&$\ell\pm1$ &--        \\
&& & &  $\ell\pm3$         &   $\red{\ell\pm3}$     & $\ell\pm3$  && &       \\
\hline
\multirow{2}{*}{$[0200]=\bold{300}$}   &-- &$\ell\pm1$ &$\red{\ell\pm1}$ & $\ell\pm1,$          & $\red{\ell\pm1},$         & $\ell\pm1,$  &$\red{\ell\pm1}$&$\ell\pm1$&--       \\
& & & & $\ell\pm3$          & $\red{\ell\pm3}$         & $\ell\pm3$  &&&       \\
\hline
{$[0120]=\bold{567}_c$}    &-- &--&$\ell,\ell\pm2$        & $\red \ell,\red{\ell\pm2}$         & $\ell,\ell\pm2$   &$\red \ell,\red{\ell\pm2}$&$\ell,\ell\pm2$&--&--      \\   
\hline
{$[0040]=\bold{294}_c$ }   &-- &-- &--  &$\ell\pm1$     & $\red{\ell\pm1}$   &$\ell\pm1$&--&--&--        \\
 \hline
{$[0300]=\bold{1925}$   }  &-- &--        & $\ell$     & $\red \ell$   &$\ell,\ell\pm2$&$\red \ell$  &$\ell$&--&--       \\
\hline
{$[0220]=\bold{4312}_c$  }   &-- & --        & --        & $\ell\pm1$   &$\red{\ell\pm1}$&$\ell\pm1$  &-- &--&--      \\
\hline
{$[0140]=\bold{3696}_c$ }    &-- & --     & --        &--  &$\ell$&-- &--&--&--       \\
\hline
\end{tabular}
\caption{All possible conformal primaries in $\cO_{\mathbf{294}_c}\times \cO_{\mathbf{294}_c}$ corresponding to the $(\Delta,\ell)^{[0100]}_{(A,0)}$ superconformal multiplet.}\label{A00100}
\end{table}

\pagebreak

\section{1d correlation functions}

\subsection{Gaussian correlators}
\label{1dGaussianCorrelators}
In this section, we list the correlation functions of the 1d theory at fixed $\sigma$ for any value of $N$.  These are used to compute OPE coefficients of short multiplets in ABJM$^\text{int}_{N,1}$, listed in Table~\ref{ABJMvalues}.  All operators $\cO(\vphi_i,y_i)$ are ordered on $S^1$ in ascending fashion with $\vphi_1 < \vphi_2 < \cdots < \vphi_n$. For notational simplicity, we strip off the polarization dependence from all expressions, which can be restored through \eqref{TwoPointGen} and \eqref{ThreePtGen}, and we work in units where $\ell =1$. It follows that all correlators can be subsequently written in terms of ten quantities,
\small
\begin{align}
f_1 &= \sum _{\alpha _1=1}^N \sum _{\alpha _2=1}^N 
   t_{\alpha _1,\alpha _2}^2 \,, &
f_2 &= \sum _{\alpha _1=1}^N \sum _{\alpha _2=1}^N 
   t_{\alpha _1,\alpha _2}^4 \,,  \nonumber \\
f_3 &= \sum
   _{\alpha _1=1}^N \sum _{\alpha _2=1}^N \sum
   _{\alpha _3=1}^N  t_{\alpha _1,\alpha _2}
   t_{\alpha _1,\alpha _3} \,, &
f_4 &= \sum _{\alpha _1=1}^N \sum _{\alpha _2=1}^N \sum
   _{\alpha _3=1}^N t_{\alpha _1,\alpha _2}^2
   t_{\alpha _1,\alpha _3}^2 \,,  \nonumber \\
f_5 &= \sum _{\alpha
   _1=1}^N \sum _{\alpha _2=1}^N \sum _{\alpha
   _3=1}^N t_{\alpha _1,\alpha
   _2}^2 t_{\alpha _1,\alpha _3} t_{\alpha
   _2,\alpha _3} \,, &
f_6 &= \sum _{\alpha _1=1}^N \sum _{\alpha _2=1}^N \sum
   _{\alpha _3=1}^N t^3_{\alpha _1,\alpha _2} t_{\alpha _1,\alpha _3} \,,   \\
f_7 &= \sum
   _{\alpha _1=1}^N \sum _{\alpha _2=1}^N  \sum
   _{\alpha _3=1}^N \sum _{\alpha _4=1}^N t_{\alpha _1,\alpha _3} t_{\alpha _1,\alpha _4}
   t_{\alpha _2,\alpha _3} t_{\alpha _2,\alpha _4} \,,  &
f_8 &= \sum
   _{\alpha _1=1}^N \sum _{\alpha _2=1}^N  \sum
   _{\alpha _3=1}^N \sum _{\alpha _4=1}^N t_{\alpha _1,\alpha _2}^2 t_{\alpha _1,\alpha _3}
   t_{\alpha _1,\alpha _4} \,,  \nonumber \\
f_9 &=  \sum
   _{\alpha _1=1}^N \sum _{\alpha _2=1}^N \sum
   _{\alpha _3=1}^N \sum
   _{\alpha _4=1}^N t_{\alpha _1,\alpha _2}^2 t_{\alpha
   _1,\alpha _3} t_{\alpha _2,\alpha _4} \,, &
f_{10} &= \sum
   _{\alpha _1=1}^N \sum _{\alpha _2=1}^N \sum _{\alpha _3=1}^N  \sum _{\alpha _4=1}^N  t_{\alpha _1,\alpha _2}^2 t_{\alpha _1,\alpha _3}
   t_{\alpha _1,\alpha _4} t_{\alpha _2,\alpha _3}
   t_{\alpha _2,\alpha _4} \nonumber \,,
\end{align}
\normalsize
where $t_{\alpha, \beta} \equiv \tanh(\pi(\sigma_\alpha - \sigma_\beta))$.

For example, the two-point functions of external operators \eqref{sampleABJMCorr} can be rewritten as
\es{sampleABJMCorrRewritten}{
\langle \cO_2 \cO_2 \rangle_\sigma &= \frac{N^2-1}{2 \ell ^2} - \frac{1}{2\ell^2} f_1  \,, \\
\langle \cO_3 \cO_3 \rangle_\sigma &= \frac{3
   \left(N^4-5 N^2+4\right)}{8 N \ell ^3} + \frac{9}{4 N \ell ^3} f_1 -\frac{9}{8 \ell ^3} f_3 \,.
}

We now list the correlation functions according to type. Those related to constructing an orthogonal basis of operators are:
\small
\es{02001pt}{
\langle \cO''_4 \rangle_\sigma &= 
B_{4,0}'' -\frac{3}{2 \ell ^2} f_1 
\,, \\
\langle \cO_2 \cO''_6 \rangle_\sigma &= \frac{\left(N^2-1\right)
   B_{6,2}''}{2 \ell ^2}-\frac{ \left(N \ell ^2
   B_{6,2}''-5\right)}{2 N \ell ^4}f_1  -\frac{5 }{4 \ell ^4} f_3 + \frac{5 }{4 \ell ^4} f_5 
\,, \\
\langle \cO_{4,1} \cO_{4,2} \rangle_\sigma &= 
\frac{\left(N^2-1\right) \left(\left(N^3+N\right)
   B_{4,4}+2 N^2-3\right)}{2 N \ell ^4} -\frac{\left(\left(N^3+N\right) B_{4,4}+2
   N^2-3\right)}{N \ell ^4}f_1 \\
&+ \frac{B_{4,4}}{\ell ^4}f_2 + \frac{1}{\ell ^4} f_4 + \frac{B_{4,4}}{2 \ell ^4} f_1^2 
\,.
}
The two-point functions of internal operators with $\Delta=j_F$ are:
\es{2ptDeltaEqualsjF}{
\langle \cO_{4,1} \cO_{4,1} \rangle_\sigma &=\frac{N^4-1}{2 \ell
   ^4} - \frac{N^2+1}{\ell^4} f_1 + \frac{1}{\ell^4}f_2 + \frac{1}{2 \ell ^4} f_1^2 \,, \\
\langle \cO_{4,2} \cO_{4,2} \rangle_\sigma &= \frac{\left(N^2-1\right) \left(2 N B_{4,4}
   \left(\left(N^3+N\right) B_{4,4}+4
   N^2-6\right)+N^4-6 N^2+18\right)}{4 N^2 \ell ^4} \\
&-\frac{2 N B_{4,4} \left(\left(N^3+N\right)
   B_{4,4}+4 N^2-6\right)+3 \left(N^2+6\right)}{2 N^2 \ell ^4}f_1 + \frac{4 B_{4,4}^2+1}{4 \ell ^4}f_2
\\&-\frac{\left(N^2-9\right)}{N \ell ^4} f_3 + \frac{2 B_{4,4}}{\ell ^4} f_4 + \frac{B_{4,4}^2}{2 \ell ^4}f_1^2 + \frac{1}{4 \ell ^4}f_7 \,, \\
\langle \cO_5 \cO_5 \rangle_\sigma &= \frac{3
   \left(N^6-21 N^2+20\right)}{16 N \ell ^5} 
   -\frac{3 \left(N^2-10\right) \left(N^2+5\right)}{16 N
   \ell ^5} f_1 -\frac{9}{2 N \ell ^5} f_2
   -\frac{9 \left(N^2+5\right)}{16 \ell ^5} f_3 \\&+ \frac{9}{8 \ell ^5} f_4
   +\frac{9}{4 \ell ^5} f_6 + \frac{9}{8 \ell ^5} f_5 -\frac{9}{4 N \ell ^5} f_1^2 + \frac{9}{16 \ell ^5} f_1 f_3 \,, \\
\langle \cO_6 \cO_6 \rangle_\sigma &= 
\frac{9 \left(N^2-4\right)^2 \left(N^4+7
   N^2-8\right)}{32 N^2 \ell ^6} + \frac{27 \left(7 N^4+16 N^2-128\right)}{32 N^2 \ell
   ^6} f_1  + \frac{81 \left(N^2+16\right)}{32 N^2 \ell ^6}  f_2 \\&-\frac{27 \left(N^4+7 N^2-32\right)}{16 N \ell ^6} f_3 -\frac{81 \left(N^2+4\right)}{16 N \ell ^6} f_5 -\frac{81}{4 N \ell ^6} f_4 
   -\frac{81}{2 N \ell ^6} f_6 +\frac{81}{4 N^2 \ell ^6}f_1^2 \\&- \frac{81 }{16 \ell ^6} f_9 + \frac{81}{16 \ell ^6} f_8 +\frac{81 }{32 \ell ^6} f_7   + \frac{81 }{32 \ell ^6} f_{10}  + \frac{81}{32 \ell ^6} f_3^2 -\frac{81}{8 N \ell
   ^6}f_1 f_3 
\,.
}
\normalsize
The single two-point function of internal operators with $\Delta = j_F -1$ is:
\small
\es{2ptDeltaEqualsjFMinusOne}{
\langle \cO'_5 \cO'_5 \rangle_\sigma &= \frac{5 N \left(N^4-5 N^2+4\right)}{32 \ell ^5} -\frac{5 N \left(N^2-5\right)}{32 \ell ^5} f_1 -\frac{5}{8 N \ell ^5} f_2 + \frac{5 \left(10-3 N^2\right)}{32 \ell ^5} f_3 + \frac{15}{8 \ell ^5} f_6 \\&-\frac{35}{16 \ell ^5} f_4 + \frac{65}{16 \ell ^5} f_5 -\frac{5 }{16 N \ell ^5} f_1^2 + \frac{15}{32 \ell ^5} f_1 f_3 \,.\\
}
\normalsize
Those with $\Delta = j_F - 2$ are:
\small
\es{2ptDeltaEqualsjFMinusTwo}{
\langle \cO''_4 \cO''_4 \rangle_\sigma &= \left(B_{4,0}''\right){}^2 + \frac{3 \left(N^4-3
   N^2+2\right)}{2 \ell ^4}  - \frac{3 \left(\ell ^2 B_{4,0}''+N^2+1\right)}{\ell
   ^4} f_1 + \frac{15}{2 \ell ^4} f_2 + \frac{15}{4\ell^4} f_1^2 \,, \\
\langle \cO''_5 \cO''_5 \rangle_\sigma &= \frac{3 \left(N^6-8 N^4+19 N^2-12\right)}{8 N \ell
   ^5} -\frac{3 \left(N^4-5 N^2-18\right)}{8 N \ell ^5} f_1 -\frac{15}{N \ell ^5} f_2 + \frac{3 \left(1-3 N^2\right)}{8 \ell ^5} f_3 \\
&+ \frac{9}{2 \ell ^5} f_6 + \frac{33}{4 \ell ^5} f_4 -\frac{15}{4 \ell ^5} f_5  -\frac{15}{2 N \ell ^5} f_1^2 + \frac{9}{8 \ell ^5} f_1 f_3 
\,, \\
\langle \cO''_6 \cO''_6 \rangle_\sigma &= 
\frac{5 \left(N^2-4\right)^2 \left(N^4-3
   N^2+2\right)}{16 N^2 \ell ^6} + \frac{\left(N^2-1\right)
   \left(B_{6,2}''\right){}^2}{2 \ell ^2} + \frac{45 \left(N^2+16\right)}{16 N^2 \ell ^6} f_2
 \\&+ \frac{5 \left(N^4+8 N^2-64\right)-8 N \ell ^2
   B_{6,2}'' \left(N \ell ^2 B_{6,2}''-10\right)}{16
   N^2 \ell ^6} f_1 + \frac{45}{2 N^2 \ell ^6} f_1^2 +\frac{45}{16 \ell ^6} f_7 \\
&-\frac{5  \left(4 N \ell ^2 B_{6,2}''+3 N^4-19
   N^2+24\right)}{8 N \ell ^6} f_3 -\frac{45}{8 \ell ^6} f_9 + \frac{45}{8 \ell ^6} f_8 -\frac{105}{16 \ell ^6} f_{10} + \frac{45}{16 \ell ^6} f_3^2 \\
&-\frac{45}{4 N \ell ^6} f_1 f_3 -\frac{45}{2 N \ell ^6}f_4 -\frac{45}{N \ell ^6} f_6 + \frac{5 \left(4 N \ell ^2 B_{6,2}''+11
   N^2-36\right)}{8 N \ell ^6} f_5
\,.
}
\normalsize 

Lastly we list the three-point functions. Those involving $\cO_2 \times \cO_2$ are:
\small
\es{3ptO2O2}{
\langle \cO_2 \cO_2 \cO_2 \rangle_\sigma &= 
\frac{1-N^2}{\ell ^3} + \frac{1}{\ell^3} f_1 
\,, \\
\langle \cO_2 \cO_2 \cO_{4,1} \rangle_\sigma &=
\frac{N^4-1}{2 \ell^4} + \frac{1}{\ell^4} f_2 - \frac{1+N^2}{\ell^4} f_1 + \frac{1}{2 \ell ^4} f_1^2  
\,, \\
\langle \cO_2 \cO_2 \cO''_4 \rangle_\sigma &=
\frac{\left(N^2-1\right) \left(\ell ^2
   B_{4,0}''+N^2-2\right)}{2 \ell ^4} -\frac{\left(2 \ell ^2
   B_{4,0}''+7
   N^2+1\right)}{4 \ell ^4} f_1 + \frac{10}{4 \ell ^4} f_2 +  \frac{5}{4 \ell
   ^4}  f_1^2 
\,.
}
\normalsize
Those involving $\cO_2 \times \cO_3$ are:
\small
\es{3ptO2O3}{
\langle \cO_2 \cO_3 \cO_3 \rangle_\sigma &= 
-\frac{9 \left(N^4-5 N^2+4\right)}{8 N \ell ^4} + \frac{9
   \left(N^2-6\right)}{8 N \ell ^4} f_1 + \frac{9}{8 \ell^4} f_3 - \frac{9}{8 \ell^4} f_5 
\,, \\
\langle \cO_2 \cO_3 \cO_5 \rangle_\sigma &= 
\frac{3 \left(N^6-21 N^2+20\right)}{16 N \ell ^5} 
-\frac{3 \left(N^2-10\right) \left(N^2+5\right)}{16 N \ell ^5} f_1 - \frac{9}{2
   N \ell ^5} f_2 
\\&-\frac{9 \left(N^2+5\right)}{16 \ell ^5} f_3
   + \frac{9}{4 \ell ^5} f_6 + \frac{9}{8 \ell ^5} f_4 + \frac{9}{8 \ell ^5} f_5
-\frac{9}{4 N \ell ^5} f_1^2 + \frac{9}{16 \ell ^5} f_1 f_3 
\,, \\
\langle \cO_2 \cO_3 \cO'_5 \rangle_\sigma &= 
\frac{3 N \left(N^4-5 N^2+4\right)}{16 \ell ^5} -\frac{3 N \left(N^2-20\right)}{16 \ell ^5} f_1 -\frac{3}{4 N \ell ^5} f_2 -\frac{3 \left(3 N^2+20\right) }{16 \ell ^5} f_3 \\&+ \frac{9}{4 \ell ^5} f_6 -\frac{21}{8 \ell ^5} f_4 + \frac{33}{16 \ell ^5} f_5  -\frac{3}{8 N \ell ^5} f_1^2 +\frac{9}{16 \ell ^5} f_1 f_3 
\,, \\
\langle \cO_2 \cO_3 \cO''_5 \rangle_\sigma &= 
\frac{3 \left(N^6-8 N^4+19 N^2-12\right)}{16 N \ell
   ^5} -\frac{3 \left(N^4-5 N^2-18\right)}{16 N \ell ^5} f_1 -\frac{15}{2 N \ell ^5} f_2 \\&+  \frac{3 \left(1-3 N^2\right)}{16 \ell ^5} f_3 + \frac{9}{4 \ell ^5} f_6 + \frac{33}{8 \ell ^5} f_4 -\frac{15}{8 \ell ^5} f_5 -\frac{15}{4 N \ell ^5} f_1^2 +\frac{9}{16 \ell ^5} f_1 f_3 
 \,.
}
\normalsize
Those involving $\cO_3 \times \cO_3$ are:
\small
\es{3ptO3O3}{
\langle \cO_3 \cO_3 \cO_2 \rangle_\sigma &= 
-\frac{9 \left(N^4-5 N^2+4\right)}{8 N \ell ^4} + \frac{9
   \left(N^2-6\right) }{8 N
   \ell ^4} f_1 +\frac{9}{8 \ell
   ^4} f_3 - \frac{9}{8 \ell
   ^4} f_5 
\,, \\
\langle \cO_3 \cO_3 \cO_{4,1} \rangle_\sigma &=
\frac{9 \left(N^4-5
   N^2+4\right)}{4 N \ell ^5} -\frac{9(N^2-4)}{2 N \ell ^5} f_1 -\frac{9}{2 N \ell ^5} f_2 + \frac{9}{2 \ell ^5} f_4 -\frac{9}{4 N \ell ^5} f_1^2 
\,, \\
\langle \cO_3 \cO_3 \cO_{4,2} \rangle_\sigma &= 
\frac{9 \left(N^4-5 N^2+4\right) \left(2 N
   B_{4,4}+N^2-6\right)}{8 N^2 \ell ^5} + \frac{9 \left(-2 \left(N^2-4\right) N
   B_{4,4}+N^2-24\right)}{4
   N^2 \ell ^5} f_1 \\
&+ \frac{9 \left(N-4 B_{4,4}\right)}{8 N \ell ^5} f_2 -\frac{9 \left(N^2-9\right)}{2 N \ell ^5} f_3 + \frac{9 \left(N B_{4,4}-1\right)}{2 N \ell ^5} f_4 -\frac{9 B_{4,4}}{4 N \ell ^5} f_1^2 + \frac{9}{8 \ell ^5} f_7 
\,, \\
\langle \cO_3 \cO_3 \cO''_4 \rangle_\sigma &= 
\frac{3 \left(N^4-5
   N^2+4\right) \left(\ell ^2 B_{4,0}''-3\right)}{8 N
   \ell ^5}
   + \frac{9 \left(4 \ell ^2
   B_{4,0}''-N^4+9 N^2+4\right)}{16 N \ell ^5} f_1 
   \\&-\frac{45}{4 N \ell ^5} f_2   
   -\frac{9  \left(\ell ^2 B_{4,0}''+6\right)}{8 \ell ^5} f_3 + \frac{9}{8 \ell ^5} f_4 + \frac{27}{4 \ell ^5} f_6
   - \frac{45}{8N \ell^5} f_1^2 +\frac{27}{16 \ell ^5} f_1 f_3
\,, \\
\langle \cO_3 \cO_3 \cO_6 \rangle_\sigma &= 
-\frac{9 \left(N^2-4\right)^2 \left(N^4+7
   N^2-8\right)}{32 N^2 \ell ^6}-\frac{27 \left(7 N^4+16 N^2-128\right)}{32 N^2 \ell ^6} f_1
   -\frac{81 \left(N^2+16\right)}{32 N^2 \ell ^6} f_2 \\&+ \frac{27 \left(N^4+7 N^2-32\right)}{16 N \ell
   ^6} f_3 + \frac{81}{2 N \ell ^6} f_6 + \frac{81}{4 N \ell ^6} f_4 -\frac{81}{16 \ell ^6} f_8 + \frac{81}{16 \ell ^6} f_9 + \frac{81 \left(N^2+4\right)}{16 N \ell ^6} f_5
   \\&-\frac{81}{4 N^2 \ell ^6} f_1^2 -\frac{81}{32 \ell ^6} f_7 + \frac{81}{8 N \ell ^6} f_1 f_3 -\frac{81}{32 \ell ^6} f_3^2 -\frac{81}{32 \ell ^6} f_{10} 
\,, \\
\langle \cO_3 \cO_3 \cO''_6 \rangle_\sigma &= 
-\frac{9 \left(N^4-5 N^2+4\right) \left(4 N \ell ^2
   B_{6,2}''+N^4-6 N^2+8\right)}{32 N^2 \ell ^6}-\frac{81 \left(N^2+16\right)}{32 N^2 \ell ^6} f_2 \\&-\frac{9 \left(-4
   \left(N^2-6\right) N \ell ^2 B_{6,2}''+N^4+28
   N^2-144\right)}{32 N^2 \ell ^6} f_1 + \frac{81}{2 N \ell ^6} f_6 + \frac{81}{4 N \ell ^6} f_4 \\&+ \frac{9 \left(2 N \ell ^2 B_{6,2}''+3 N^4-14
   N^2+4\right)}{16 N \ell ^6} f_3 -\frac{81}{4 N^2 \ell ^6} f_1^2
   -\frac{81}{32 \ell ^6} f_7 + \frac{81}{16 \ell ^6} f_9 \\&- \frac{81}{16 \ell ^6} f_8 + \frac{189}{32 \ell ^6} f_{10} + \frac{81}{8 N \ell ^6} f_1 f_3 -\frac{81}{32 \ell ^6} f_3^2 -\frac{9 \left(N \ell ^2 B_{6,2}''+8 N^2-28\right)}{8
   N \ell ^6} f_5
\,.
}
\normalsize

\subsection{Additional relations} 
\label{extraRel}

Correlation functions in the 1d sector of ABJM$^\text{int}_{N,1}$, related to the Gaussian correlators from the previous section by \eqref{1dCorrelators}, satisfy additional relations that can be derived without explicit integration. In particular, all of the three-point functions in the mixed correlator, except for $\langle \cO_3 \cO_3 \cO_{4,1} \rangle$, can be written in terms of associated two-point functions. Up to the $\mathfrak{su}(2)_F$ prefactor, these constraints are given by
\begin{align}\label{extra1dRelations}
\langle \cO_2 \cO_2 \cO_2 \rangle &= -\frac{2}{\ell} \langle \cO_2 \cO_2 \rangle \,, &
\langle \cO_2 \cO_2 \cO_{4,1} \rangle &= \langle \cO_{4,1} \cO_{4,1} \rangle  \,, &
\langle \cO_2 \cO_2 \cO''_4 \rangle &= \frac13 \langle \cO''_4 \cO''_4 \rangle \,, \nonumber \\
\langle \cO_3 \cO_3 \cO_2 \rangle &= -\frac{3}{\ell} \langle \cO_3 \cO_3 \rangle \,, &
\langle \cO_3 \cO_3 \cO_{4,2} \rangle &= \frac{9}{2\ell} \langle \cO_{4,2} \cO_{4,2} \rangle \,, &
\langle \cO_3 \cO_3 \cO''_4 \rangle &= -\frac{3}{8\ell} \langle \cO''_4 \cO''_4 \rangle \,, \nonumber \\
\langle \cO_3 \cO_3 \cO_6 \rangle &= - \langle \cO_6 \cO_6 \rangle \,, &
\langle \cO_3 \cO_3 \cO''_6 \rangle &= \langle \cO''_6 \cO''_6 \rangle \,, &
\langle \cO_2 \cO_3 \cO_3 \rangle &= -\frac{3}{\ell} \langle \cO_3 \cO_3 \rangle \,, \nonumber \\
\langle \cO_2 \cO_3 \cO_5 \rangle &= \langle \cO_5 \cO_5 \rangle \,, & \langle \cO_2 \cO_3 \cO'_5 \rangle &= \frac{12}{10} \langle \cO'_5 \cO'_5 \rangle \,, &
\langle \cO_2 \cO_3 \cO''_5 \rangle &= \frac12 \langle \cO''_5 \cO''_5 \rangle \,.
\end{align}

Along with the 1d crossing equations \eqref{1dCrossing}, the constraints in \eqref{extra1dRelations} can be used to rewrite all of the 3d OPE coefficients in terms of $\lambda_1 \equiv \lambda^2_{22(B,+)_{1,0}^{[0020]}}$ and $\lambda_2 \equiv \lambda^2_{22(B,+)^{[0040]}_{2,0}}$, which were computed to all orders in $1/N$ in \cite{Agmon:2017xes}, as well as three correlators, which we can choose to be $b_3 \equiv B_{\cO_3}$, $b_6 \equiv B_{\cO_6}$, and $c_4 \equiv C_{\cO_3 \cO_3 \cO_{4,1}}$. In terms of these quantities, the OPE coefficients can be written as
\es{3dOPEcoeffsSimp}{
\lambda^2_{22(B,2)_{2,0}^{[0200]}}  &= 5 \lambda _2^2-4 \left(\lambda _1^2+4\right) \,, \\
\lambda^2_{33(B,+)_{1,0}^{[0020]}}  &= \frac{9 \lambda _1^2}{4} \,, \\
\lambda^2_{33(B,+)_{2,0}^{[0040],1}}  &= \frac{c_4^2 \lambda _1^4}{9 b_3^2 \lambda _2^2} \,, \\
\lambda^2_{33(B,+)_{2,0}^{[0040],2}}&=\frac{-144 b_3^2 \lambda _2^2 \lambda _1^4+72 b_6
   \lambda _2^2 \lambda _1^4-c_4^2 \lambda _1^8+2916
   \lambda _2^2 \lambda _1^2-3645 \lambda _2^4+11664
   \lambda _2^2}{9 b_3^2 \lambda _1^4 \lambda _2^2} \,, \\
\lambda^2_{33(B,2)_{2,0}^{[0200]}} &= -\frac{81 \left(4 \lambda _1^2-5 \lambda
   _2^2+16\right)}{b_3^2 \lambda _1^4} \,, \\
\lambda^2_{33(B,+)_{3,0}^{[0060]}} &= \frac{16 b_6}{5 b_3^2} \,, \\
\lambda^2_{33(B,2)_{3,0}^{[0220]}} &= -\frac{36 \left(5 b_3^2 \lambda _1^2-4 b_6\right)}{5
   b_3^2} \,, \\
\lambda^2_{23(B,+)_{\frac32,0}^{[0030]}}  &= 3 \lambda _1^2 \,, \\
\lambda^2_{23(B,+)_{\frac52,0}^{[0050]}} &=\frac{18 b_3 \lambda _1^4+48 b_3 \lambda _1^2+c_4
   \lambda _1^4-108 \lambda _1^2+135 \lambda
   _2^2-432}{15 b_3 \lambda _1^2} \,, \\
\lambda^2_{23(B,2)_{\frac52,0}^{[0130]}} &= \frac{4 \left(33 b_3 \lambda _1^4+48 b_3 \lambda
   _1^2-4 c_4 \lambda _1^4-108 \lambda _1^2+135
   \lambda _2^2-432\right)}{15 b_3 \lambda _1^2} \,, \\
\lambda^2_{23(B,2)_{\frac52,0}^{[0210]}} &= \frac{-30 b_3 \lambda _1^4+48 b_3 \lambda _1^2+5 c_4
   \lambda _1^4-108 \lambda _1^2+135 \lambda
   _2^2-432}{3 b_3 \lambda _1^2} \,.
}

\section{Crossing functions}
\label{Vs}

Here we give the explicit crossing function $\vec V$ obtained by equating the 4-point functions in \eqref{crossings} and using the expressions for  $Y^{0,0}_{nm}(\sigma,\tau)$, $Y^{-1,-1}_{nm}(\sigma,\tau)$, and $Y^{1,1}_{nm}(\sigma,\tau)$ in Appendix \ref{Ys}. These crossing equations are written in the $A^{k_{12},k_{34}}_{nm}(U,V)$ basis in \eqref{Ybasis} using the crossing functions $F_{\pm,nm}^{k_{1}k_2k_{3}k_4}(U,V)$ defined in \eqref{Fs}, and take the form:

 \abovedisplayskip=-0.7cm
   
\es{V}{
\vspace{-.5cm}
\vec V=\begin{pmatrix} F^{2222}_{-,2,0}+ F^{2222}_{-,2,1}+ F^{2222}_{-,2,2}\\
 F^{2222}_{-,1,1}+\frac
   {4}{3}  F^{2222}_{-,2,1}+\frac{8}{3}
    F^{2222}_{-,2,2}\\
    F^{2222}_{-,1,0}+\frac{3}{5}  F^{2222}_{-,2,1}+3
    F^{2222}_{-,2,2}\\
    F^{2222}_{-,0,0}-\frac{12}{7}
    F^{2222}_{-,2,1}+\frac{24}{35}
    F^{2222}_{-,2,2}\\
    F^{2222}_{+,1,0}+ F^{2222}_{+,1,1}+\frac{5}{3}
    F^{2222}_{+,2,0}-\frac{2}{5}  F^{2222}_{+,2,1}-\frac{14}{3}
    F^{2222}_{+,2,2}\\
    F^{2222}_{+,0,0}-\frac{1}{4}
    F^{2222}_{+,1,1}-\frac{20}{21}
    F^{2222}_{+,2,0}+ F^{2222}_{+,2,1}-\frac{14}{15}
    F^{2222}_{+,2,2}\\
    F^{3333}_{-,3,0}+ F^{3333}_{-,3,1}+ F^{3333}_{-,3,2}+F^{3333}_{-,3,3}\\
    F^{3333}_{-,2,1}+3  F^{3333}_{-,2,2}+\frac{25}{12}
    F^{3333}_{-,3,1}+\frac{165}{28}  F^{3333}_{-,3,2}+\frac{75}{8}
    F^{3333}_{-,3,3}\\
    F^{3333}_{-,2,0}-2  F^{3333}_{-,2,2}-\frac{1}{9}
    F^{3333}_{-,3,1}+\frac{9}{4}  F^{3333}_{-,3,3}\\
    F^{3333}_{-,1,1}+\frac{5}{3}
    F^{3333}_{-,2,2}-\frac{125}{81}  F^{3333}_{-,3,1}+\frac{375}{56}
    F^{3333}_{-,3,3}\\
    F^{3333}_{-,1,0}-\frac{9}{5}  F^{3333}_{-,2,2}-\frac{5}{3}
    F^{3333}_{-,3,1}-\frac{11}{3}  F^{3333}_{-,3,2}+\frac{95}{24}
    F^{3333}_{-,3,3}\\
    F^{3333}_{-,0,0}+\frac{81}{140}
    F^{3333}_{-,2,2}+\frac{45}{28}  F^{3333}_{-,3,1}+\frac{795}{224}
    F^{3333}_{-,3,3}\\
    F^{3333}_{+,2,0}+ F^{3333}_{+,2,1}+ F^{3333}_{+,2,2}+\frac{9}{
   4}  F^{3333}_{+,3,0}+\frac{2}{9}  F^{3333}_{+,3,1}-\frac{27}{7}
    F^{3333}_{+,3,2}-\frac{81}{8}  F^{3333}_{+,3,3}\\
    F^{3333}_{+,1,1}+\frac{4}{3}
    F^{3333}_{+,2,1}-\frac{1}{3}  F^{3333}_{+,2,2}+\frac{100}{81}
    F^{3333}_{+,3,1}-\frac{15}{7}  F^{3333}_{+,3,2}-\frac{675}{56}
    F^{3333}_{+,3,3}\\
    F^{3333}_{+,1,0}-\frac{7}{5}
    F^{3333}_{+,2,1}-\frac{25}{12}  F^{3333}_{+,3,0}+4
    F^{3333}_{+,3,2}\\
    F^{3333}_{+,0,0}+\frac{2}{7}
    F^{3333}_{+,2,1}-\frac{9}{140}  F^{3333}_{+,2,2}+\frac{25}{28}
    F^{3333}_{+,3,0}-\frac{5}{7}  F^{3333}_{+,3,1}+\frac{127}{196}
    F^{3333}_{+,3,2}-\frac{795}{224}
    F^{3333}_{+,3,3}\\
    F^{2323}_{-,\frac{5}{2},\frac{1}{2}}+F^{2323}_{-,\frac{5}{2},\frac{3}{2}}+ F^{2323}_{-,\frac{5}{2
   },\frac{5}{2}}\\
    F^{2323}_{-,\frac{3}{2},\frac{3}{2}}+
   \frac{32}{27}
    F^{2323}_{-,\frac{5}{2},\frac{3}{2}}+\frac{128}{63}
    F^{2323}_{-,\frac{5}{2},\frac{5}{2}}\\
    F^{2323}_{-,\frac{3}{2},\frac{1}{2}}+\frac{47}{27}
    F^{2323}_{-,\frac{5}{2},\frac{3}{2}}+\frac{152}{27}
    F^{2323}_{-,\frac{5}{2},\frac{5}{2}}\\
    F^{2323}_{-,\frac{1}{2},\frac{1}{2}}-\frac{81}{35}
    F^{2323}_{-,\frac{5}{2},\frac{3}{2}}+\frac{129}{35}
    F^{2323}_{-,\frac{5}{2},\frac{5}{2}}\\
    F^{2323}_{+,\frac{3}{2},\frac{1}{2}}+ F^{2323}_{+,\frac{3}{2},\frac{3}{2}}+\frac{40}{27}
    F^{2323}_{+,\frac{5}{2},\frac{1}{2}}-\frac{43}{27}
    F^{2323}_{+,\frac{5}{2},\frac{3}{2}}-\frac{48}{7}
    F^{2323}_{+,\frac{5}{2},\frac{5}{2}}\\
    F^{2323}_{+,\frac{1}{2},\frac{1}{2}}-\frac{9}{35}
    F^{2323}_{+,\frac{3}{2},\frac{3}{2}}-\frac{5}{7}
    F^{2323}_{+,\frac{5}{2},\frac{1}{2}}+\frac{38}{21}
    F^{2323}_{+,\frac{5}{2},\frac{3}{2}}-\frac{188}{49}
    F^{2323}_{+,\frac{5}{2},\frac{5}{2}}\\
    F^{2233}_{-,2,2}+\frac{
   1}{6}  F^{3223}_{-,\frac{1}{2},\frac{1}{2}}-\frac{4}{21}
    F^{3223}_{-,\frac{3}{2},\frac{1}{2}}+\frac{4}{15}
    F^{3223}_{-,\frac{3}{2},\frac{3}{2}}+\frac{25}{162}
    F^{3223}_{-,\frac{5}{2},\frac{1}{2}}-\frac{20}{81}
    F^{3223}_{-,\frac{5}{2},\frac{3}{2}}+\frac{5}{21}
    F^{3223}_{-,\frac{5}{2},\frac{5}{2}}\\
    F^{2233}_{-,2,1}-\frac{
   1}{2}  F^{3223}_{-,\frac{1}{2},\frac{1}{2}}+\frac{1}{14}
    F^{3223}_{-,\frac{3}{2},\frac{1}{2}}+\frac{1}{5}
    F^{3223}_{-,\frac{3}{2},\frac{3}{2}}+\frac{25}{54}
    F^{3223}_{-,\frac{5}{2},\frac{1}{2}}-\frac{55}{54}
    F^{3223}_{-,\frac{5}{2},\frac{3}{2}}+\frac{10}{7}
    F^{3223}_{-,\frac{5}{2},\frac{5}{2}}\\
    F^{2233}_{-,2,0}+\frac{
   1}{3}  F^{3223}_{-,\frac{1}{2},\frac{1}{2}}+\frac{5}{42}
    F^{3223}_{-,\frac{3}{2},\frac{1}{2}}-\frac{7}{15}
    F^{3223}_{-,\frac{3}{2},\frac{3}{2}}+\frac{31}{81}
    F^{3223}_{-,\frac{5}{2},\frac{1}{2}}-\frac{119}{162}
    F^{3223}_{-,\frac{5}{2},\frac{3}{2}}+\frac{4}{3}
    F^{3223}_{-,\frac{5}{2},\frac{5}{2}}\\
    F^{2233}_{-,1,1}+\frac{
   2}{9}  F^{3223}_{-,\frac{1}{2},\frac{1}{2}}+\frac{26}{63}
    F^{3223}_{-,\frac{3}{2},\frac{1}{2}}+\frac{47}{90}
    F^{3223}_{-,\frac{3}{2},\frac{3}{2}}-\frac{250}{243}
    F^{3223}_{-,\frac{5}{2},\frac{1}{2}}-\frac{50}{243}
    F^{3223}_{-,\frac{5}{2},\frac{3}{2}}+\frac{200}{63}
    F^{3223}_{-,\frac{5}{2},\frac{5}{2}}\\
    F^{2233}_{-,1,0}-\frac{
   1}{5}  F^{3223}_{-,\frac{1}{2},\frac{1}{2}}-\frac{33}{70}
    F^{3223}_{-,\frac{3}{2},\frac{1}{2}}-\frac{21}{50}
    F^{3223}_{-,\frac{3}{2},\frac{3}{2}}-\frac{5}{9}
    F^{3223}_{-,\frac{5}{2},\frac{1}{2}}+\frac{7}{18}
    F^{3223}_{-,\frac{5}{2},\frac{3}{2}}+4
    F^{3223}_{-,\frac{5}{2},\frac{5}{2}}\\
    F^{2233}_{-,0,0}+\frac{
   1}{35}  F^{3223}_{-,\frac{1}{2},\frac{1}{2}}+\frac{27}{245}
    F^{3223}_{-,\frac{3}{2},\frac{1}{2}}+\frac{27}{200}
    F^{3223}_{-,\frac{3}{2},\frac{3}{2}}+\frac{5}{7}
    F^{3223}_{-,\frac{5}{2},\frac{1}{2}}+ F^{3223}_{-,\frac{5}{2},\frac{3}{2}}+\frac{12}{7}
    F^{3223}_{-,\frac{5}{2},\frac{5}{2}}\\
    F^{2233}_{+,2,2}-\frac{
   1}{6}  F^{3223}_{+,\frac{1}{2},\frac{1}{2}}+\frac{4}{21}
    F^{3223}_{+,\frac{3}{2},\frac{1}{2}}-\frac{4}{15}
    F^{3223}_{+,\frac{3}{2},\frac{3}{2}}-\frac{25}{162}
    F^{3223}_{+,\frac{5}{2},\frac{1}{2}}+\frac{20}{81}
    F^{3223}_{+,\frac{5}{2},\frac{3}{2}}-\frac{5}{21}
    F^{3223}_{+,\frac{5}{2},\frac{5}{2}}\\
    F^{2233}_{+,2,1}+\frac{
   1}{2}  F^{3223}_{+,\frac{1}{2},\frac{1}{2}}-\frac{1}{14}
    F^{3223}_{+,\frac{3}{2},\frac{1}{2}}-\frac{1}{5}
    F^{3223}_{+,\frac{3}{2},\frac{3}{2}}-\frac{25}{54}
    F^{3223}_{+,\frac{5}{2},\frac{1}{2}}+\frac{55}{54}
    F^{3223}_{+,\frac{5}{2},\frac{3}{2}}-\frac{10}{7}
    F^{3223}_{+,\frac{5}{2},\frac{5}{2}}\\
    F^{2233}_{+,2,0}-\frac{
   1}{3}  F^{3223}_{+,\frac{1}{2},\frac{1}{2}}-\frac{5}{42}
    F^{3223}_{+,\frac{3}{2},\frac{1}{2}}+\frac{7}{15}
    F^{3223}_{+,\frac{3}{2},\frac{3}{2}}-\frac{31}{81}
    F^{3223}_{+,\frac{5}{2},\frac{1}{2}}+\frac{119}{162}
    F^{3223}_{+,\frac{5}{2},\frac{3}{2}}-\frac{4}{3}
    F^{3223}_{+,\frac{5}{2},\frac{5}{2}}\\
    F^{2233}_{+,1,1}-\frac{
   2}{9}  F^{3223}_{+,\frac{1}{2},\frac{1}{2}}-\frac{26}{63}
    F^{3223}_{+,\frac{3}{2},\frac{1}{2}}-\frac{47}{90}
    F^{3223}_{+,\frac{3}{2},\frac{3}{2}}+\frac{250}{243}
    F^{3223}_{+,\frac{5}{2},\frac{1}{2}}+\frac{50}{243}
    F^{3223}_{+,\frac{5}{2},\frac{3}{2}}-\frac{200}{63}
    F^{3223}_{+,\frac{5}{2},\frac{5}{2}}\\
    F^{2233}_{+,1,0}+\frac{
   1}{5}  F^{3223}_{+,\frac{1}{2},\frac{1}{2}}+\frac{33}{70}
    F^{3223}_{+,\frac{3}{2},\frac{1}{2}}+\frac{21}{50}
    F^{3223}_{+,\frac{3}{2},\frac{3}{2}}+\frac{5}{9}
    F^{3223}_{+,\frac{5}{2},\frac{1}{2}}-\frac{7}{18}
    F^{3223}_{+,\frac{5}{2},\frac{3}{2}}-4
    F^{3223}_{+,\frac{5}{2},\frac{5}{2}}\\
    F^{2233}_{+,0,0}-\frac{
   1}{35}  F^{3223}_{+,\frac{1}{2},\frac{1}{2}}-\frac{27}{245}
    F^{3223}_{+,\frac{3}{2},\frac{1}{2}}-\frac{27}{200}
    F^{3223}_{+,\frac{3}{2},\frac{3}{2}}-\frac{5}{7}
    F^{3223}_{+,\frac{5}{2},\frac{1}{2}}- F^{3223}_{+,\frac{5}{2},\frac{3}{2}}-\frac{12}{7}
    F^{3223}_{+,\frac{5}{2},\frac{5}{2}}
   \end{pmatrix}.
   }
   
   \abovedisplayskip=0.43cm

The crossing equations for the superconformal block decomposition of the 4-point functions are obtained by replacing each $A_{nm}^{k_{12},k_{34}}$ in $F_{\pm,nm}^{k_{1}k_2k_{3}k_4}(U,V)$ by the linear combination of conformal blocks that appear for the superblock $\mathfrak{G}^{k_{12},k_{34}}_{\mathcal{M}}$ for each supermultiplet $\mathcal{M}$. For instance, we use the explicit expression for the stress tensor multiplet $(B,+)_{1,0}^{[0020]}$ superblock in \eqref{stressBlock} to write $\vec V_{(B,+)_{1,0}^{[0020]}}$ as 
\es{stressV}{
\vec {\bf V}_{(B,+)_{1,0}^{[0020]}}=\begin{pmatrix}
0\\
 {\bf F}^{2222}_{-,1, 0}\\
  -{\bf F}^{2222}_{-,2, 1}\\
   \frac14 {\bf F}^{2222}_{-,3, 2}\\ 
 {\bf F}^{2222}_{+,1, 0} - {\bf F}^{2222}_{+,2, 1}\\
  -\frac14 {\bf F}^{2222}_{+,1, 0} + \frac14 {\bf F}^{2222}_{+,3, 2}\\
   \vec 0_3\\
 {\bf F}^{3333}_{-,1, 0}\\
  -{\bf F}^{3333}_{-,2, 1}\\
   \frac14 {\bf F}^{3333}_{-,3, 2}\\
    0\\
     {\bf F}^{3333}_{+,1, 0}\\
      -{\bf F}^{3333}_{+,2, 1}\\ 
 \frac14 {\bf F}^{3333}_{+,3, 2}\\
  \vec 0_9\\
   {\bf F}^{2233}_{-,1, 0}\\
    -{\bf F}^{2233}_{-,2, 1}\\
 \frac14 {\bf F}^{2233}_{-,3, 2}\\
  \vec 0_3\\
   {\bf F}^{2233}_{+,1, 0}\\
    -{\bf F}^{2233}_{+,2, 1}\\
     \frac14 {\bf F}^{2233}_{+,3, 2}\\
 \end{pmatrix}\,,
}
where $\vec 0_a$ denotes vectors of length $a$ with entries equal to the $2\times 2$ zero matrix, and we define
 \es{bfFDef}{
  {\bf F}^{2222}_{\pm nm} 
  \equiv \begin{pmatrix}
   {\cal F}^{2222}_{\pm nm} & 0 \\
   0 & 0 
  \end{pmatrix} \,, \qquad
   {\bf F}^{3333}_{\pm nm} 
  \equiv \begin{pmatrix}
   0 & 0 \\
   0 & {\cal F}^{3333}_{\pm nm}
  \end{pmatrix} \,, \qquad
   {\bf F}^{2233}_{\pm nm} 
  \equiv \frac 12 \begin{pmatrix}
   0 & {\cal F}^{2233}_{\pm nm} \\
   {\cal F}^{2233}_{\pm nm} & 0
  \end{pmatrix}
 }
with $\mathcal{F}_{\pm,nm}^{k_1k_2 k_3 k_4}(U,V)$ written in terms of conformal blocks as
\es{Fs2}{
\mathcal{F}_{\pm,\Delta,\ell}^{k_{1}k_2k_{3}k_4}(U,V)\equiv V^{\frac{k_2+k_3}{2}}G_{\Delta,\ell}^{k_{12},k_{34}}(U,V)\pm U^{\frac{k_2+k_3}{2}}G_{\Delta,\ell}^{k_{12},k_{34}}(V,U)\,.
}

In the main text, we then used the 1d crossing relations and the conformal Ward identity constraints \eqref{opetoCT} to write all the short OPE coefficients in terms of $1/c_T$, $\lambda^2_{kk(B,2)_{2,0}^{[0200]}}$, $\lambda^2_{(B,+)_{3,0}^{[0060]}}$, $\lambda^2_{(B,+)_{\frac52,0}^{[0050]}}$, $\lambda^2_{(B,2)_{\frac52,0}^{[0130]}}$. The resulting explicit crossing functions are then:

\es{matToscal}{
\vec V_\text{Id}&\equiv \begin{pmatrix} 1&1\end{pmatrix} \vec{{\bf V}}_\text{Id}\begin{pmatrix}1\\1 \end{pmatrix}+\begin{pmatrix} \frac{16}{5}&-16\end{pmatrix} \vec{{\bf V}}_{(B,2)_{2,0}^{[0200]}}\begin{pmatrix}-16\\-16 \end{pmatrix}\,,\\
 \vec V_{(B,+)_{1,0}^{[0020]}}&\equiv \begin{pmatrix}16& 24\end{pmatrix}\vec {\bf V}_{(B,+)_{1,0}^{[0020]}} \begin{pmatrix} 16 \\  24 \end{pmatrix}+ \begin{pmatrix}\frac{1024}{5}& 256\end{pmatrix}\vec {\bf V}_{(B,+)_{2,0}^{[0040]}} \begin{pmatrix} 256 \\  0 \end{pmatrix}\\
 &-1792 \begin{pmatrix}0& 1\end{pmatrix}\vec {\bf V}_{(B,2)_{2,0}^{[0200]}} \begin{pmatrix} 1 \\  0 \end{pmatrix}+768 \vec V_{(B,+)_{\frac32,0}^{[0030]}}-9216 \vec V_{(B,2)_{3,0}^{[0220]}}-3072 \vec V_{(B,2)_{\frac52,0}^{[0210]}}\,,\\
   \vec V^{2222}_{(B,2)_{2,0}^{[0200]}}&\equiv \begin{pmatrix}1& 0\end{pmatrix}\vec {\bf V}_{(B,2)_{2,0}^{[0200]}} \begin{pmatrix} 0 \\  0 \end{pmatrix}+\frac15 \begin{pmatrix}1& 0\end{pmatrix}\vec {\bf V}_{(B,+)_{2,0}^{[0040]}} \begin{pmatrix} 0 \\  0 \end{pmatrix}\,,\\
     \vec V^{3333}_{(B,2)_{2,0}^{[0200]}}&\equiv \begin{pmatrix}0& 0\end{pmatrix}\vec {\bf V}_{(B,2)_{2,0}^{[0200]}} \begin{pmatrix} 0 \\  1 \end{pmatrix}- \begin{pmatrix}1& 0\end{pmatrix}\vec {\bf V}_{(B,+)_{2,0}^{[0040]}} \begin{pmatrix} 0 \\  0 \end{pmatrix}\,,\\
       \vec V'_{(B,+)_{3,0}^{[0060]}}&\equiv \vec { V}_{(B,+)_{3,0}^{[0060]}}+9\vec { V}_{(B,2)_{3,0}^{[0220]}}+\frac52\begin{pmatrix}0& 0\end{pmatrix}\vec {\bf V}_{(B,+)_{2,0}^{[0040]}} \begin{pmatrix} 0 \\  1 \end{pmatrix}\,,\\
         \vec V'_{(B,+)_{\frac52,0}^{[0050]}}&\equiv \vec { V}_{(B,+)_{\frac52,0}^{[0050]}}+9\vec { V}_{(B,2)_{\frac52,0}^{[0210]}}+\begin{pmatrix}0& 1\end{pmatrix}\vec {\bf V}_{(B,+)_{2,0}^{[0040]}} \begin{pmatrix} 1 \\  0 \end{pmatrix}+4\begin{pmatrix}0& 1\end{pmatrix}\vec {\bf V}_{(B,2)_{2,0}^{[0200]}} \begin{pmatrix} 1 \\  0 \end{pmatrix}\,,\\
             \vec V'_{(B,2)_{\frac52,0}^{[0130]}}&\equiv \vec { V}_{(B,2)_{\frac52,0}^{[0130]}}-\vec { V}_{(B,2)_{\frac52,0}^{[0210]}}-\frac14\begin{pmatrix}0& 1\end{pmatrix}\vec {\bf V}_{(B,+)_{2,0}^{[0040]}} \begin{pmatrix} 1 \\  0 \end{pmatrix}+\frac14\begin{pmatrix}0& 1\end{pmatrix}\vec {\bf V}_{(B,2)_{2,0}^{[0200]}} \begin{pmatrix} 1 \\  0 \end{pmatrix}\,.\\
}
 
\bibliographystyle{ssg}
\bibliography{BootstrapMix}

\begingroup\raggedright\begin{thebibliography}{10}

\bibitem{Polyakov:1974gs}
A.~Polyakov, ``{Nonhamiltonian approach to conformal quantum field theory},''
  {\em Zh.Eksp.Teor.Fiz.} {\bf 66} (1974) 23--42.

\bibitem{Ferrara:1973yt}
S.~Ferrara, A.~Grillo, and R.~Gatto, ``{Tensor representations of conformal
  algebra and conformally covariant operator product expansion},'' {\em Annals
  Phys.} {\bf 76} (1973) 161--188.

\bibitem{Mack:1975jr}
G.~Mack, ``{Duality in Quantum Field Theory},'' {\em Nucl.Phys.} {\bf B118}
  (1977) 445.

\bibitem{Rattazzi:2008pe}
R.~Rattazzi, V.~S. Rychkov, E.~Tonni, and A.~Vichi, ``{Bounding scalar operator
  dimensions in 4D CFT},'' {\em JHEP} {\bf 0812} (2008) 031,
  \href{http://xxx.lanl.gov/abs/0807.0004}{{\tt 0807.0004}}.

\bibitem{Rychkov:2016iqz}
S.~Rychkov, {\em {EPFL Lectures on Conformal Field Theory in D>= 3
  Dimensions}}.
\newblock SpringerBriefs in Physics. 2016.

\bibitem{Simmons-Duffin:2016gjk}
D.~Simmons-Duffin, ``{The Conformal Bootstrap},'' in {\em {Proceedings,
  Theoretical Advanced Study Institute in Elementary Particle Physics: New
  Frontiers in Fields and Strings (TASI 2015): Boulder, CO, USA, June 1-26,
  2015}}, pp.~1--74, 2017.
\newblock \href{http://xxx.lanl.gov/abs/1602.07982}{{\tt 1602.07982}}.

\bibitem{Poland:2018epd}
D.~Poland, S.~Rychkov, and A.~Vichi, ``{The Conformal Bootstrap: Theory,
  Numerical Techniques, and Applications},'' {\em Rev. Mod. Phys.} {\bf 91}
  (2019), no.~1 15002, \href{http://xxx.lanl.gov/abs/1805.04405}{{\tt
  1805.04405}}. [Rev. Mod. Phys.91,015002(2019)].

\bibitem{Chester:2019wfx}
S.~M. Chester, ``{Weizmann Lectures on the Numerical Conformal Bootstrap},''
  \href{http://xxx.lanl.gov/abs/1907.05147}{{\tt 1907.05147}}.

\bibitem{Qualls:2015qjb}
J.~D. Qualls, ``{Lectures on Conformal Field Theory},''
  \href{http://xxx.lanl.gov/abs/1511.04074}{{\tt 1511.04074}}.

\bibitem{Kos:2014bka}
F.~Kos, D.~Poland, and D.~Simmons-Duffin, ``{Bootstrapping Mixed Correlators in
  the 3D Ising Model},'' {\em JHEP} {\bf 11} (2014) 109,
  \href{http://xxx.lanl.gov/abs/1406.4858}{{\tt 1406.4858}}.

\bibitem{Simmons-Duffin:2015qma}
D.~Simmons-Duffin, ``{A Semidefinite Program Solver for the Conformal
  Bootstrap},'' {\em JHEP} {\bf 06} (2015) 174,
  \href{http://xxx.lanl.gov/abs/1502.02033}{{\tt 1502.02033}}.

\bibitem{Kos:2016ysd}
F.~Kos, D.~Poland, D.~Simmons-Duffin, and A.~Vichi, ``{Precision Islands in the
  Ising and $O(N)$ Models},'' {\em JHEP} {\bf 08} (2016) 036,
  \href{http://xxx.lanl.gov/abs/1603.04436}{{\tt 1603.04436}}.

\bibitem{Kos:2015mba}
F.~Kos, D.~Poland, D.~Simmons-Duffin, and A.~Vichi, ``{Bootstrapping the O(N)
  Archipelago},'' {\em JHEP} {\bf 11} (2015) 106,
  \href{http://xxx.lanl.gov/abs/1504.07997}{{\tt 1504.07997}}.

\bibitem{Rong:2018okz}
J.~Rong and N.~Su, ``{Bootstrapping minimal $\mathcal{N}=1$ superconformal
  field theory in three dimensions},''
  \href{http://xxx.lanl.gov/abs/1807.04434}{{\tt 1807.04434}}.

\bibitem{Atanasov:2018kqw}
A.~Atanasov, A.~Hillman, and D.~Poland, ``{Bootstrapping the Minimal 3D
  SCFT},'' {\em JHEP} {\bf 11} (2018) 140,
  \href{http://xxx.lanl.gov/abs/1807.05702}{{\tt 1807.05702}}.

\bibitem{Li:2017kck}
Z.~Li and N.~Su, ``{3D CFT Archipelago from Single Correlator Bootstrap},''
  \href{http://xxx.lanl.gov/abs/1706.06960}{{\tt 1706.06960}}.

\bibitem{Li:2016wdp}
Z.~Li and N.~Su, ``{Bootstrapping Mixed Correlators in the Five Dimensional
  Critical O(N) Models},'' {\em JHEP} {\bf 04} (2017) 098,
  \href{http://xxx.lanl.gov/abs/1607.07077}{{\tt 1607.07077}}.

\bibitem{Kousvos:2018rhl}
S.~R. Kousvos and A.~Stergiou, ``{Bootstrapping Mixed Correlators in
  Three-Dimensional Cubic Theories},'' {\em SciPost Phys.} {\bf 6} (2019),
  no.~3 035, \href{http://xxx.lanl.gov/abs/1810.10015}{{\tt 1810.10015}}.

\bibitem{Poland:2011ey}
D.~Poland, D.~Simmons-Duffin, and A.~Vichi, ``{Carving Out the Space of 4D
  CFTs},'' {\em JHEP} {\bf 1205} (2012) 110,
  \href{http://xxx.lanl.gov/abs/1109.5176}{{\tt 1109.5176}}.

\bibitem{Bobev:2015jxa}
N.~Bobev, S.~El-Showk, D.~Mazac, and M.~F. Paulos, ``{Bootstrapping SCFTs with
  Four Supercharges},'' {\em JHEP} {\bf 08} (2015) 142,
  \href{http://xxx.lanl.gov/abs/1503.02081}{{\tt 1503.02081}}.

\bibitem{Baggio:2017mas}
M.~Baggio, N.~Bobev, S.~M. Chester, E.~Lauria, and S.~S. Pufu, ``{Decoding a
  Three-Dimensional Conformal Manifold},'' {\em JHEP} {\bf 02} (2018) 062,
  \href{http://xxx.lanl.gov/abs/1712.02698}{{\tt 1712.02698}}.

\bibitem{Chester:2015lej}
S.~M. Chester, L.~V. Iliesiu, S.~S. Pufu, and R.~Yacoby, ``{Bootstrapping
  $O(N)$ Vector Models with Four Supercharges in $3 \leq d \leq4$},'' {\em
  JHEP} {\bf 05} (2016) 103, \href{http://xxx.lanl.gov/abs/1511.07552}{{\tt
  1511.07552}}.

\bibitem{Berkooz:2014yda}
M.~Berkooz, R.~Yacoby, and A.~Zait, ``{Bounds on $\mathcal{N}=1$ Superconformal
  Theories with Global Symmetries},''
  \href{http://xxx.lanl.gov/abs/1402.6068}{{\tt 1402.6068}}.

\bibitem{Li:2017ddj}
D.~Li, D.~Meltzer, and A.~Stergiou, ``{Bootstrapping mixed correlators in 4D $
  \mathcal{N} $ = 1 SCFTs},'' {\em JHEP} {\bf 07} (2017) 029,
  \href{http://xxx.lanl.gov/abs/1702.00404}{{\tt 1702.00404}}.

\bibitem{Chang:2017xmr}
C.-M. Chang and Y.-H. Lin, ``{Carving Out the End of the World or
  (Superconformal Bootstrap in Six Dimensions)},'' {\em JHEP} {\bf 08} (2017)
  128, \href{http://xxx.lanl.gov/abs/1705.05392}{{\tt 1705.05392}}.

\bibitem{Chang:2017cdx}
C.-M. Chang, M.~Fluder, Y.-H. Lin, and Y.~Wang, ``{Spheres, Charges,
  Instantons, and Bootstrap: A Five-Dimensional Odyssey},''
  \href{http://xxx.lanl.gov/abs/1710.08418}{{\tt 1710.08418}}.

\bibitem{Beem:2015aoa}
C.~Beem, M.~Lemos, L.~Rastelli, and B.~C. van Rees, ``{The (2, 0)
  superconformal bootstrap},'' {\em Phys. Rev.} {\bf D93} (2016), no.~2 025016,
  \href{http://xxx.lanl.gov/abs/1507.05637}{{\tt 1507.05637}}.

\bibitem{Chester:2015qca}
S.~M. Chester, S.~Giombi, L.~V. Iliesiu, I.~R. Klebanov, S.~S. Pufu, and
  R.~Yacoby, ``{Accidental Symmetries and the Conformal Bootstrap},'' {\em
  JHEP} {\bf 01} (2016) 110, \href{http://xxx.lanl.gov/abs/1507.04424}{{\tt
  1507.04424}}.

\bibitem{Beem:2014zpa}
C.~Beem, M.~Lemos, P.~Liendo, L.~Rastelli, and B.~C. van Rees, ``{The $
  \mathcal{N}=2 $ superconformal bootstrap},'' {\em JHEP} {\bf 03} (2016) 183,
  \href{http://xxx.lanl.gov/abs/1412.7541}{{\tt 1412.7541}}.

\bibitem{Lemos:2015awa}
M.~Lemos and P.~Liendo, ``{Bootstrapping $ \mathcal{N}=2 $ chiral
  correlators},'' {\em JHEP} {\bf 01} (2016) 025,
  \href{http://xxx.lanl.gov/abs/1510.03866}{{\tt 1510.03866}}.

\bibitem{Cornagliotto:2017snu}
M.~Cornagliotto, M.~Lemos, and P.~Liendo, ``{Bootstrapping the $(A_1,A_2)$
  Argyres-Douglas theory},'' {\em JHEP} {\bf 03} (2018) 033,
  \href{http://xxx.lanl.gov/abs/1711.00016}{{\tt 1711.00016}}.

\bibitem{Liendo:2018ukf}
P.~Liendo, C.~Meneghelli, and V.~Mitev, ``{Bootstrapping the half-BPS line
  defect},'' {\em JHEP} {\bf 10} (2018) 077,
  \href{http://xxx.lanl.gov/abs/1806.01862}{{\tt 1806.01862}}.

\bibitem{Gimenez-Grau:2019hez}
A.~Gimenez-Grau and P.~Liendo, ``{Bootstrapping line defects in $\mathcal{N}=2$
  theories},'' \href{http://xxx.lanl.gov/abs/1907.04345}{{\tt 1907.04345}}.

\bibitem{Beem:2013sza}
C.~Beem, M.~Lemos, P.~Liendo, W.~Peelaers, L.~Rastelli, and B.~C. van Rees,
  ``{Infinite Chiral Symmetry in Four Dimensions},'' {\em Commun. Math. Phys.}
  {\bf 336} (2015), no.~3 1359--1433,
  \href{http://xxx.lanl.gov/abs/1312.5344}{{\tt 1312.5344}}.

\bibitem{Beem:2013qxa}
C.~Beem, L.~Rastelli, and B.~C. van Rees, ``{The $\mathcal N=4$ Superconformal
  Bootstrap},'' {\em Phys.Rev.Lett.} {\bf 111} (2013), no.~7 071601,
  \href{http://xxx.lanl.gov/abs/1304.1803}{{\tt 1304.1803}}.

\bibitem{Beem:2016wfs}
C.~Beem, L.~Rastelli, and B.~C. van Rees, ``{More ${\mathcal N}=4$
  superconformal bootstrap},'' {\em Phys. Rev.} {\bf D96} (2017), no.~4 046014,
  \href{http://xxx.lanl.gov/abs/1612.02363}{{\tt 1612.02363}}.

\bibitem{Lemos:2016xke}
M.~Lemos, P.~Liendo, C.~Meneghelli, and V.~Mitev, ``{Bootstrapping
  $\mathcal{N}=3$ superconformal theories},'' {\em JHEP} {\bf 04} (2017) 032,
  \href{http://xxx.lanl.gov/abs/1612.01536}{{\tt 1612.01536}}.

\bibitem{Chester:2014mea}
S.~M. Chester, J.~Lee, S.~S. Pufu, and R.~Yacoby, ``{Exact Correlators of BPS
  Operators from the 3d Superconformal Bootstrap},'' {\em JHEP} {\bf 03} (2015)
  130, \href{http://xxx.lanl.gov/abs/1412.0334}{{\tt 1412.0334}}.

\bibitem{Chester:2014fya}
S.~M. Chester, J.~Lee, S.~S. Pufu, and R.~Yacoby, ``{The $ \mathcal{N}=8 $
  superconformal bootstrap in three dimensions},'' {\em JHEP} {\bf 09} (2014)
  143, \href{http://xxx.lanl.gov/abs/1406.4814}{{\tt 1406.4814}}.

\bibitem{Dedushenko:2016jxl}
M.~Dedushenko, S.~S. Pufu, and R.~Yacoby, ``{A one-dimensional theory for Higgs
  branch operators},'' \href{http://xxx.lanl.gov/abs/1610.00740}{{\tt
  1610.00740}}.

\bibitem{VanRaamsdonk:2008ft}
M.~Van~Raamsdonk, ``{Comments on the Bagger-Lambert theory and multiple
  M2-branes},'' {\em JHEP} {\bf 0805} (2008) 105,
  \href{http://xxx.lanl.gov/abs/0803.3803}{{\tt 0803.3803}}.

\bibitem{Bandres:2008vf}
M.~A. Bandres, A.~E. Lipstein, and J.~H. Schwarz, ``{${\cal N} = 8$
  Superconformal Chern-Simons Theories},'' {\em JHEP} {\bf 0805} (2008) 025,
  \href{http://xxx.lanl.gov/abs/0803.3242}{{\tt 0803.3242}}.

\bibitem{Bagger:2007vi}
J.~Bagger and N.~Lambert, ``{Comments on multiple M2-branes},'' {\em JHEP} {\bf
  0802} (2008) 105, \href{http://xxx.lanl.gov/abs/0712.3738}{{\tt 0712.3738}}.

\bibitem{Bagger:2007jr}
J.~Bagger and N.~Lambert, ``{Gauge symmetry and supersymmetry of multiple
  M2-branes},'' {\em Phys.Rev.} {\bf D77} (2008) 065008,
  \href{http://xxx.lanl.gov/abs/0711.0955}{{\tt 0711.0955}}.

\bibitem{Bagger:2006sk}
J.~Bagger and N.~Lambert, ``{Modeling Multiple M2's},'' {\em Phys.Rev.} {\bf
  D75} (2007) 045020, \href{http://xxx.lanl.gov/abs/hep-th/0611108}{{\tt
  hep-th/0611108}}.

\bibitem{Gustavsson:2007vu}
A.~Gustavsson, ``{Algebraic structures on parallel M2-branes},'' {\em
  Nucl.Phys.} {\bf B811} (2009) 66--76,
  \href{http://xxx.lanl.gov/abs/0709.1260}{{\tt 0709.1260}}.

\bibitem{Aharony:2008ug}
O.~Aharony, O.~Bergman, D.~L. Jafferis, and J.~Maldacena, ``{${\cal N}=6$
  superconformal Chern-Simons-matter theories, M2-branes and their gravity
  duals},'' {\em JHEP} {\bf 0810} (2008) 091,
  \href{http://xxx.lanl.gov/abs/0806.1218}{{\tt 0806.1218}}.

\bibitem{Bashkirov:2011pt}
D.~Bashkirov and A.~Kapustin, ``{Dualities between ${\cal N} = 8$
  superconformal field theories in three dimensions},'' {\em JHEP} {\bf 1105}
  (2011) 074, \href{http://xxx.lanl.gov/abs/1103.3548}{{\tt 1103.3548}}.

\bibitem{Aharony:2008gk}
O.~Aharony, O.~Bergman, and D.~L. Jafferis, ``{Fractional M2-branes},'' {\em
  JHEP} {\bf 0811} (2008) 043, \href{http://xxx.lanl.gov/abs/0807.4924}{{\tt
  0807.4924}}.

\bibitem{Agmon:2017lga}
N.~B. Agmon, S.~M. Chester, and S.~S. Pufu, ``{A New Duality Between
  $\mathcal{N}=8$ Superconformal Field Theories in Three Dimensions},''
  \href{http://xxx.lanl.gov/abs/1708.07861}{{\tt 1708.07861}}.

\bibitem{Lambert:2010ji}
N.~Lambert and C.~Papageorgakis, ``{Relating $U(N) \times U(N)$ to $SU(N)
  \times SU(N)$ Chern-Simons Membrane theories},'' {\em JHEP} {\bf 1004} (2010)
  104, \href{http://xxx.lanl.gov/abs/1001.4779}{{\tt 1001.4779}}.

\bibitem{Bashkirov:2012rf}
D.~Bashkirov, ``{BLG theories at low values of Chern-Simons coupling},''
  \href{http://xxx.lanl.gov/abs/1211.4887}{{\tt 1211.4887}}.

\bibitem{Gang:2011xp}
D.~Gang, E.~Koh, K.~Lee, and J.~Park, ``{ABCD of 3d ${\cal N}=8$ and 4
  Superconformal Field Theories},''
  \href{http://xxx.lanl.gov/abs/1108.3647}{{\tt 1108.3647}}.

\bibitem{Closset:2012ru}
C.~Closset, T.~T. Dumitrescu, G.~Festuccia, and Z.~Komargodski,
  ``{Supersymmetric Field Theories on Three-Manifolds},'' {\em JHEP} {\bf 1305}
  (2013) 017, \href{http://xxx.lanl.gov/abs/1212.3388}{{\tt 1212.3388}}.

\bibitem{Imamura:2011wg}
Y.~Imamura and D.~Yokoyama, ``{${\cal N}=2$ supersymmetric theories on squashed
  three-sphere},'' {\em Phys.Rev.} {\bf D85} (2012) 025015,
  \href{http://xxx.lanl.gov/abs/1109.4734}{{\tt 1109.4734}}.

\bibitem{Agmon:2017xes}
N.~B. Agmon, S.~M. Chester, and S.~S. Pufu, ``{Solving M-theory with the
  Conformal Bootstrap},'' {\em JHEP} {\bf 06} (2018) 159,
  \href{http://xxx.lanl.gov/abs/1711.07343}{{\tt 1711.07343}}.

\bibitem{Kapustin:2009kz}
A.~Kapustin, B.~Willett, and I.~Yaakov, ``{Exact Results for Wilson Loops in
  Superconformal Chern-Simons Theories with Matter},'' {\em JHEP} {\bf 1003}
  (2010) 089, \href{http://xxx.lanl.gov/abs/0909.4559}{{\tt 0909.4559}}.

\bibitem{Marino:2011eh}
M.~Marino and P.~Putrov, ``{ABJM theory as a Fermi gas},'' {\em J. Stat. Mech.}
  {\bf 1203} (2012) P03001, \href{http://xxx.lanl.gov/abs/1110.4066}{{\tt
  1110.4066}}.

\bibitem{Nosaka:2015iiw}
T.~Nosaka, ``{Instanton effects in ABJM theory with general R-charge
  assignments},'' {\em JHEP} {\bf 03} (2016) 059,
  \href{http://xxx.lanl.gov/abs/1512.02862}{{\tt 1512.02862}}.

\bibitem{Poland:2010wg}
D.~Poland and D.~Simmons-Duffin, ``{Bounds on 4D Conformal and Superconformal
  Field Theories},'' {\em JHEP} {\bf 1105} (2011) 017,
  \href{http://xxx.lanl.gov/abs/1009.2087}{{\tt 1009.2087}}.

\bibitem{ElShowk:2012hu}
S.~El-Showk and M.~F. Paulos, ``{Bootstrapping Conformal Field Theories with
  the Extremal Functional Method},'' {\em Phys.Rev.Lett.} {\bf 111} (2013),
  no.~24 241601, \href{http://xxx.lanl.gov/abs/1211.2810}{{\tt 1211.2810}}.

\bibitem{El-Showk:2014dwa}
S.~El-Showk, M.~F. Paulos, D.~Poland, S.~Rychkov, D.~Simmons-Duffin, and
  A.~Vichi, ``{Solving the 3d Ising Model with the Conformal Bootstrap II.
  c-Minimization and Precise Critical Exponents},'' {\em J. Stat. Phys.} {\bf
  157} (2014) 869, \href{http://xxx.lanl.gov/abs/1403.4545}{{\tt 1403.4545}}.

\bibitem{Zhou:2017zaw}
X.~Zhou, ``{On Superconformal Four-Point Mellin Amplitudes in Dimension
  $d>2$},'' {\em JHEP} {\bf 08} (2018) 187,
  \href{http://xxx.lanl.gov/abs/1712.02800}{{\tt 1712.02800}}.

\bibitem{Chester:2018lbz}
S.~M. Chester, ``{AdS$_{4}$/CFT$_{3}$ for unprotected operators},'' {\em JHEP}
  {\bf 07} (2018) 030, \href{http://xxx.lanl.gov/abs/1803.01379}{{\tt
  1803.01379}}. [JHEP01,030(2018)].

\bibitem{Nirschl:2004pa}
M.~Nirschl and H.~Osborn, ``{Superconformal Ward identities and their
  solution},'' {\em Nucl.Phys.} {\bf B711} (2005) 409--479,
  \href{http://xxx.lanl.gov/abs/hep-th/0407060}{{\tt hep-th/0407060}}.

\bibitem{Hogervorst:2013sma}
M.~Hogervorst and S.~Rychkov, ``{Radial Coordinates for Conformal Blocks},''
  {\em Phys.Rev.} {\bf D87} (2013), no.~10 106004,
  \href{http://xxx.lanl.gov/abs/1303.1111}{{\tt 1303.1111}}.

\bibitem{Dolan:2004mu}
F.~A. Dolan, L.~Gallot, and E.~Sokatchev, ``{On four-point functions of 1/2-BPS
  operators in general dimensions},'' {\em JHEP} {\bf 0409} (2004) 056,
  \href{http://xxx.lanl.gov/abs/hep-th/0405180}{{\tt hep-th/0405180}}.

\bibitem{Dolan:2008vc}
F.~Dolan, ``{On Superconformal Characters and Partition Functions in Three
  Dimensions},'' {\em J.Math.Phys.} {\bf 51} (2010) 022301,
  \href{http://xxx.lanl.gov/abs/0811.2740}{{\tt 0811.2740}}.

\bibitem{Maldacena:2011jn}
J.~Maldacena and A.~Zhiboedov, ``{Constraining Conformal Field Theories with A
  Higher Spin Symmetry},'' {\em J.Phys.} {\bf A46} (2013) 214011,
  \href{http://xxx.lanl.gov/abs/1112.1016}{{\tt 1112.1016}}.

\bibitem{Ferrara:2001uj}
S.~Ferrara and E.~Sokatchev, ``{Universal properties of superconformal OPEs for
  1/2 BPS operators in $3 \leq d \leq 6$},'' {\em New J.Phys.} {\bf 4} (2002)
  2, \href{http://xxx.lanl.gov/abs/hep-th/0110174}{{\tt hep-th/0110174}}.

\bibitem{Osborn:1993cr}
H.~Osborn and A.~Petkou, ``{Implications of conformal invariance in field
  theories for general dimensions},'' {\em Annals Phys.} {\bf 231} (1994)
  311--362, \href{http://xxx.lanl.gov/abs/hep-th/9307010}{{\tt
  hep-th/9307010}}.

\bibitem{Beem:2016cbd}
C.~Beem, W.~Peelaers, and L.~Rastelli, ``{Deformation quantization and
  superconformal symmetry in three dimensions},'' {\em Commun. Math. Phys.}
  {\bf 354} (2017), no.~1 345--392,
  \href{http://xxx.lanl.gov/abs/1601.05378}{{\tt 1601.05378}}.

\bibitem{Dedushenko:2017avn}
M.~Dedushenko, Y.~Fan, S.~S. Pufu, and R.~Yacoby, ``{Coulomb Branch Operators
  and Mirror Symmetry in Three Dimensions},'' {\em JHEP} {\bf 04} (2018) 037,
  \href{http://xxx.lanl.gov/abs/1712.09384}{{\tt 1712.09384}}.

\bibitem{Dedushenko:2018icp}
M.~Dedushenko, Y.~Fan, S.~S. Pufu, and R.~Yacoby, ``{Coulomb Branch
  Quantization and Abelianized Monopole Bubbling},''
  \href{http://xxx.lanl.gov/abs/1812.08788}{{\tt 1812.08788}}.

\bibitem{Chester:2018aca}
S.~M. Chester, S.~S. Pufu, and X.~Yin, ``{The M-Theory S-Matrix From ABJM:
  Beyond 11D Supergravity},'' {\em JHEP} {\bf 08} (2018) 115,
  \href{http://xxx.lanl.gov/abs/1804.00949}{{\tt 1804.00949}}.

\bibitem{Binder:2018yvd}
D.~J. Binder, S.~M. Chester, and S.~S. Pufu, ``{Absence of $D^4 R^4$ in
  M-Theory From ABJM},'' \href{http://xxx.lanl.gov/abs/1808.10554}{{\tt
  1808.10554}}.

\bibitem{Binder:2019mpb}
D.~J. Binder, S.~M. Chester, and S.~S. Pufu, ``{AdS$_4$/CFT$_3$ from Weak to
  Strong String Coupling},'' \href{http://xxx.lanl.gov/abs/1906.07195}{{\tt
  1906.07195}}.

\end{thebibliography}\endgroup

\end{document}